\documentclass[twocolumn]{aa}  
\graphicspath{{./}{figures/}}

\usepackage{graphicx}
\usepackage{txfonts}
\usepackage{lscape}
\usepackage{booktabs}
\usepackage{multirow}
\usepackage[figuresright]{rotating}
\usepackage{subfig}
\usepackage{caption}
\usepackage{booktabs}
\usepackage{longtable}
\usepackage{float}
\usepackage[colorlinks=true, linkcolor=blue, citecolor=blue]{hyperref}

\begin{document}


\title{Large-scale Velocity-coherent Filaments in the SEDIGISM Survey: 
Association with Spiral Arms and Fraction of Dense Gas}

   \author{Y.\,Ge
          \inst{1,2},
          K.\,Wang\inst{2}
          \fnmsep\thanks{Corresponding author.},
          A.\,Duarte-Cabral\inst{3},
          A.\,R.\,Pettitt\inst{4},
          C.\,L.\,Dobbs\inst{5},
          \'A.\,S\'anchez-Monge\inst{6},
          K.\,R.\,Neralwar\inst{7},
          J.\,S.\,Urquhart\inst{8},
          D.\,Colombo\inst{7},
          E.\,Durán-Camacho\inst{3},
          H.\,Beuther\inst{9},
          L.\,Bronfman\inst{10},
          A.\,J.\,Rigby\inst{11},
          D.\,Eden\inst{12},
          S.\,Neupane\inst{7},
          P.\,Barnes\inst{13,14},
          T.\,Henning\inst{9},
          A.\,Y.\,Yang\inst{15,7}
          }

   \institute{Department of Astronomy, School of Physics, Peking University, 5 Yiheyuan Road, Haidian District, Beijing 100871, China
         \and
             Kavli Institute for Astronomy and Astrophysics, Peking University, 5 Yiheyuan Road, Haidian District, Beijing 100871, China\\
             \email{kwang.astro@pku.edu.cn}
         \and
            School of Physics \& Astronomy, Cardiff University, Queen’s building, The parade, Cardiff, CF24 3AA, U.K.  
        \and 
            Department of Physics and Astronomy, California State University, Sacramento, 6000 J Street, Sacramento, CA 95819-6041, USA
        \and 
            School of Physics and Astronomy, University of Exeter, Stocker Road, Exeter, EX4 4QL, UK
        \and 
            Observatorio Astron\'omico Nacional (OAN, IGN), Calle Alfonso XII 3, 28014, Madrid, Spain
        \and 
            Max-Planck-Institut f\"ur Radioastronomie, Auf dem H\"ugel 69, 53121 Bonn, Germany
        \and 
            Centre for Astrophysics and Planetary Science, University of Kent, Canterbury, CT2\,7NH, UK
        \and 
            Max Planck Institute for Astronomy, K\"onigstuhl 17, 69117 Heidelberg, Germany
        \and 
            Departamento de Astronomía, Universidad de Chile, Casilla 36-D, Santiago, Chile
        \and 
            School of Physics and Astronomy, University of Leeds, Leeds LS2 9JT, UK
        \and 
            Armagh Observatory and Planetarium, College Hill, Armagh, BT61 9DB
        \and 
            Space Science Institute, 4765 Walnut St. Suite B, Boulder CO 80301, USA
        \and 
            School of Science and Technology, University of New England, Armidale NSW 2351, Australia
        \and 
            National Astronomical Observatories, Chinese Academy of Sciences, A20 Datun
            Road, Chaoyang District, Beijing 100101, People’s Republic of China
             }




 
  \abstract
   {Filamentary structures in the interstellar medium are closely related to star formation. Dense gas mass fraction (DGMF) or clump formation efficiency in large-scale filaments possibly determine their hosting star formation activities.}
   {We aim to automatically identify large-scale filaments, characterize them, investigate their association with Galactic structures, and study their DGMFs.}
   {We use a modified minimum spanning tree (MST) algorithm to chain parsec-scale $^{13}$CO clumps previously extracted from the SEDIGISM (Structure, Excitation, and Dynamics of the Inner Galactic InterStellar Medium) survey. The MST connects nodes in a graph such that the sum of edge lengths is minimum. Modified MST also ensures velocity coherence between nodes, so the identified filaments are coherent in position-position-velocity (PPV) space.}
   {We generate a catalog of 88 large-scale ($>10$ pc) filaments in the inner Galactic plane (with $-60^\circ <l<18^\circ$ and $|b|<0.5^\circ$). These SEDIGISM filaments are larger and less dense than MST filaments previously identified from the BGPS and ATLASGAL surveys. We find that eight of the filaments run along spiral arms and can be regarded as ``bones'' of the Milky Way. We also find three bones associated with the Local Spur in PPV space. By compiling 168 large-scale filaments with available DGMF across the Galaxy, an order of magnitude more than previously investigated, we find that DGMFs do not correlate with Galactic location, but bones have higher DGMFs than other filaments. 
   }
   {}


  \titlerunning{SEDIGISM Filaments}
  \authorrunning{Ge et al.}
  
   \maketitle

\section{Introduction}
   The interstellar medium (ISM) has a highly filamentary nature over a wide range of scales \citep[e.g.][]{Andre2010,Wang2015,Li2016,Arzoumanian2019,Schisano2020,Palmeirim2013,Soler2022}, and such filaments may play a vital role in star formation \citep[e.g.][]{LiuHB2012,Peretto2013,Peretto2014,Liu2016,YuanJH2018_G22,ZhangM2019,Xu2023,Zhang2023}. Theoretical works on filamentary structures can be traced back to \citet{Ostriker1964}, who employed a semi-analytic model to characterise an idealized infinite isothermal cylinder under hydrostatic equilibrium. Early observations on filaments, limited by resolution and sensitivity, focused mainly on nearby regions such as Taurus \citep{Schneider1979}. The Herschel Space Observatory \citep{Pilbratt2010} with higher sensitivity, spatial dynamic range, and angular resolution revolutionized the detailed study of filaments and revealed the ubiquity of filamentary structures throughout the ISM at near \citep[e.g.,][]{Andre2010,Arzoumanian2011} and far \citep[e.g.,][]{Molinari2010, Wang2015}.\\

   Among the different identified filamentary structures, the large-scale (80 pc) filament seen as a chain of infrared dark clouds (IRDC) ``Nessie'' was firstly reported by \citet{Jackson2010}. Follow-up observations of Nessie in a CO emission line revealed that it is potentially part of a larger 430 pc structure in the position-position-velocity (PPV) space lying close to the physical Galactic mid-plane \citep{Goodman2014}. The latter suggest that the Nessie IRDC may be a dense ``spine'' or ``bone'' of a section of the Scutum–Centaurus Arm. Since then, the existence of many more large-scale massive filaments has been discerned. Their lengths are tens to hundreds of parsecs with masses of up to the order of $10^6$ M$_\odot$. These filaments might be formed by large-scale phenomenon such as galactic shear, shock in spiral arms, and supernova (SN) feedback \citep[e.g.][]{Smith2014,Duarte2016,Smith2020}. Several systematic studies of large-scale filaments have been conducted in the past \citep{Ragan2014,Wang2015,Wang2016,Abreu2016,Zucker2015,Colombo2021,Ge2022}. The filaments are identified in a variety of wavelengths, from near-infrared, mid-infrared, far-infrared to submillimeter. Most of these filaments are detected in Galactic plane unbiased surveys, but they are usually broken into smaller structures by the finding algorithms of those studies \citep[e.g.][]{Schisano2014,Schisano2020,Koch2015,Li2016,Mattern2018}. The first identifications of such large-scale filaments were done mostly by-eye \citep{Ragan2014,Wang2015,Zucker2015,Abreu2016}, but more recently, more automated approaches have been developed. For instance, \citet{Wang2016,Ge2022} adopted the MST to automatically identify filaments by ``chaining'' dense parsec-scale clumps. \citet{Colombo2021} use a dendrogram analysis \citep{Rosolowsky2008} to isolate coherent structures and then consider long elongated ``trunks'' of the dendrogram as large-scale filaments. \citet{Zucker2018} analysed the properties of large-scale filaments in the inner Galaxy from different surveys homogeneously, and classify them to sub-samples which may indicate different formation mechanisms or histories. \citet{ZhangM2019} also studied large-scale filaments homogeneously, focusing on star-forming content, and found that the star formation rate (SFR) surface density and the star formation efficiency (SFE) in large-scale filaments are similar to those found in molecular clouds in general. In addition to filament catalogs, there are also studies on individual large-scale filaments such as Orion \citep[e.g.][]{Johnstone1999}, Taurus L1495 filament \citep[e.g.][]{Schmalzl2010}, Musca filament \citep[e.g.][]{Cox2016}, and other large-scale filamentary clouds or IRDCs \citep[e.g.][]{Battersby2014,Wang2014,Du2017,Sokolov2017,Sokolov2018,Sokolov2019,
   Hong-Li2018,Watkins2019,Tang2019,Lin2020,Guo2022,Veena2021,Clarke2023,Chen2023}.\\
  
   Star formation takes place in dense regions of molecular clouds, known as clumps. The fraction of dense gas in the molecular clouds has been linked to the SFR \citep[e.g.][]{Lada2012}. This fraction is often referred to as the dense gas mass fraction (DGMF) or clump formation efficiency (CFE). DGMF does not show any specific correlation with the total cloud mass, and the lack of dense gas will cause inefficiency of star formation \citep{Battisti2014}. \citet{Eden2013} find that the CFE of molecular clouds in the Milky Way shows no difference between the inter-arm and spiral-arm regions. They further infer that outside the Galactic center region, Galactic-scale structures do not play a significant role in the formation of dense, potentially star-forming structures within molecular clouds. In contrast, \citet{Torii2019} find that the molecular clouds within spiral arms have a higher fraction of dense gas than inter-arm clouds. For filaments, from the study of 163 large-scale filaments, \citet{Ge2022} find that DGMF in the spiral arms has no significant distinction from inter-arm filaments. \citet{ZhangM2019} also find that the star formation activity per gas mass in spiral arm and inter-arm environments are similar. However, some other studies find that DGMFs of filaments in spiral arms are higher than those of inter-arm ones \citep{Ragan2014,Abreu2016}. They suggest that star-formation activity in large-scale filaments depends on their location with respect to spiral arms. Note that these results are derived based on small numbers of filaments (9 and 16), and should be taken with caution, as pointed out by the authors \citep{Ragan2014,Abreu2016}. DGMFs of large-scale filaments are also found to have similar values compared to those in nearby galaxies \citep{Wang2020}.\\
 
   A minimum spanning tree (MST) is the unique set of straight lines (``edges'') connecting a given set of points (``nodes'') without closed loops, such that the sum of the edge lengths is minimum. MST has so far been widely used in astrophysics. It is employed to find large-scale distribution of galaxy or galaxy clusters, and filamentary features have been found \citep[e.g.][]{Barrow1985,AdamiMazure1999,Doroshkevich2004,Colberg2007,ParkLee2009,Alpaslan2014,Naidoo2020,Pereyra2020}. MST has also been used to identify star clusters \citep[e.g.][]{Cartwright2004,Schmeja2006,Gutermuth2009,Wu2017} and quantify core separations and mass segregation \citep[e.g.][]{Sanhueza2019,Dib2019}. The main advantage of the MST adopted by \citet{Wang2016} is that it considers an additional dimension, the velocity. Therefore, it is a clustering process in 3D (PPV) space, rather than a procedure purely based on 2D morphological characteristics. Using the MST method, \citet{Wang2016} identified and characterized 54 large-scale velocity-coherent filaments in the Bolocam Galactic Plane Survey (BGPS), and \citet{Ge2022} built a catalog of 163 filaments in the ATLASGAL survey. These works greatly increased the number of known large-scale filaments, and more importantly, introduced a physically driven definition of filaments \citep{Wang2016}, making it possible for statistically significant results to be derived, e.g., association with Galactic spiral arms, dense gas fraction, among other parameters. Both BGPS and ATLASGAL are dust continuum and thus primarily trace relatively dense gas. The recently published SEDIGISM cloud catalog extracted from $^{13}$CO emission \citep{DuarteCabral2021} provides us with high-quality nodes for MST. As such we are able to search for more diffuse large-scale filaments with comparatively lower densities. Moreover, the survey includes the region near the Galactic center (Galactic longitude $|l|<5^\circ$), which is not included in the works of \citet{Wang2016} and \citet{Ge2022} due to coverage of the input PPV catalogs they used.\\
    
   In this paper, we will identify large-scale filaments in the Galactic plane,investigate their association with Galactic structures, and study their DGMFs. The paper is structured as follows. We describe the data and methodology that we used to identify large-scale filaments in Sect. \ref{sec:method}. In Sect. \ref{sec:results}, we present our results, including physical properties of large-scale filaments and statistics. Then we compare our SEDIGISM filaments with two other MST filament catalogs (from BGPS and ATLASGAL) and other previously known large-scale filaments in Sect. \ref{discussion}. We also investigate the association between filaments and Galactic spiral arms, and examine DGMFs of filaments in different Galactic location as well. Finally, we summarize our results in Sect. \ref{sec:summary}.


\section{Data and Method}\label{sec:method}

\subsection{SEDIGISM molecular clouds} \label{sec_SEDIGISM}
\noindent
   The SEDIGISM survey used the APEX telescope to map the Galactic plane in $300^\circ <l< 18^\circ$ with $|b|\leq 0.5$ continuously and then a small region around W43 ($29^\circ <l< 31^\circ$ with $|b|\leq 0.5$) in several molecular transitions, including $^{13}$CO(2–1) and C$^{18}$O(2–1) with an angular resolution of 30 arcseconds \citep{Schuller2021}. Using $^{13}$CO(2–1) emission in the SEDIGISM spectral-line survey, \citet{DuarteCabral2021} extract the molecular cloud population with a large dynamic range in spatial scales. They use the Spectral Clustering for Interstellar Molecular Emission Segmentation (SCIMES) algorithm \citep{Colombo2015,Colombo2019} and compile a cloud catalog with a total of 10663 molecular clouds. In brief, firstly, they construct dendrograms\footnote{They use ASTRODENDRO based on the original IDL procedures from \citet{Rosolowsky2008}} from the preprocessed\footnote{They enhance the signal-to-noise ratio of the data set by smoothing the data in velocity and mask the datacubes using the local noise level.} SEDIGISM data cubes. Secondly, they identify gas clusters (cloud candidates) within the dendrograms using the SCIMES algorithm. Thirdly, they handle clouds in overlapping regions and remove spurious sources. Although they have performed a clustering analysis, the lack of large clouds nearby $(d<2.5$ kpc) indicates that they might be breaking nearby clouds into smaller substructures. Besides, despite the fact that \citet{Neralwar2022} classifies these clouds as filaments and non filaments, large filaments are found to be outliers. This may be due to the noise in the long filaments. That is, if a filament itself falls below the noise threshold in sections (that means this long filament is divided into pieces by several noisy parts), it is not continuous in emission. Then it is not part of the same dendrogram structure, and therefore SCIMES could never connect the two sections together. So limited by the data, if we intend to study large filamentary clouds, an alternative method is required. The MST method to be described in the next section provides us a way to deal with this limitation. In the SEDIGISM cloud catalog, most of the clouds (84\%, or 8945/10663) contain only a single dendrogram leaf \footnote{Prior to the clustering analysis of SCIMES, the dendrograms are constructed from the SEDIGISM datacubes. The dendrogram is made up of three different kinds of structures: the trunk, which is at the bottom of the hierarchy (i.e., it has no parent structure) and contains all branches and leaves; branches, which split into multiple substructures; and leaves, which are at the top of the hierarchy and contain no substructure because they are connected to local peaks of emission.}. A chain of such clouds (or leaves) in PPV space could be a structure on a grander scale. This catalog of SEDIGISM clouds, containing only one leaf (hereafter ``SEDIGISM leaves''), is an excellent data set for searching velocity-coherent filaments. \\
   
\subsection{MST Filament Identification} \label{sec_MST}
\noindent
  We identify large-scale velocity-coherent filaments using the modified minimum spanning tree algorithm\footnote{The code is available at \citet{Wang2021}, \href{https://ascl.net/2102.002}{https://ascl.net/2102.002}} \citep{Wang2016}. A spanning tree is defined as a network connecting all nodes on a graph. A MST connects all nodes so that the sum of edges is minimal and there are no loops on it. In the modified MST by \citet{Wang2016}, an additional dimension, velocity, is considered. When a node is to join the tree, it will be accepted only when it has a velocity similar to a nearby node in the tree. We consider SEDIGISM leaves described in Sect. \ref{sec_SEDIGISM}, as nodes in the algorithm. The criteria for MST matching and filament selection follow \citet{Wang2016,Ge2022}:
    \begin{itemize}
    \item[(1)]The accepted MST must contain at least five SEDIGISM leaves: $N \geq 5$.
    \item[(2)]Only edges shorter than a maximum length (cut-off length) can be connected $(\Delta L < 0.^\circ1 )$.
    \item[(3)]For any two leaves to be connected, the difference in line-of-sight velocity (matching velocity $\Delta v$) must be less than 2\,km\,s$^{-1}$.
    \item[(4)]Linearity $f_L > 1.6$. Here linearity is defined to quantify the degree of similarity between the target shape with a straight line shape.
    \item[(5)]Projected length (sum of edges) $L_{sum}\geq  10$ pc. We only focus on large-scale filaments in this work.
    \end{itemize}
   The value for N refers to the pruning level of MSTs in \citet{ParkLee2009} and \citet{Pereyra2020}. In their work, when a branch of an MST has fewer than five nodes, it is thought to be a minor branch and should be removed from the tree. The cut-off length is chosen because the observed angle for a filament with a length of 10 pc at a distance of 5 kpc is about $0^\circ .1$. The velocity difference between the two ends of a filament has an order of 1 km s$^{-1}$ if we treat the velocity gradient as 0.1 km s$^{-1}$ pc$^{-1}$. This order of velocity gradient is estimated by averaging the global velocity gradients of the filaments from several large-scale filament catalogs \citep{Ragan2014,Wang2015,Zucker2015,Abreu2016}. We have tested 1, 2, 3, and 5 km s$^{-1}$ for matching velocity. 1 km s$^{-1}$  is too strict to connect leaves. Results obtained for 2, 3, and 5 km s$^{-1}$ do not vary too much. This is reasonable considering the low possibility that a physically isolated source is observed exactly within a structure. Then we choose the relatively strict matching velocity, 2 km s$^{-1}$. Linearity is the ratio between the spread (standard deviation) of leaves along the major axis and the spread perpendicular to the major axis. To define the major axis of a filament, we plot all the leaves belonging to this filament in the projected sky (Galactic longitude as $x$ and Galactic latitude as $y$) and fit a line with principle component analysis \citep[PCA,][]{Pearson1901} as the major axis of this filament. We slightly increase the critical linearity compared to \citet{Wang2016,Ge2022} (1.5), where a few MSTs are more likely to be a collection of clumps within a larger cloud-like environment. A collection of clumps has lower linearity and aspect ratio than a filament. These dense structures have the potential to develop within the network of filaments within large molecular clouds, and can grow in size by preferentially gathering gas from the greater gravo-turbulent molecular cloud environment rather than forming through sausage instability of a single large-scale filament \citep{Zucker2018}. Identifying them is still useful but beyond the scope of this paper. We test various critical linearity values and found that 1.6 is the best value to remove collections of leaves and at the same time, to avoid getting rid of real filaments.\\
  
\subsection{Large-scale filaments in the SEDIGISM cloud catalog}\label{sec_SEDfl}
\noindent
  In Sect. \ref{sec_MST}, we have described how to identify large-scale filaments through the use of the SEDIGISM leaves. However, there are also a number of clouds (16\%, or 1718/10663) in the SEDIGISM cloud catalog \citep{DuarteCabral2021} that contain more than one dendrogram leaf. We select large elongated objects from these clouds to enlarge our filament sample. Specifically, a cloud can be regarded as a large-scale filament only when it satisfies:
  \begin{itemize}
    \item[i.]The cloud must contain at least five dendrogram leaves: $N \geq 5$.
    \item[ii.]Aspect ratio > 3.5. It is the ratio between the major and minor axis.
    \item[iii.]Medial axis length $l_{MA}\ge 10$ pc.
   \end{itemize}
  The criteria are chosen to be identical to those of the MST analysis when it is possible. The required number of nodes is the same as what we do in the MST. The aspect ratio is derived from the second moment of the emission in 2D, weighted by the intensity \citep{DuarteCabral2021}. The aspect ratio threshold is chosen because if we apply this value to MST, it will select similar structures as our linearity threshold does. That is, when we replace the criterion linearity>1.6 to aspect ratio>3.5 for the MST, most of the filaments retain. The skeleton of a cloud is derived from the medial axis transform, which calculates the position of the skeleton by computing the minimum distance of each pixel in the structure to pixel outside of it. The result of this method is reducing a structure to a single-pixel wide skeleton. The medial axis is the longest running skeleton along the 2D-projected cloud mask. It is farthest away from the external edges \citep[see Fig. 5 in][for an example]{DuarteCabral2021}.  As we are interested in the large-scale filaments, the length threshold is 10 pc, same as that for the MST filaments. J-plot is a method to classify pixelated structure to different morphology \citep{Jaffa2018}. Two J moments ($J_1$ and $J_2$) are calculate with principal moments of inertia along the two principal axes of the structures. \citet{Neralwar2022} use J-plot to classify clouds into three types: centrally concentrated disks (cores) with $J_1$>0 and $J_2$>0; elongated ellipses (filaments) with $J_1$>0 and $J_2$<0; rings (limb-brightened bubbles) with $J_1$<0 and $J_2$<0. We find all of the clouds selected following the above three criteria are also classified as filaments by \citet{Neralwar2022}.

\section{Results}\label{sec:results}

\begin{figure*}[!t]
\centering
\includegraphics[width=.33\textwidth]{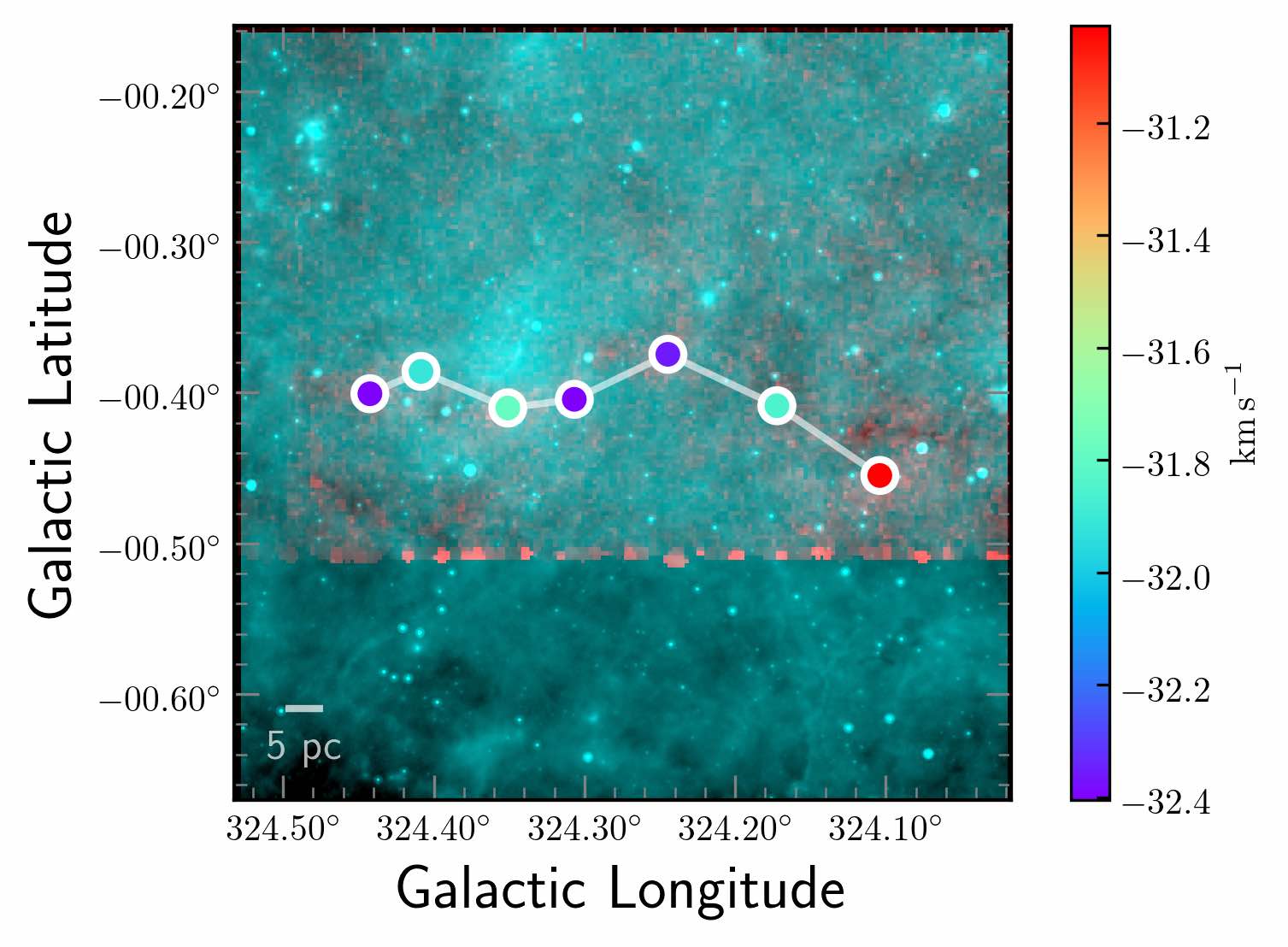}
\includegraphics[width=.33\textwidth]{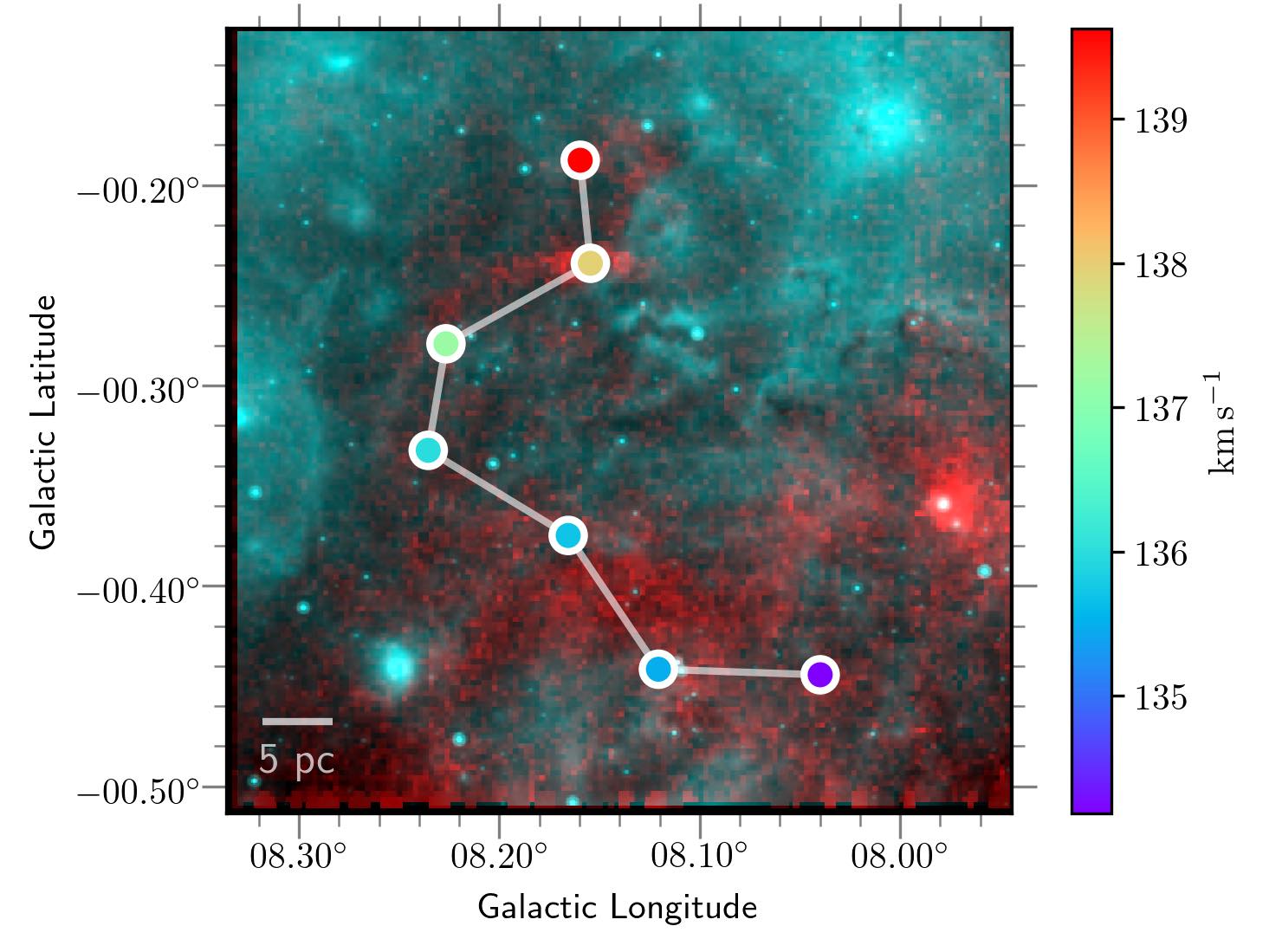}
\includegraphics[width=.33\textwidth]{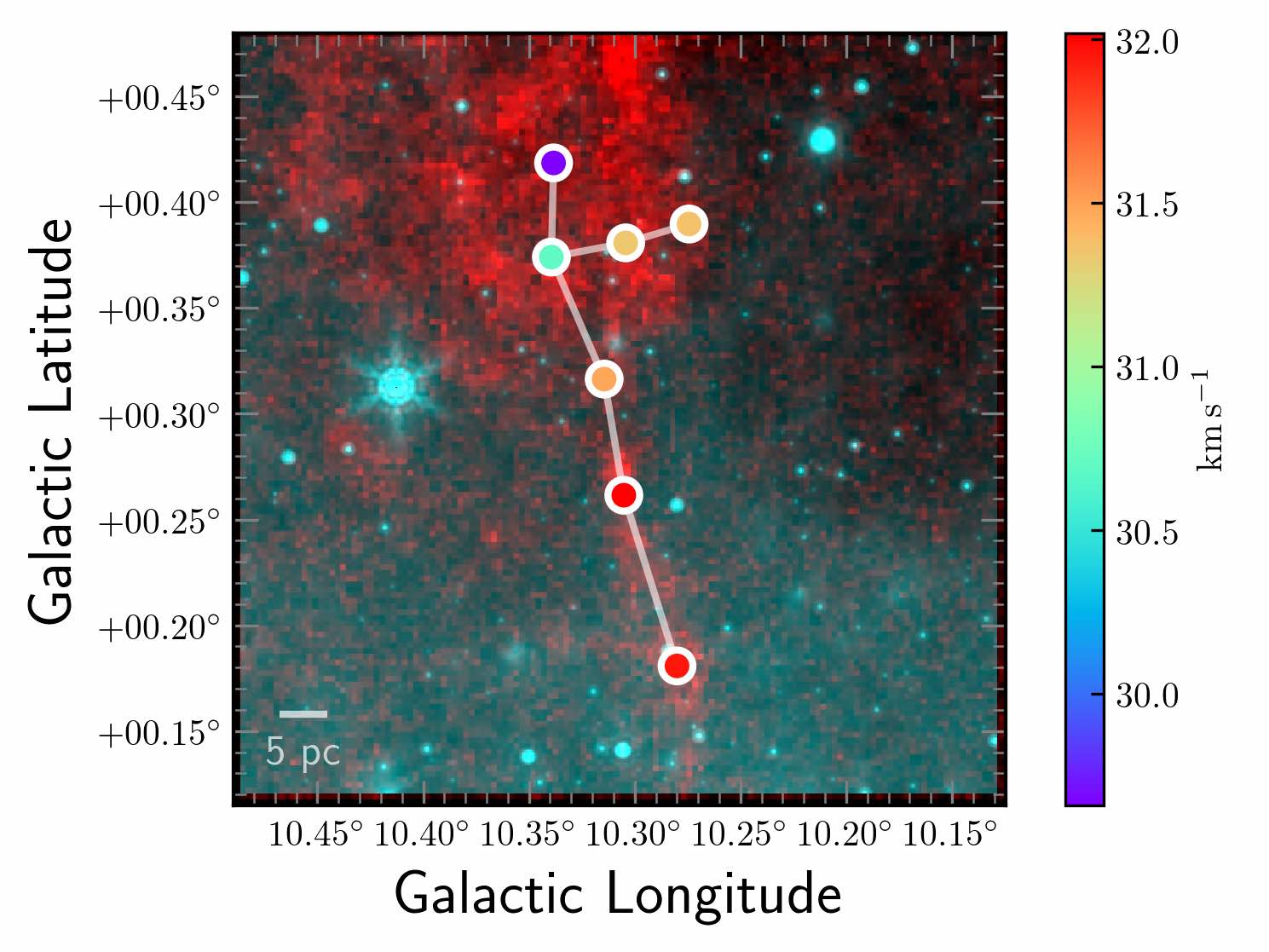}\\
\includegraphics[width=.33\textwidth]{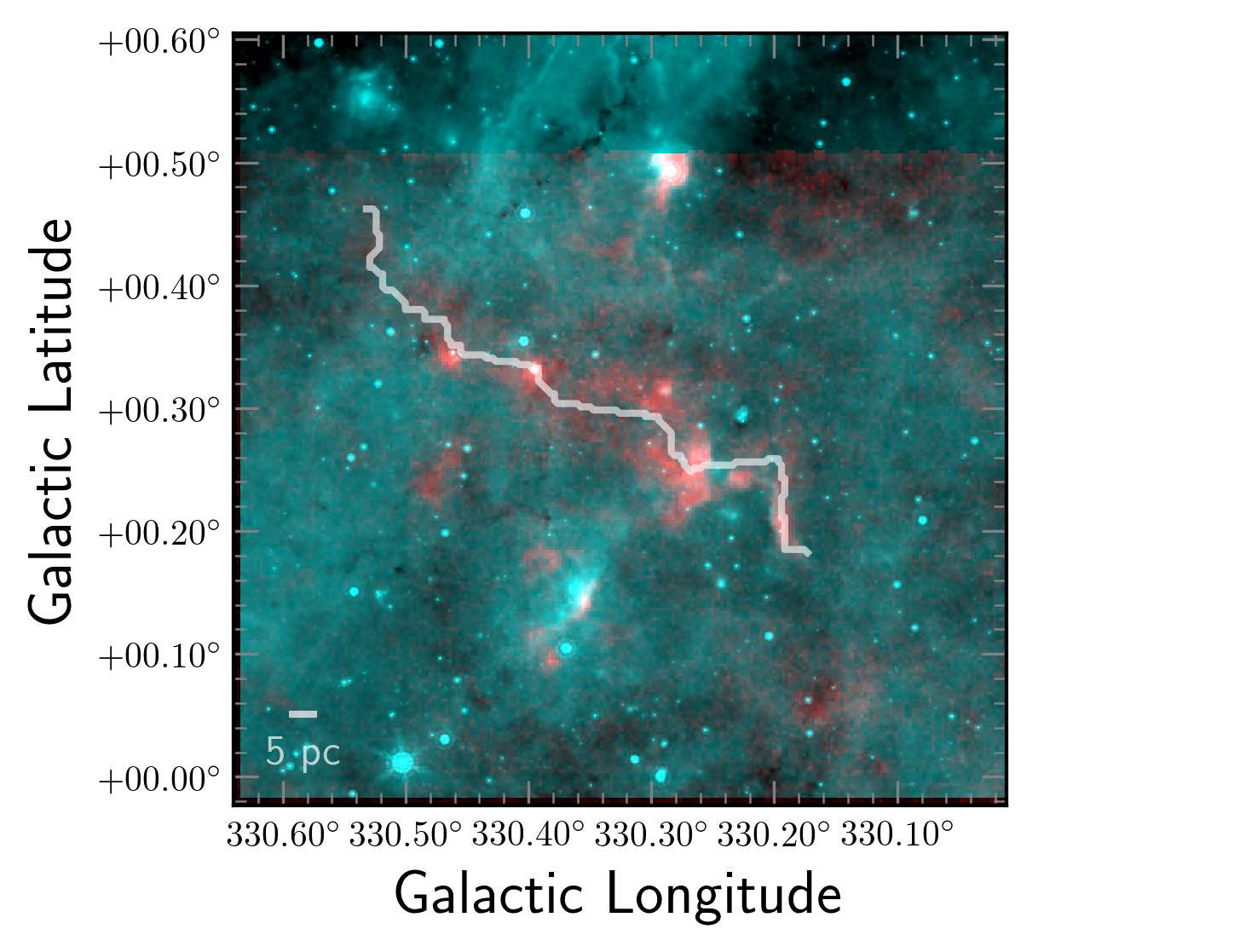}
\includegraphics[width=.33\textwidth]{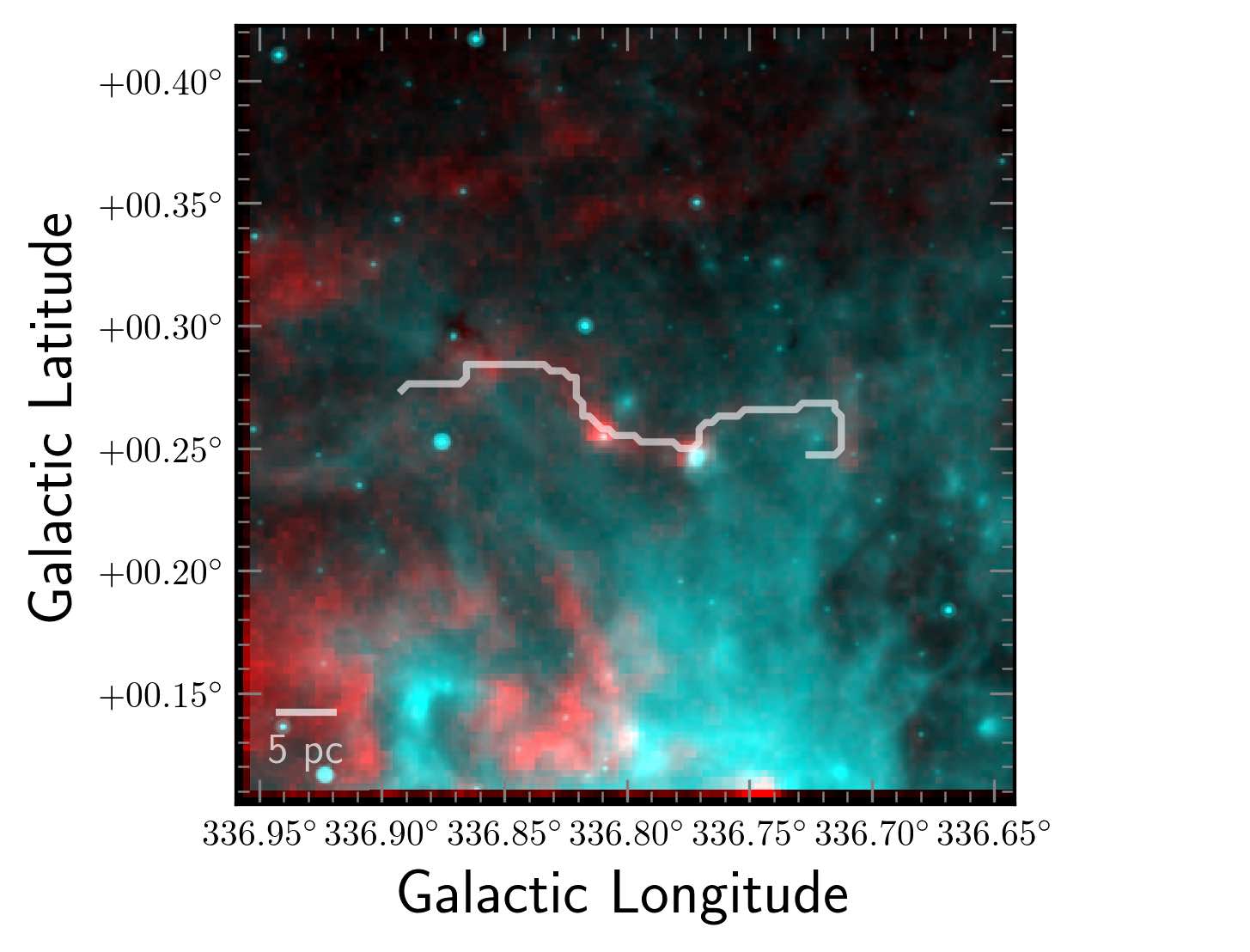}
\includegraphics[width=.33\textwidth]{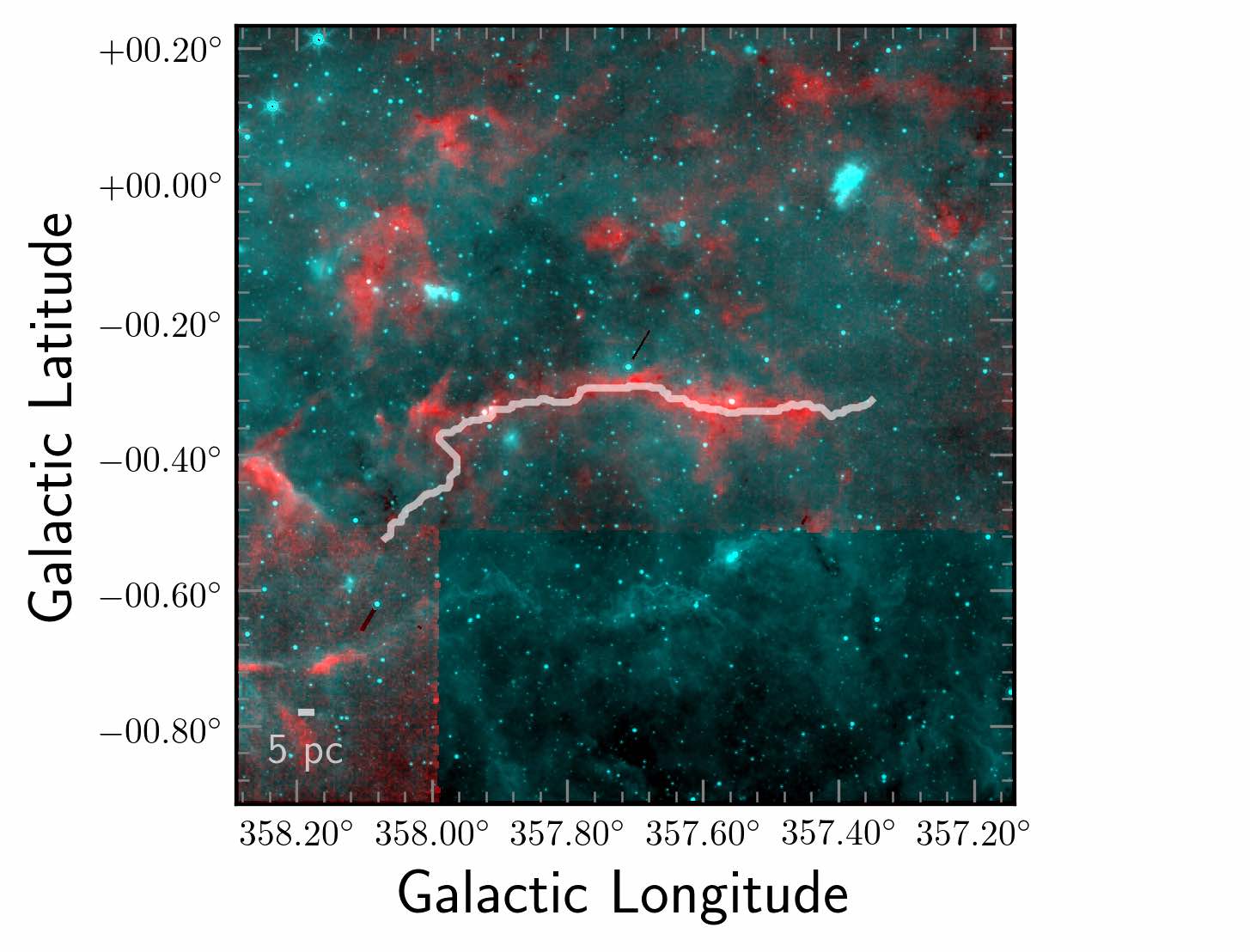}\\
\caption{Two-color composite images of six example large-scale filaments. In all panels, red shows integrated $^{13}$CO(2-1) emission in linear scale and cyan represents MIPSGAL mid-infrared 24 $\mu$m emission \citep{Carey2009} in logarithmic scale. The top 3 panels show filaments identified from SEDIGISM leaves with the MST approach (M7, M42, and M45, respectively). The color-coded circles denote leaves in filaments with different velocities and the white line segments are the edges. The color bars show radial velocities of the leaves. The bottom 3 panels show large elongated filaments selected from the SEDIGISM cloud catalog (S63, S69, and S80, respectively). White curves are their medial axis.  Two-color composite images for all the 88 filaments are shown in Appendix \ref{sec:ap_twocolor}.}
\label{twocolor}
\end{figure*}

\noindent
  We identify 55 large-scale filaments through SEDIGISM leaves with the MST approach (method described in Sect. \ref{sec_MST}) and select 33 large elongated filaments from the SEDIGISM cloud catalog (criteria described in Sect. \ref{sec_SEDfl}). These two samples compose our large-scale filament catalog. In summary, this large-scale filament catalog contains 88 filaments in the inner Galactic plane (with $-60^\circ <l<18^\circ$ and $|b|<0.^\circ 5$). The two color composite images of some of the filaments are shown in Fig. \ref{twocolor}. The CO emission is from SEDIGISM \citep{Schuller2021} and the 24 $\mu$m emission from \citet{Carey2009} is used to examine whether a filament is infrared dark. The Multi-band Imaging Photometer Galactic Plane Survey (MIPSGAL) is an infrared survey with a resolution of $6^{\prime \prime}$ at 24 $\mu$m.\\

  In order to discern which filaments are new identifications, we compare our catalog with MST filaments from BGPS \citep{Wang2016}, ATLASGAL \citep{Ge2022}, and other previously known large-scale filaments. We find 11 of our filaments are the same or overlap with those from ATLASGAL, and only one of our filaments has overlap with a filament from BGPS, as listed in the last column of Table \ref{t1}. The small number of our filaments overlapping with BGPS filaments is due to the survey coverage difference. That is, most of our filaments are identified in the Southern sky ($-60^\circ<l<18^\circ$) while BGPS filaments are in Northern sky ($7^\circ.5<l<194^\circ$). In the common coverage of SEDIGISM and ATLASGAL ($-60^\circ<l<18^\circ$ and $-0^\circ.5<b<0^\circ.5$), 77 SEDIGISM filaments are not identified in the ATLASGAL filament catalogue. The main cause for this may be the different tracer that is used. ATLASGAL clumps are extracted from 870 $\mu$m dust continuum which traces relatively dense regions, while SEDIGISM clouds are extracted from $^{13}$CO emission, which can trace more diffuse, lower density structures. So some filamentary structures that are not conspicuous in the dust continuum can be identified from $^{13}$CO emission. Similarly, \citet{DuarteCabral2021} find that only a small portion ($\sim16\%$) of SEDIGISM clouds are associated with dense clumps as traced by ATLASGAL.\\
  
  As for the comparison to other known large-scale filaments, 13 of our filaments have overlap with Milky Way bones \citep{Zucker2015,Jackson2010} and GMFs \citep{Abreu2016}, as reanalysed and termed by \citet{Zucker2018}. The corresponding associations and references are listed in the last column of Table \ref{t1}. In the common Galactic longitude range ($-60^\circ<l<18^\circ$) where Milky Way bones, GMFs, and our SEDIGISM filaments reside, there are 14 Milky Way bones or GMFs. Eight of them are identified in our SEDIGISM large-scale filament catalog (though three of them are split into multiple filaments). After comparison with BGPS filaments, ATLASGAL filaments and other known large-scale filaments, we find 66 of our filaments are newly identified large-scale filaments.
  \\
\subsection{Galactic location and physical properties} 

\noindent
  The physical properties of the 88 filaments are shown in Table\;\ref{t1}. Filaments with ID M1-M55 are those identified from SEDIGISM leaves via MST, while S56-S88 correspond to filaments selected from the SEDIGISM cloud catalog (see Sect. \ref{sec_SEDfl} for selection criteria). Some of the physical properties are derived differently for filaments identified through leaves with MST (M1-M55) and filaments selected from SEDIGISM clouds (S56-S88). If so, this physical property will be described individually for the two categories. Apart from a small portion of clouds associated with masers, distances of most clouds in the SEDIGISM cloud catalog are kinematic distances determined after solving for distance ambiguities \citep[Sect. 4 in ][]{DuarteCabral2021}. The distances of filaments listed in Col. (6) are the median of leaf distances for M1-M55, while they are taken from the SEDIGISM cloud catalog for S56-S88. For M1-M55, the aspect ratio in Col. (13) is the ratio of the area between the circle enclosing the filament and the concave hull (the minimum-area concave polygon that contains all the leaves in the MST). For an approximately elliptical distribution, $f_{A}$ is very similar to the aspect ratio of the ellipse when the number of nodes is large \citep{Gutermuth2009}. For S56-S88, the aspect ratio is the ratio between the major and minor axis of the ellipse. For M1-M55, $|\theta|$ in Col. (17) is derived from PCA (orientation of the major axis described in Sect.\ref{sec_MST}). For S56-S88, $|\theta|$ is the position angle of the major axis, with 0 being along the $l$ axis.\\
 \onecolumn
\begin{landscape}
\tiny
\centering
\tabcolsep=2pt
\begin{longtable}{ccccccccccccccccccccl}
\caption{\label{t1} Physical properties of large-scale filaments}\\
\toprule
\multicolumn{1}{c}{(1)} & \multicolumn{1}{c}{(2)} & \multicolumn{1}{c}{(3)} & \multicolumn{1}{c}{(4)} & \multicolumn{1}{c}{(5)} & \multicolumn{1}{c}{(6)} & \multicolumn{1}{c}{(7)} & \multicolumn{1}{c}{(8)} & \multicolumn{1}{c}{(9)} & \multicolumn{1}{c}{(10)} & \multicolumn{1}{c}{(11)} & \multicolumn{1}{c}{(12)} & \multicolumn{1}{c}{(13)} & \multicolumn{1}{c}{(14)} & \multicolumn{1}{c}{(15)} & \multicolumn{1}{c}{(16)} & \multicolumn{1}{c}{(17)} & \multicolumn{1}{c}{(18)} &
\multicolumn{1}{c}{(19)} & \multicolumn{1}{c}{(20)} & \multicolumn{1}{c}{(21)}\\
   ID & Origin &         $l_{wt}$ &        $b_{wt}$ &        $v_{wt}$ &           $d$ &   N &  $L_{sum}$ &  $L_{end}$ &    Mass &  $M_{line}$ &  $N_{H_2}$ &  $f_A$ &  $f_L$ &  $R_{gc}$ &             $z$ &    $|\theta|$ &   DGMF & Morph. & Arm & \multicolumn{1}{c}{Ref.} \\
      & &  ($^\circ$) &   ($^\circ$) & (km$\,{\rm s}^{-1}$) &  (kpc) &   &  (pc) &  (pc) &   ($10^3 M_\odot$) &  ($M_\odot$ pc$^{-1}$) &  ($10^{22}$ cm$^{-2}$) &   &   &  (kpc) &  (pc) & ($^\circ$) &   & class  &  & \\
\midrule
\endhead
\midrule
\multicolumn{21}{r}{{Continued on next page}} \\
\midrule
\endfoot

\bottomrule
\endlastfoot
   M1 &    MST &  300.85$\pm$0.01 &   0.27$\pm$0.03 &   -36.1$\pm$0.3 &   4.3$\pm$0.1 &   6 &       21.3 &       14.6 &     1.5 &        68.5 &        0.2 &   4.10 &   2.50 &       7.2 &    42.1$\pm$2.0 &  77.5$\pm$3.0 &     - &  C,X &                &                       \\
   M2 &    MST &  302.79$\pm$0.03 &   0.15$\pm$0.02 &   -34.8$\pm$0.4 &   5.7$\pm$0.5 &   6 &       35.4 &       21.0 &     8.8 &       247.7 &        0.5 &   2.90 &   2.16 &       7.1 &    35.6$\pm$2.0 &  28.2$\pm$2.8 &     - &    C &                &                       \\
   M3 &    MST &  305.07$\pm$0.02 &   0.14$\pm$0.04 &   -38.2$\pm$0.6 &   6.1$\pm$0.5 &   7 &       55.6 &       35.0 &     6.3 &       113.4 &        0.4 &   8.14 &   3.91 &       7.0 &    34.6$\pm$3.9 &  70.5$\pm$1.5 &  0.15 &  C,X &                &  GMF307 (1); F163 (5) \\
   M4 &    MST &  305.50$\pm$0.02 &   0.09$\pm$0.03 &   -36.2$\pm$0.3 &   6.6$\pm$0.5 &   6 &       35.0 &       24.5 &     8.8 &       250.3 &        0.4 &   4.32 &   1.84 &       7.0 &    29.7$\pm$3.3 &  75.1$\pm$3.5 &     - &    C &                &            GMF307 (1) \\
   M5 &    MST &  320.64$\pm$0.01 &   0.09$\pm$0.02 &    -5.2$\pm$0.7 &  12.6$\pm$0.1 &   6 &       56.1 &       35.9 &    17.0 &       303.2 &        0.4 &   7.13 &   3.70 &       8.1 &    25.1$\pm$5.0 &  86.0$\pm$2.8 &  0.04 &    L &                &                       \\
   M6 &    MST &  324.19$\pm$0.02 &  -0.08$\pm$0.01 &   -56.5$\pm$0.2 &  10.1$\pm$1.0 &  10 &       70.7 &       42.9 &     8.7 &       123.8 &        0.2 &   4.78 &   2.05 &       5.9 &    -5.2$\pm$2.4 &  23.1$\pm$2.3 &     - &    X &  Scu-Cen, Bone &                       \\
   M7 &    MST &  324.28$\pm$0.04 &  -0.39$\pm$0.01 &   -32.1$\pm$0.2 &  11.5$\pm$1.6 &   7 &       75.5 &       68.9 &    19.4 &       257.1 &        0.4 &   8.75 &   6.10 &       6.8 &   -73.0$\pm$1.8 &   6.9$\pm$1.3 &     - &    S &                &            GMF324 (1) \\
   M8 &    MST &  326.86$\pm$0.03 &  -0.10$\pm$0.02 &   -62.6$\pm$0.3 &  10.2$\pm$1.0 &   6 &       63.8 &       48.1 &    20.4 &       318.8 &        0.5 &   5.87 &   3.91 &       5.6 &    -9.4$\pm$4.0 &  31.8$\pm$1.8 &  0.27 &    H &                &              F142 (5) \\
   M9 &    MST &  327.03$\pm$0.02 &  -0.17$\pm$0.02 &   -56.5$\pm$0.4 &  10.6$\pm$0.7 &   9 &       69.4 &       39.9 &    11.6 &       167.1 &        0.3 &   3.48 &   2.64 &       5.8 &   -24.2$\pm$2.9 &  30.3$\pm$2.0 &     - &    H &        Scu-Cen &              F141 (5) \\
  M10 &    MST &  334.07$\pm$0.04 &  -0.40$\pm$0.02 &   -47.2$\pm$0.2 &  11.9$\pm$0.8 &  10 &      128.9 &       75.7 &    33.6 &       260.5 &        0.3 &   4.19 &   2.89 &       5.7 &   -80.5$\pm$3.6 &  16.1$\pm$1.1 &     - &  S,X &        Nor-Out &                       \\
  M11 &    MST &  334.95$\pm$0.03 &  -0.34$\pm$0.02 &   -90.0$\pm$0.4 &   5.3$\pm$0.9 &   7 &       36.1 &       21.2 &     4.3 &       120.5 &        0.2 &   3.87 &   1.77 &       4.2 &   -16.5$\pm$1.6 &  21.1$\pm$3.3 &     - &    X &        Nor-Out &              Fil9 (2) \\
  M12 &    MST &  336.47$\pm$0.01 &  -0.05$\pm$0.02 &  -127.1$\pm$0.6 &   7.6$\pm$0.1 &   7 &       44.8 &       25.1 &     3.0 &        66.8 &        0.1 &   7.70 &   3.96 &       3.3 &     3.5$\pm$3.0 &  86.4$\pm$2.3 &     - &  L,X &                &                       \\
  M13 &    MST &  337.25$\pm$0.02 &  -0.21$\pm$0.02 &  -114.6$\pm$0.4 &   8.4$\pm$0.2 &  10 &       62.4 &       35.7 &     7.2 &       115.2 &        0.2 &   3.88 &   2.09 &       3.3 &   -22.1$\pm$3.1 &  55.1$\pm$2.2 &     - &    X &                &                       \\
  M14 &    MST &  337.12$\pm$0.01 &   0.11$\pm$0.03 &   -75.6$\pm$0.3 &  10.7$\pm$0.1 &   6 &       55.9 &       47.2 &     7.1 &       126.6 &        0.2 &  10.05 &   4.80 &       4.4 &    24.6$\pm$6.1 &  72.6$\pm$1.9 &     - &    S &                &                       \\
  M15 &    MST &  337.22$\pm$0.03 &  -0.04$\pm$0.01 &   -52.3$\pm$0.3 &  11.9$\pm$1.6 &   6 &       54.9 &       42.7 &     9.9 &       180.7 &        0.3 &   6.37 &   2.98 &       5.3 &    -6.1$\pm$3.0 &  20.3$\pm$2.4 &     - &    C &                &                       \\
  M16 &    MST &  336.61$\pm$0.02 &   0.44$\pm$0.02 &   -33.0$\pm$0.2 &  12.9$\pm$1.6 &   6 &       67.5 &       40.6 &    11.7 &       172.9 &        0.2 &   3.96 &   1.75 &       6.2 &    98.0$\pm$5.1 &  55.8$\pm$4.4 &     - &    S &                &                       \\
  M17 &    MST &  338.26$\pm$0.03 &  -0.06$\pm$0.02 &   -44.0$\pm$0.6 &  12.3$\pm$1.4 &   6 &       72.3 &       51.1 &    20.7 &       286.4 &        0.3 &   5.14 &   2.76 &       5.5 &   -12.2$\pm$4.1 &  34.6$\pm$2.6 &     - &    X &        Scu-Cen &                       \\
  M18 &    MST &  338.37$\pm$0.03 &  -0.16$\pm$0.01 &   -90.5$\pm$0.5 &  10.1$\pm$0.6 &   7 &       48.0 &       40.3 &    14.4 &       300.6 &        0.4 &   8.43 &   5.23 &       3.9 &   -23.3$\pm$2.1 &  21.1$\pm$1.8 &     - &  L,H &                &                       \\
  M19 &    MST &  336.22$\pm$0.03 &   0.34$\pm$0.01 &   -70.3$\pm$0.5 &  10.9$\pm$1.3 &   6 &       49.8 &       41.3 &     4.3 &        86.6 &        0.2 &  20.82 &   5.46 &       4.7 &    67.5$\pm$1.9 &  15.7$\pm$2.1 &     - &  L,X &                &                       \\
  M20 &    MST &  339.43$\pm$0.04 &  -0.16$\pm$0.02 &  -119.0$\pm$0.5 &   8.4$\pm$0.2 &   6 &       59.6 &       38.5 &     9.3 &       155.4 &        0.2 &   4.44 &   2.74 &       3.0 &   -15.2$\pm$2.3 &  12.7$\pm$2.1 &     - &    X &                &                       \\
  M21 &    MST &  343.92$\pm$0.01 &   0.32$\pm$0.02 &   -67.5$\pm$0.5 &   5.0$\pm$1.1 &   7 &       25.6 &       14.3 &     0.8 &        32.6 &        0.1 &   4.81 &   1.83 &       3.8 &    43.1$\pm$1.6 &  85.7$\pm$3.8 &     - &    X &        Nor-Out &                       \\
  M22 &    MST &  344.67$\pm$0.02 &  -0.43$\pm$0.02 &   -21.9$\pm$0.8 &  13.9$\pm$1.8 &   6 &       58.8 &       37.2 &     7.1 &       120.7 &        0.2 &   6.08 &   2.01 &       6.3 &  -108.9$\pm$3.8 &  44.5$\pm$4.7 &     - &    X &        Nor-Out &                       \\
  M23 &    MST &  325.73$\pm$0.02 &   0.42$\pm$0.03 &   -41.1$\pm$0.2 &   2.5$\pm$1.7 &   6 &       13.0 &        8.6 &     1.0 &        74.5 &        0.2 &   4.15 &   2.11 &       6.4 &    39.5$\pm$1.1 &  57.3$\pm$3.1 &     - &  S,X &                &                       \\
  M24 &    MST &  349.91$\pm$0.02 &  -0.25$\pm$0.01 &   -24.1$\pm$0.4 &   3.3$\pm$1.7 &   9 &       24.4 &       14.0 &     1.8 &        72.4 &        0.1 &   4.20 &   1.85 &       5.3 &     3.6$\pm$0.8 &  16.5$\pm$2.7 &     - &    X &  Perseus, Bone &                       \\
  M25 &    MST &  352.36$\pm$0.01 &  -0.20$\pm$0.02 &   -84.9$\pm$0.2 &   8.2$\pm$0.2 &   8 &       42.7 &       22.1 &     4.8 &       112.6 &        0.2 &   2.63 &   1.95 &       1.3 &   -22.1$\pm$2.9 &  75.9$\pm$3.5 &     - &    C &                &                       \\
  M26 &    MST &  353.02$\pm$0.03 &  -0.07$\pm$0.01 &   -17.6$\pm$0.2 &  13.4$\pm$1.4 &  12 &      125.0 &       72.4 &    26.4 &       211.3 &        0.3 &   4.00 &   2.92 &       5.3 &   -21.5$\pm$2.2 &   2.3$\pm$1.2 &     - &  S,X &        Scu-Cen &                       \\
  M27 &    MST &  353.20$\pm$0.02 &  -0.34$\pm$0.01 &   -17.6$\pm$0.3 &  13.4$\pm$0.1 &   9 &       68.3 &       38.9 &    11.2 &       164.4 &        0.2 &   3.95 &   2.07 &       5.2 &   -84.3$\pm$2.4 &  15.1$\pm$3.3 &     - &    C &        Scu-Cen &                       \\
  M28 &    MST &  353.75$\pm$0.01 &   0.06$\pm$0.03 &     1.9$\pm$0.5 &  18.1$\pm$0.2 &   6 &      111.2 &       63.6 &    18.4 &       165.7 &        0.1 &   6.11 &   2.82 &       9.8 &     5.4$\pm$8.6 &  71.7$\pm$2.6 &     - &  L,X &        Scu-Cen &                       \\
  M29 &    MST &  356.87$\pm$0.02 &   0.06$\pm$0.03 &     4.9$\pm$0.3 &  21.0$\pm$0.1 &   8 &      173.7 &       77.2 &    40.4 &       232.7 &        0.2 &   2.04 &   1.82 &      12.7 &    2.8$\pm$10.3 &  76.8$\pm$2.7 &     - &  C,X &        Scu-Cen &            GMF358 (1) \\
  M30 &    MST &  358.28$\pm$0.04 &  -0.69$\pm$0.01 &   -21.4$\pm$0.3 &   9.7$\pm$0.5 &   6 &       63.3 &       42.3 &     3.4 &        54.1 &        0.1 &   4.55 &   2.70 &       1.4 &  -113.2$\pm$2.4 &   5.4$\pm$2.0 &     - &    S &                &                       \\
  M31 &    MST &  358.87$\pm$0.01 &  -0.26$\pm$0.03 &    -3.2$\pm$0.6 &  15.9$\pm$0.7 &   7 &      106.4 &       56.2 &    21.1 &       197.9 &        0.2 &   3.51 &   2.86 &       7.6 &   -81.3$\pm$7.6 &  77.5$\pm$2.4 &  0.47 &    X &        Scu-Cen &                       \\
  M32 &    MST &  358.84$\pm$0.02 &  -0.06$\pm$0.01 &  -193.2$\pm$0.4 &   8.4$\pm$0.1 &   6 &       26.2 &       21.0 &     1.7 &        64.0 &        0.1 &   5.44 &   4.88 &       0.2 &    -2.4$\pm$1.0 &  12.3$\pm$3.0 &     - &    S &                &                       \\
  M33 &    MST &    0.80$\pm$0.03 &   0.29$\pm$0.01 &   170.0$\pm$0.5 &   8.4$\pm$0.1 &   7 &       50.1 &       30.0 &     2.5 &        49.9 &        0.1 &  10.89 &   4.82 &       0.2 &    49.7$\pm$0.8 &   0.6$\pm$1.4 &     - &  C,X &                &                       \\
  M34 &    MST &    0.95$\pm$0.01 &   0.34$\pm$0.01 &   160.7$\pm$0.8 &   8.4$\pm$0.1 &   6 &       31.5 &       14.1 &     1.7 &        54.2 &        0.1 &   6.05 &   4.23 &       0.2 &    56.1$\pm$2.2 &  61.9$\pm$4.1 &     - &    X &                &                       \\
  M35 &    MST &    3.12$\pm$0.02 &   0.14$\pm$0.01 &   134.7$\pm$0.4 &   8.3$\pm$0.1 &   6 &       37.7 &       23.2 &     4.2 &       111.3 &        0.2 &   4.14 &   1.65 &       0.5 &    27.3$\pm$2.1 &  28.7$\pm$5.6 &     - &    X &                &                       \\
  M36 &    MST &    6.12$\pm$0.02 &  -0.18$\pm$0.02 &    13.6$\pm$1.0 &  13.2$\pm$1.2 &  14 &      203.2 &       74.0 &    49.5 &       243.8 &        0.3 &   5.84 &   3.07 &       5.3 &   -44.6$\pm$3.5 &  31.8$\pm$1.2 &  0.02 &    X &        Scu-Cen &                F2 (5) \\
  M37 &    MST &    5.96$\pm$0.03 &  -0.33$\pm$0.01 &     7.3$\pm$0.2 &  15.0$\pm$1.9 &  11 &      140.7 &       80.7 &    85.7 &       608.7 &        0.9 &   5.28 &   2.02 &       6.7 &   -94.5$\pm$3.8 &   8.1$\pm$1.7 &     - &    C &        Perseus &                       \\
  M38 &    MST &    6.69$\pm$0.02 &  -0.08$\pm$0.03 &    19.1$\pm$0.8 &  12.6$\pm$1.7 &   7 &       97.3 &       51.4 &    22.4 &       230.0 &        0.3 &   3.67 &   2.37 &       4.8 &   -20.4$\pm$6.1 &  63.3$\pm$2.3 &     - &    X &        Scu-Cen &                       \\
  M39 &    MST &    6.66$\pm$0.02 &  -0.14$\pm$0.03 &     5.8$\pm$0.4 &  15.4$\pm$1.9 &   7 &       95.8 &       70.5 &    46.5 &       485.3 &        0.6 &   4.13 &   2.83 &       7.2 &   -46.8$\pm$7.9 &  63.9$\pm$2.0 &  0.03 &    C &        Sag-Car &                       \\
  M40 &    MST &    7.01$\pm$0.03 &   0.22$\pm$0.02 &    15.4$\pm$0.4 &  13.8$\pm$0.1 &   6 &      100.3 &       62.8 &    19.1 &       190.4 &        0.2 &   3.79 &   1.94 &       5.6 &    49.1$\pm$4.5 &  18.4$\pm$2.9 &     - &    X &        Scu-Cen &                       \\
  M41 &    MST &    7.59$\pm$0.02 &  -0.29$\pm$0.01 &   149.4$\pm$0.4 &   8.3$\pm$0.1 &   6 &       33.0 &       18.5 &     1.2 &        35.2 &        0.1 &   3.12 &   2.22 &       1.1 &   -34.5$\pm$1.6 &  26.3$\pm$4.7 &     - &    S &                &                       \\
  M42 &    MST &    8.13$\pm$0.02 &  -0.32$\pm$0.03 &   136.6$\pm$0.6 &   8.3$\pm$0.1 &   7 &       62.2 &       40.8 &     3.6 &        57.7 &        0.1 &   3.28 &   2.06 &       1.2 &   -39.4$\pm$5.0 &  64.2$\pm$2.1 &     - &    S &                &                       \\
  M43 &    MST &    8.63$\pm$0.03 &  -0.10$\pm$0.02 &    32.4$\pm$0.4 &  12.3$\pm$1.0 &   7 &       74.7 &       49.1 &    55.3 &       740.6 &        0.6 &   6.69 &   2.55 &       4.2 &   -22.1$\pm$3.7 &  27.4$\pm$2.3 &     - &    C &        Sag-Car &                       \\
  M44 &    MST &   10.50$\pm$0.03 &  -0.33$\pm$0.01 &    13.2$\pm$0.7 &  14.5$\pm$2.4 &   6 &       89.9 &       56.1 &    12.2 &       135.9 &        0.2 &   4.15 &   2.74 &       6.5 &   -88.2$\pm$2.8 &   3.0$\pm$2.1 &     - &    S &        Perseus &                F5 (4) \\
  M45 &    MST &   10.31$\pm$0.01 &   0.37$\pm$0.03 &    31.1$\pm$0.3 &  12.7$\pm$1.5 &   7 &       69.7 &       54.1 &    22.9 &       328.4 &        0.3 &   8.22 &   3.87 &       4.7 &    80.7$\pm$6.6 &  81.3$\pm$2.0 &     - &  L,H &        Sag-Car &                       \\
  M46 &    MST &   13.71$\pm$0.02 &  -0.37$\pm$0.02 &    40.0$\pm$0.4 &  12.4$\pm$1.2 &  13 &      176.0 &       74.2 &   104.2 &       592.2 &        0.5 &   2.76 &   1.89 &       4.9 &   -80.7$\pm$3.3 &  20.4$\pm$1.7 &  0.01 &    X &        Sag-Car &                       \\
  M47 &    MST &   14.84$\pm$0.03 &  -0.38$\pm$0.02 &    26.0$\pm$0.3 &   2.8$\pm$2.0 &   7 &       17.9 &       12.3 &     0.3 &        17.3 &        0.0 &   6.48 &   3.38 &       5.8 &     1.2$\pm$0.8 &  24.2$\pm$1.8 &     - &    S &                &                       \\
  M48 &    MST &   14.94$\pm$0.01 &  -0.32$\pm$0.03 &    30.9$\pm$0.4 &  13.2$\pm$0.1 &   7 &       80.4 &       55.5 &    21.8 &       271.1 &        0.3 &   6.85 &   4.52 &       5.6 &   -75.4$\pm$6.2 &  64.3$\pm$1.9 &     - &    S &                &                       \\
  M49 &    MST &   14.59$\pm$0.03 &  -0.45$\pm$0.01 &    18.4$\pm$0.6 &  14.1$\pm$0.1 &  11 &      112.7 &       66.6 &    28.9 &       256.6 &        0.3 &   4.35 &   4.15 &       6.4 &  -114.6$\pm$1.8 &   4.8$\pm$1.3 &     - &    S &        Perseus &                       \\
  M50 &    MST &   14.58$\pm$0.01 &   0.12$\pm$0.02 &    35.0$\pm$0.1 &  12.9$\pm$0.1 &   9 &       91.9 &       49.0 &    19.4 &       211.0 &        0.2 &   4.22 &   1.91 &       5.2 &    25.6$\pm$4.5 &  60.0$\pm$3.0 &     - &  C,X &        Scu-Cen &                       \\
  M51 &    MST &   14.69$\pm$0.02 &   0.15$\pm$0.01 &    39.4$\pm$0.3 &  12.5$\pm$1.6 &   8 &       75.9 &       48.2 &    12.9 &       170.4 &        0.3 &   5.72 &   2.88 &       5.0 &    32.0$\pm$2.5 &  19.2$\pm$2.3 &     - &  S,X &        Scu-Cen &                       \\
  M52 &    MST &   15.09$\pm$0.02 &   0.18$\pm$0.01 &    28.4$\pm$0.6 &  13.3$\pm$1.4 &   7 &       84.9 &       39.0 &    15.4 &       181.3 &        0.2 &   3.70 &   2.48 &       5.7 &    38.5$\pm$3.0 &  29.3$\pm$3.2 &     - &    X &                &                       \\
  M53 &    MST &   16.66$\pm$0.02 &  -0.20$\pm$0.03 &    46.7$\pm$0.5 &  12.2$\pm$1.0 &   8 &       89.6 &       66.1 &    19.0 &       212.1 &        0.2 &   9.04 &   4.64 &       4.9 &   -42.7$\pm$6.8 &  61.9$\pm$1.3 &  0.19 &    S &        Sag-Car &                       \\
  M54 &    MST &   16.68$\pm$0.01 &  -0.13$\pm$0.03 &    36.9$\pm$0.7 &  12.7$\pm$1.9 &   6 &       60.1 &       40.0 &    14.6 &       243.0 &        0.4 &   5.50 &   2.54 &       5.4 &   -30.5$\pm$6.2 &  80.2$\pm$2.9 &     - &    C &                &                       \\
  M55 &    MST &   16.79$\pm$0.03 &   0.37$\pm$0.03 &    23.7$\pm$0.4 &  13.7$\pm$2.0 &   8 &      117.2 &       68.6 &    92.0 &       785.1 &        0.9 &   3.98 &   1.84 &       6.2 &    85.3$\pm$7.0 &  52.8$\pm$2.1 &     - &  C,X &        Perseus &                       \\
  S56 &    320 &           305.70 &            0.33 &   -39.2$\pm$3.2 &   3.5$\pm$1.0 &   7 &       27.9 &       12.3 &     6.3 &       227.4 &        0.6 &   3.87 &      - &       6.9 &            41.7 &          87.9 &  0.05 &    S &        Scu-Cen &            GMF307 (1) \\
  S57 &    852 &           312.62 &            0.03 &   -63.8$\pm$1.4 &   5.7$\pm$1.4 &  11 &       64.3 &       31.6 &    48.5 &       755.0 &        0.9 &   6.13 &      - &       6.1 &            21.3 &           8.0 &  0.04 &  S,X &                &                       \\
  S58 &   1421 &           317.64 &           -0.13 &   -44.4$\pm$3.0 &   2.8$\pm$0.6 &   6 &       15.0 &        6.9 &     3.6 &       242.1 &        0.8 &   4.31 &      - &       6.5 &            14.8 &          10.3 &  0.06 &    S &                &            GMF319 (1) \\
  S59 &   1772 &           320.25 &            0.36 &   -32.5$\pm$1.0 &   2.0$\pm$0.5 &   5 &       22.9 &        6.9 &     6.3 &       276.3 &        0.6 &   3.64 &      - &       6.9 &            34.7 &          68.7 &  0.24 &  L,H &                &                       \\
  S60 &   1879 &           323.49 &            0.07 &   -66.3$\pm$1.3 &   4.0$\pm$0.5 &   5 &       34.4 &       13.6 &    26.4 &       768.5 &        0.8 &   4.10 &      - &       5.6 &            23.3 &          34.7 &  0.09 &    C &                &                       \\
  S61 &   2590 &           328.05 &            0.37 &   -44.9$\pm$1.0 &   2.8$\pm$0.5 &   6 &       33.7 &       13.5 &     8.8 &       260.3 &        0.4 &   5.27 &      - &       6.2 &            37.9 &          31.1 &  0.02 &    S &                &                       \\
  S62 &   2884 &           330.53 &            0.23 &   -61.6$\pm$2.0 &   3.7$\pm$0.4 &   7 &       41.0 &       17.5 &    13.9 &       338.9 &        0.5 &   3.63 &      - &       5.4 &            33.3 &          14.3 &  0.01 &    C &        Nor-Out &                       \\
  S63 &   2942 &           330.32 &            0.29 &   -96.1$\pm$1.4 &   5.8$\pm$0.7 &   8 &       57.2 &       20.8 &    33.5 &       586.2 &        0.5 &   3.84 &      - &       4.4 &            44.1 &          27.8 &  0.06 &    S &        Nor-Out &                       \\
  S64 &   2947 &           331.69 &            0.01 &   -48.7$\pm$2.7 &   3.1$\pm$0.5 &  16 &       41.9 &       18.3 &    28.1 &       670.3 &        0.9 &   5.49 &      - &       5.8 &            20.1 &          19.2 &  0.08 &    C &        Scu-Cen &              F130 (5) \\
  S65 &   3108 &           331.52 &           -0.11 &   -87.6$\pm$2.8 &   5.1$\pm$0.5 &  17 &       60.6 &       21.8 &   213.8 &      3527.1 &        1.7 &   3.55 &      - &       4.6 &             5.8 &          17.2 &  0.27 &    S &                &              F130 (5) \\
  S66 &   3129 &           331.61 &           -0.28 &   -47.9$\pm$1.0 &   3.0$\pm$0.5 &   6 &       16.1 &        8.4 &     8.1 &       501.2 &        1.0 &   4.58 &      - &       5.9 &             4.8 &          18.7 &  0.29 &    C &  Scu-Cen, Bone &                       \\
  S67 &   3233 &           332.10 &           -0.43 &   -57.0$\pm$2.8 &   3.5$\pm$0.4 &  10 &       21.6 &        8.5 &    24.4 &      1129.4 &        1.9 &   3.85 &      - &       5.5 &            -7.6 &          24.7 &  0.26 &  C,H &  Nor-Out, Bone &              F129 (5) \\
  S68 &   3423 &           334.22 &            0.17 &   -88.8$\pm$3.9 &   5.2$\pm$0.5 &   5 &       67.9 &       25.0 &    33.7 &       496.0 &        0.4 &   3.92 &      - &       4.3 &            30.6 &          30.1 &  0.12 &    S &        Nor-Out &                       \\
  S69 &   4022 &           336.80 &            0.27 &   -75.6$\pm$2.1 &   4.7$\pm$0.4 &   6 &       19.1 &        8.4 &     6.2 &       326.6 &        0.7 &   4.06 &      - &       4.5 &            37.6 &          10.1 &  0.14 &    S &                &                       \\
  S70 &   4449 &           336.70 &            0.25 &   -40.9$\pm$1.3 &   2.9$\pm$0.5 &   7 &       23.1 &       12.3 &     4.1 &       175.5 &        0.4 &   4.05 &      - &       5.8 &            31.9 &          46.7 &  0.02 &  C,X &        Nor-Out &                       \\
  S71 &   4698 &           337.68 &           -0.40 &   -41.0$\pm$1.3 &   2.9$\pm$0.5 &  33 &       64.9 &       25.2 &    72.4 &      1115.9 &        0.8 &   3.77 &      - &       5.7 &            -1.2 &           0.1 &  0.29 &  S,H &  Scu-Cen, Bone &  Nessie (3); F113 (5) \\
  S72 &   4930 &           339.19 &           -0.42 &   -38.1$\pm$1.4 &   2.9$\pm$0.5 &   8 &       31.6 &       13.3 &    14.4 &       454.9 &        0.7 &   4.19 &      - &       5.8 &            -1.6 &           1.3 &  0.15 &    S &  Scu-Cen, Bone &            Nessie (3) \\
  S73 &   5079 &           339.57 &            0.09 &  -110.6$\pm$1.5 &   6.7$\pm$0.6 &   5 &       59.5 &       21.5 &    52.0 &       873.5 &        0.7 &   4.88 &      - &       3.1 &            22.3 &           6.3 &  0.24 &    L &                &                       \\
  S74 &   5539 &           342.58 &            0.15 &   -41.6$\pm$1.2 &   3.4$\pm$0.5 &   6 &       25.7 &       12.7 &    23.8 &       927.0 &        0.9 &   4.72 &      - &       5.2 &            26.8 &          11.3 &  0.35 &    S &                &   GMF343 (1); F99 (5) \\
  S75 &   5774 &           344.96 &           -0.24 &   -85.9$\pm$1.0 &   5.9$\pm$0.4 &   6 &       47.2 &       16.9 &    37.3 &       789.3 &        0.6 &   3.81 &      - &       3.0 &           -11.7 &          10.4 &  0.09 &    C &                &                       \\
  S76 &   6191 &           349.21 &            0.14 &   -64.0$\pm$2.7 &   5.7$\pm$0.4 &   7 &       53.6 &       21.5 &    42.0 &       783.6 &        0.6 &   4.95 &      - &       3.0 &            27.1 &           3.6 &  0.30 &    S &                &               F83 (5) \\
  S77 &   6300 &           349.86 &            0.09 &   -61.4$\pm$1.7 &   5.7$\pm$0.4 &   5 &       35.2 &       15.6 &    24.2 &       687.9 &        0.6 &   4.12 &      - &       2.9 &            22.2 &          15.6 &  0.50 &    S &                &               F82 (5) \\
  S78 &   6861 &           353.18 &           -0.01 &   -82.5$\pm$1.5 &   8.8$\pm$0.4 &   5 &       42.8 &       20.3 &    28.5 &       666.4 &        0.5 &   3.86 &      - &       1.1 &             3.6 &          19.4 &     - &    S &                &                       \\
  S79 &   7275 &           357.14 &           -0.09 &    -7.4$\pm$1.8 &   2.7$\pm$2.0 &   5 &       17.0 &        7.7 &     2.5 &       145.7 &        0.5 &   5.57 &      - &       5.6 &            14.6 &           0.9 &  0.17 &    S &        Scu-Cen &            GMF358 (1) \\
  S80 &   7443 &           357.71 &           -0.34 &     4.7$\pm$1.6 &  21.3$\pm$4.4 &  12 &      327.3 &      141.6 &  1318.4 &      4028.2 &        0.7 &   4.71 &      - &      13.0 &          -148.1 &           4.1 &  0.55 &    C &        Scu-Cen &              Fil8 (2) \\
  S81 &   7686 &           359.69 &           -0.00 &  -72.1$\pm$40.5 &   8.4$\pm$4.6 &  17 &      157.3 &       82.5 &  1295.3 &      8236.8 &        3.7 &   8.23 &      - &       0.1 &             6.1 &           3.1 &  0.07 &  L,H &                &                       \\
  S82 &   7701 &           359.66 &           -0.10 &  -30.4$\pm$12.2 &   8.2$\pm$0.1 &  11 &      102.7 &       39.1 &   592.8 &      5770.5 &        2.9 &   4.02 &      - &       0.1 &            -7.5 &           8.5 &  0.07 &  S,H &                &                       \\
  S83 &   7736 &             0.10 &           -0.07 &   42.3$\pm$23.3 &   8.4$\pm$5.4 &  64 &      137.1 &       51.6 &  3634.2 &     26504.2 &        6.1 &   3.69 &      - &       0.1 &            -3.0 &           2.1 &  0.14 &    C &                &                       \\
  S84 &   8513 &             6.44 &           -0.44 &    20.0$\pm$3.4 &   3.6$\pm$1.0 &  14 &       24.0 &       11.1 &    13.2 &       550.5 &        0.9 &   4.60 &      - &       4.8 &           -10.2 &          19.8 &  0.09 &  S,H &        Scu-Cen &                       \\
  S85 &   8591 &             5.53 &           -0.32 &    13.2$\pm$1.3 &   2.9$\pm$1.4 &   9 &       30.1 &       13.0 &     4.5 &       150.9 &        0.4 &   5.29 &      - &       5.5 &             3.1 &          18.3 &  0.57 &    S &  Scu-Cen, Bone &                       \\
  S86 &   8624 &             5.82 &           -0.48 &    17.3$\pm$1.3 &   3.4$\pm$1.2 &   7 &       22.5 &        9.1 &    16.5 &       736.4 &        1.1 &   4.44 &      - &       5.0 &           -11.2 &           0.8 &  0.02 &    S &        Scu-Cen &                       \\
  S87 &   9021 &             8.27 &           -0.43 &    17.2$\pm$1.9 &   2.7$\pm$1.0 &   5 &       29.0 &       11.3 &     5.1 &       174.6 &        0.3 &   3.63 &      - &       5.7 &            -1.0 &          28.2 &     - &    S &  Scu-Cen, Bone &                       \\
  S88 &   9707 &            12.84 &            0.03 &    32.7$\pm$0.6 &   3.4$\pm$0.6 &   5 &       19.4 &        9.6 &     3.8 &       193.5 &        0.6 &   3.67 &      - &       5.1 &            19.7 &           5.0 &     - &    C &        Scu-Cen &                       \\
  \hline
  min &        &             0.10 &           -0.69 &          -193.2 &           2.0 &   5 &       13.0 &        6.9 &     0.3 &        17.3 &        0.0 &   2.04 &   1.65 &       0.1 &          -148.1 &           0.1 &  0.01 &      &                &                       \\
  max &        &           359.69 &            0.44 &           170.0 &          21.3 &  64 &      327.3 &      141.6 &  3634.2 &     26504.2 &        6.1 &  20.82 &   6.10 &      13.0 &            98.0 &          87.9 &  0.57 &      &                &                       \\
  med &        &           327.54 &           -0.07 &           -33.9 &           8.4 &   7 &       56.6 &       33.3 &    14.4 &       242.6 &        0.4 &   4.27 &   2.70 &       5.4 &             3.3 &          23.7 &  0.13 &      &                &                       \\
 mean &        &           228.64 &           -0.05 &           -21.3 &           8.6 &   8 &       65.5 &       35.8 &    98.5 &       846.7 &        0.6 &   5.09 &   2.94 &       5.0 &            -2.6 &          32.8 &  0.17 &      &                &                       \\
  Std &        &           155.09 &            0.26 &            62.9 &           4.6 &   7 &       48.7 &       24.3 &   432.0 &      3000.9 &        0.8 &   2.38 &   1.12 &       2.4 &            47.1 &          27.0 &  0.16 &      &                &                       \\
    S &        &            -0.71 &            0.00 &             0.7 &           0.4 &   6 &        2.4 &        1.2 &     6.9 &         7.5 &        4.7 &   3.72 &   0.99 &       0.1 &            -0.8 &           0.7 &  1.10 &      &                &                       \\
    K &        &            -1.47 &           -0.91 &             1.5 &          -0.5 &  42 &        8.6 &        2.4 &    50.6 &        59.7 &       26.1 &  20.21 &   0.02 &       1.9 &             0.6 &          -0.9 &  0.35 &      &                &                       \\

\end{longtable}
\tablefoot{Columns are:\\
(1) ID: Assigned filament ID \\
(2) Origin: Identification algorithm for M1-M55 is MST, and the  SEDIGISM cloud IDs \citep{DuarteCabral2021} for leaves contained in each filament are listed in Table.\;\ref{t2}. For S56-S88, the identification algorithm is SCIMES, and the original SEDIGISM cloud IDs \citep{DuarteCabral2021}, one per filament, is listed here. \\
(3) $l_{wt}$ (deg): Flux-weighted Galactic longitude for M1-M55, and Galactic longitude of the cloud’s centroid for S56-S88\\
(4) $b_{wt}$ (deg): Flux-weighted Galactic latitude for M1-M55, and Galactic latitude of the cloud’s centroid for S56-S88\\
(5) $v_{wt}$ (km s$^{-1}$): Flux weighted velocity for M1-M55, and system velocity for S56-S88\\
(6) $d$ (kpc): Distance, the median of all leaf distances for M1-M55, and final adopted distance for S56-S88 \\
(7) N: The number of leaves in the filament \\
(8) $L_{sum}$ (pc): The sum of ``edge'' lengths (separation of each two connected leaves) for M1-M55 and medaxis length for S56-S88 \\
(9) $L_{end}$ (pc): End-to-end length along the major axis for M1-M55, and major axis for S56-S88. To define the major axis of a filament, we plot all the leaves belonging to this filament in the projected sky (Galactic longitude as x and Galactic latitude as y) and fit a line with principle component analysis \citet[PCA][]{Pearson1901} as the major axis of this filament \\
(10) Mass ($10^3 M_\odot$): Mass of the filament, derived by summing leaf mass for M1-M55, and cloud mass for S56-S88 \\
(11) $M_{line}$ ($M_\odot$ pc$^{-1}$): Line mass, mass per unit length (Mass/$L_{sum}$)\\
(12) $N_{H_2}$ ($10^{22}$ cm$^{-2}$): $H_2$ column density \\
(13) $f_{A}$: Aspect ratio\\
(14) $f_{L}$: Linearity, the ratio between spread (standard deviation) of leaves along major axis and spread along minor axis, only for M1-M55 \\
(15) $R_{gc}$ (kpc):  Galactocentric radius, median of leaf Galactocentric radius for M1-M55, and cloud Galactocentric radius for S56-S88 \\
(16) z (pc): Height above Galactic mid-plane \\
(17) $|\theta|$  (deg): Orientation angle between filaments major axes and Galactic mid-plane in the projected sky\\
(18) DGMF: Dense gas mass fraction, the ratio between dense gas mass and total mass of the filaments\\
(19) Mor.: Morphology class following \citet{Wang2015}, ``L'' represents linear straight L-shape, ``C'' bent C-shape, ``S'' quasi-sinusoidal shape, ``X'' crossing of multiple filaments, ``H'' head-tail or hub-filament system \\
(20) Arm: Nearest arm the filaments is associated. If a filament is also a bone, ``Bone'' is added after the arm name. For inter-arm filaments, the value is empty. We propose eight bones, they are M6, M24, S66, S67, S71, S72, S85, and S87.\\
(21) Ref: References to previously known filaments. (1) means \citet{Abreu2016}, (2) for \citet{Zucker2015}, (3) for \citet{Jackson2010}, (4) for \citet{Wang2016}, (5) for \citet{Ge2022} \\
Some of the properties are derived differently for filaments identified through leaves with MST (with ID of M1-M55) and filaments selected from SEDIGISM clouds (with ID of S56-S88), such as $L_{sum}$, $L_{end}$, $f_{A}$, and $|\theta|$. Most of the physical properties for S56-S88 are adopted from \citet{DuarteCabral2021}. The last seven rows show the statistics of the filament properties. Skewness measures the asymmetry of a probability distribution. For instance, the normal distribution has $S=0$. Negative skewness means that the tail is on the left side of the distribution, and on the right side for the positive skewness. Kurtosis characterizes how the distribution is shaped compared to a normal distribution. The normal distribution has $K=0$. A distribution with $K>0$ is more centrally peaked than normal distribution while for $K<0$ is flatter.
}
\end{landscape} 
 \twocolumn
  DGMFs for S56-S88 are taken from \citet{Urquhart2021} with an uncertainty of a factor of 2.5 (or 0.45 dex). They take the ATLASGAL mass of the clumps coincident with a SEDIGISM cloud in position and velocity as the value of the dense mass, and their DGMF is defined as $DGF=\Sigma M_{clump}/M_{cloud}$. Material traced by ATLASGAL have column densities about 15 times higher than clouds traced by SEDIGISM \citep{Urquhart2021}. As for M1-M55, we follow a similar approach. That is, dense gas mass can be calculated directly from DGMFs of the leaves. We take the sum of the dense gas mass of leaves in a filament as the dense gas mass of this filament ($\sum_{i=1}^{N}M_i\cdot DGF_i$, where N is the number of leaves the filament contains). With this dense mass and total mass of filaments ($\sum M_{leaves}$), DGMFs for filaments M1-M55 are calculated. The DGMFs are listed in Col. (18) of Table\;\ref{t1}. Filaments with a DGMF of `-' correspond to those do not host any ATLASGAL clumps. The statistics of the filament properties are shown in the last seven rows of Table \ref{t1}, including minimum, maximum, median, mean, standard deviation, skewness (S), and kurtosis (K).
\subsection{Distribution of Galactic location and physical properties} 

\begin{figure*}[!t]
\centering
    \subfloat[]{
    \includegraphics[width=.33\textwidth]{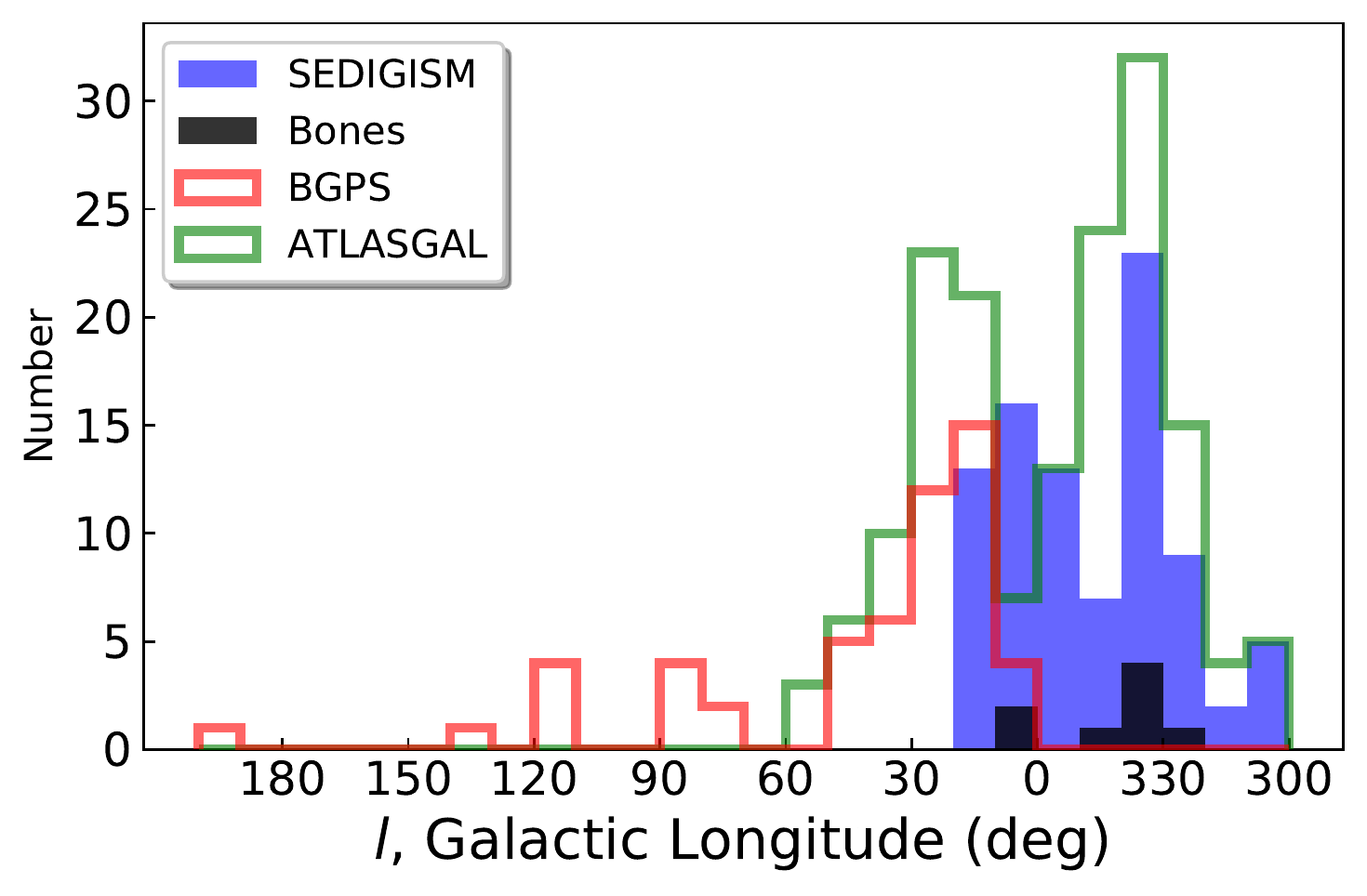}}
    \subfloat[]{
    \includegraphics[width=.33\textwidth]{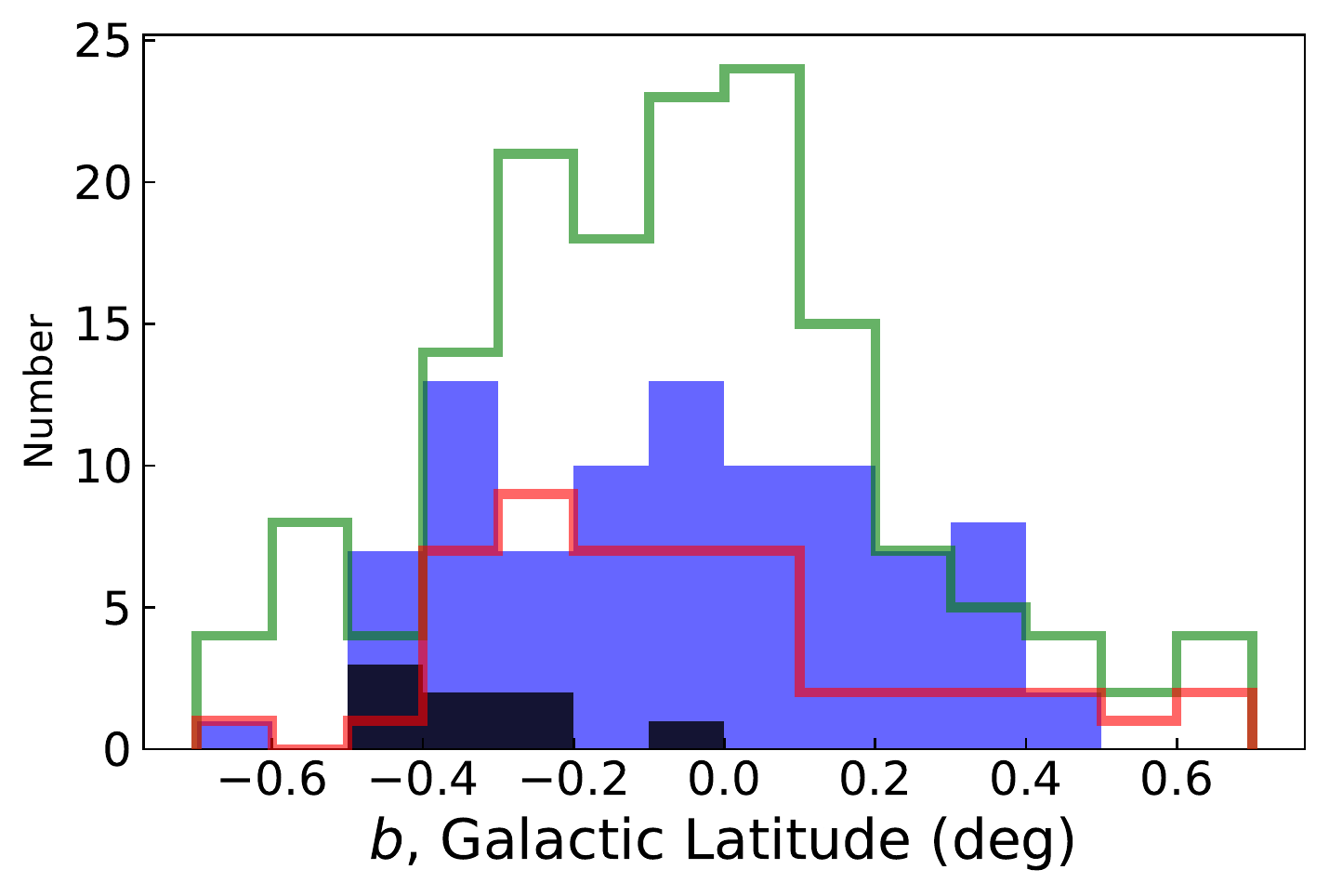}}\\
    \subfloat[]{
    \includegraphics[width=.33\textwidth]{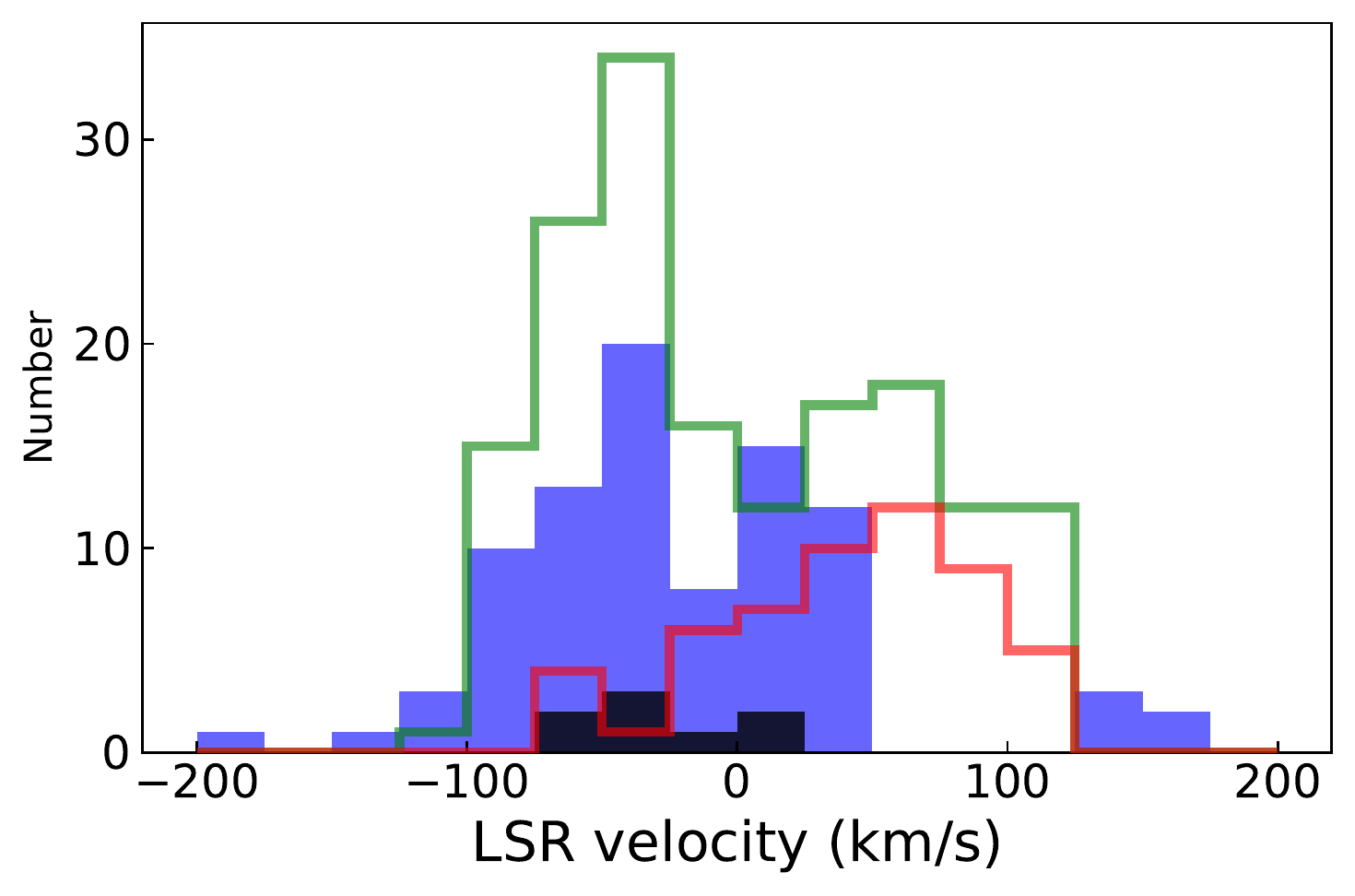}}
    \subfloat[]{
    \includegraphics[width=.33\textwidth]{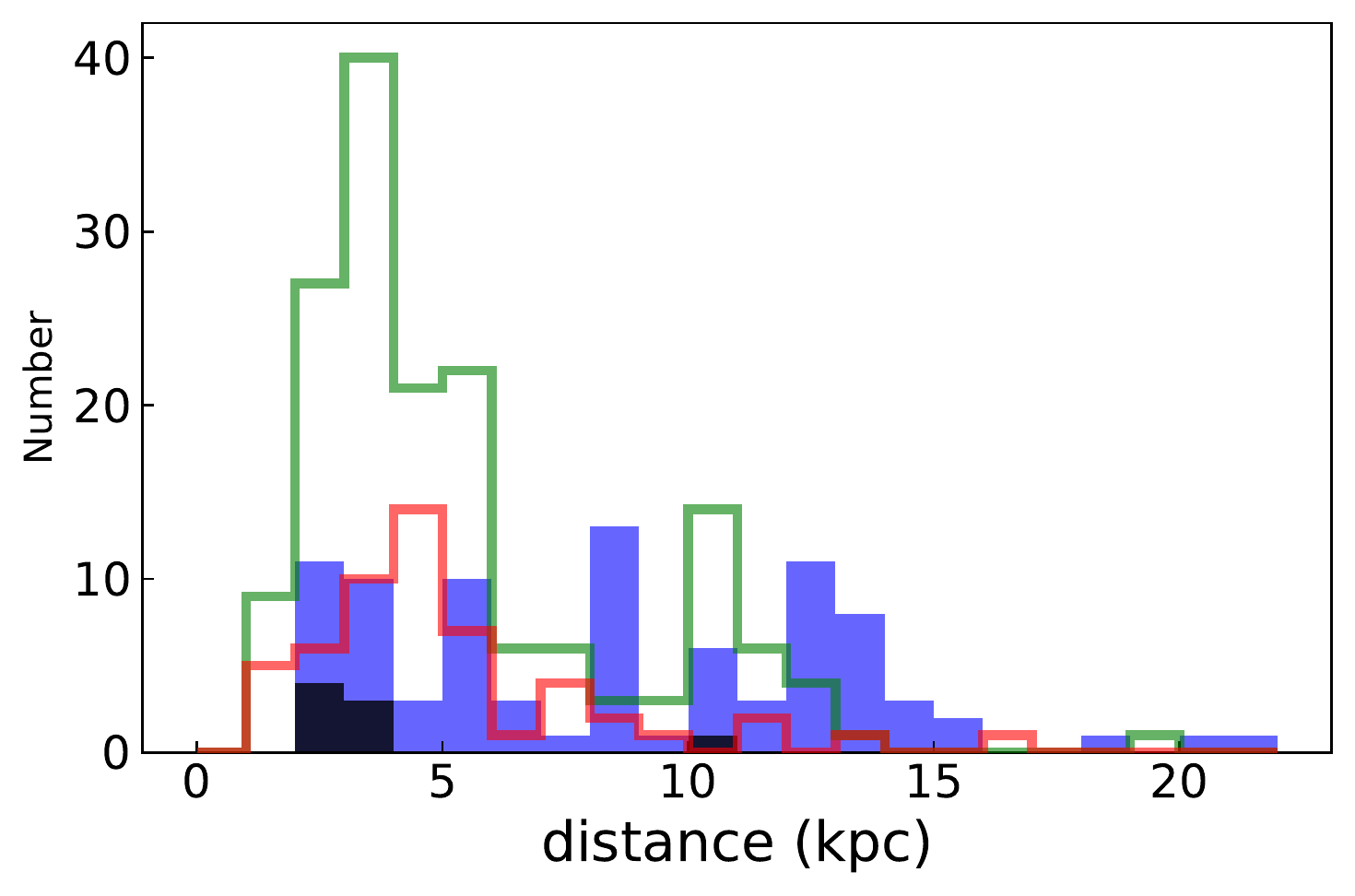}}
    \subfloat[]{
    \includegraphics[width=.33\textwidth]{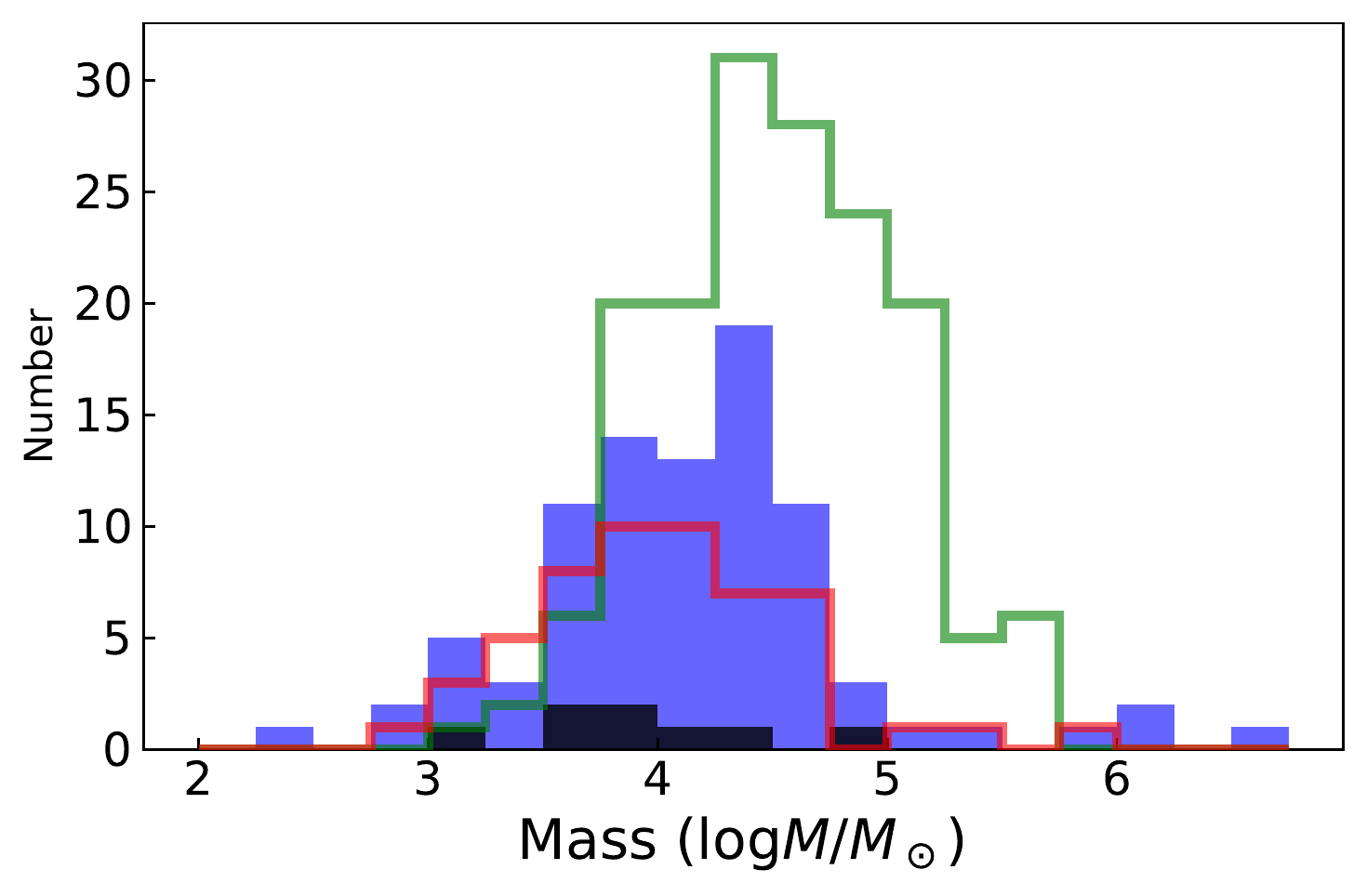}}\\
    \subfloat[]{
    \includegraphics[width=.33\textwidth]{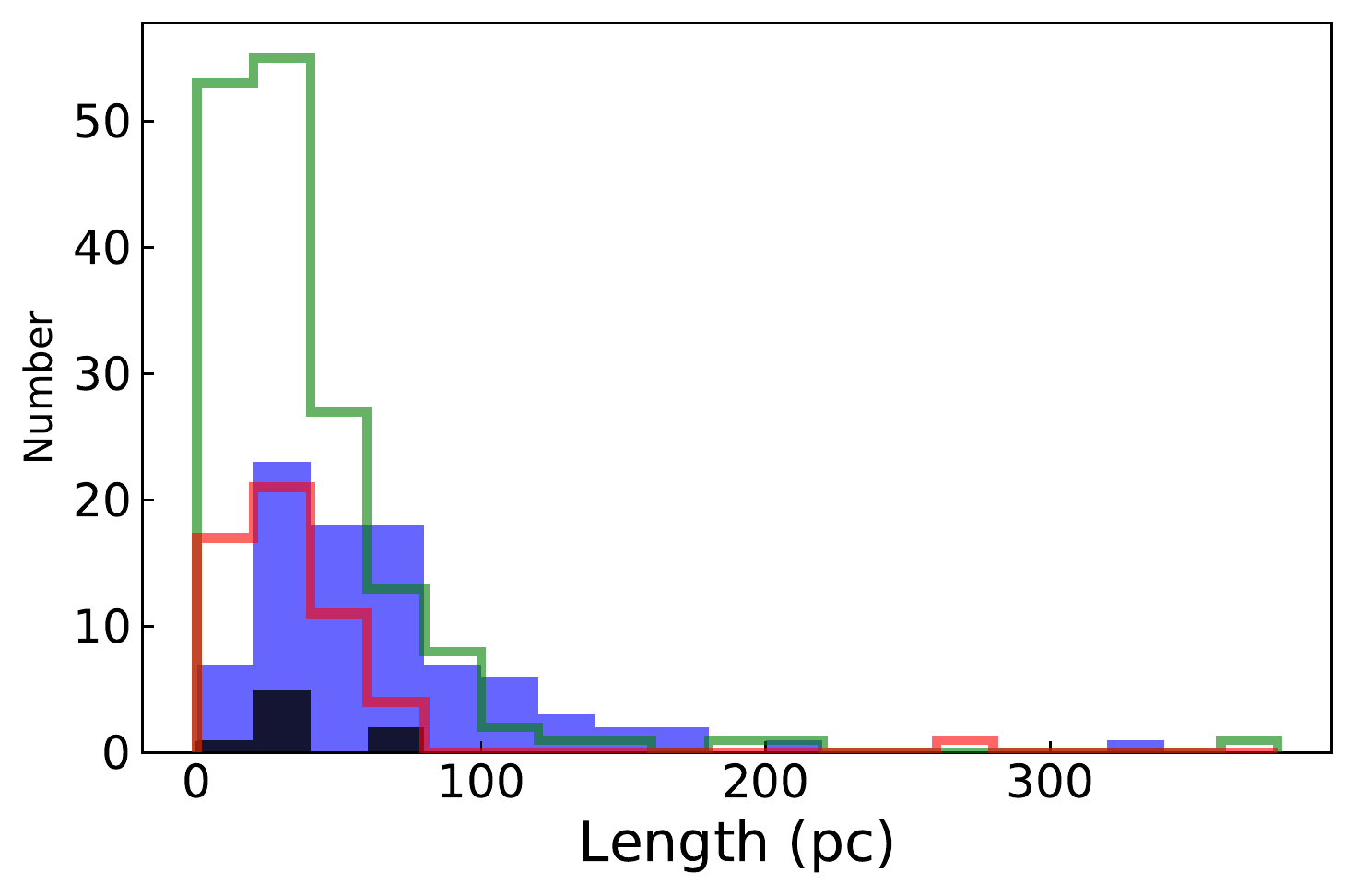}}
    \subfloat[]{
    \includegraphics[width=.33\textwidth]{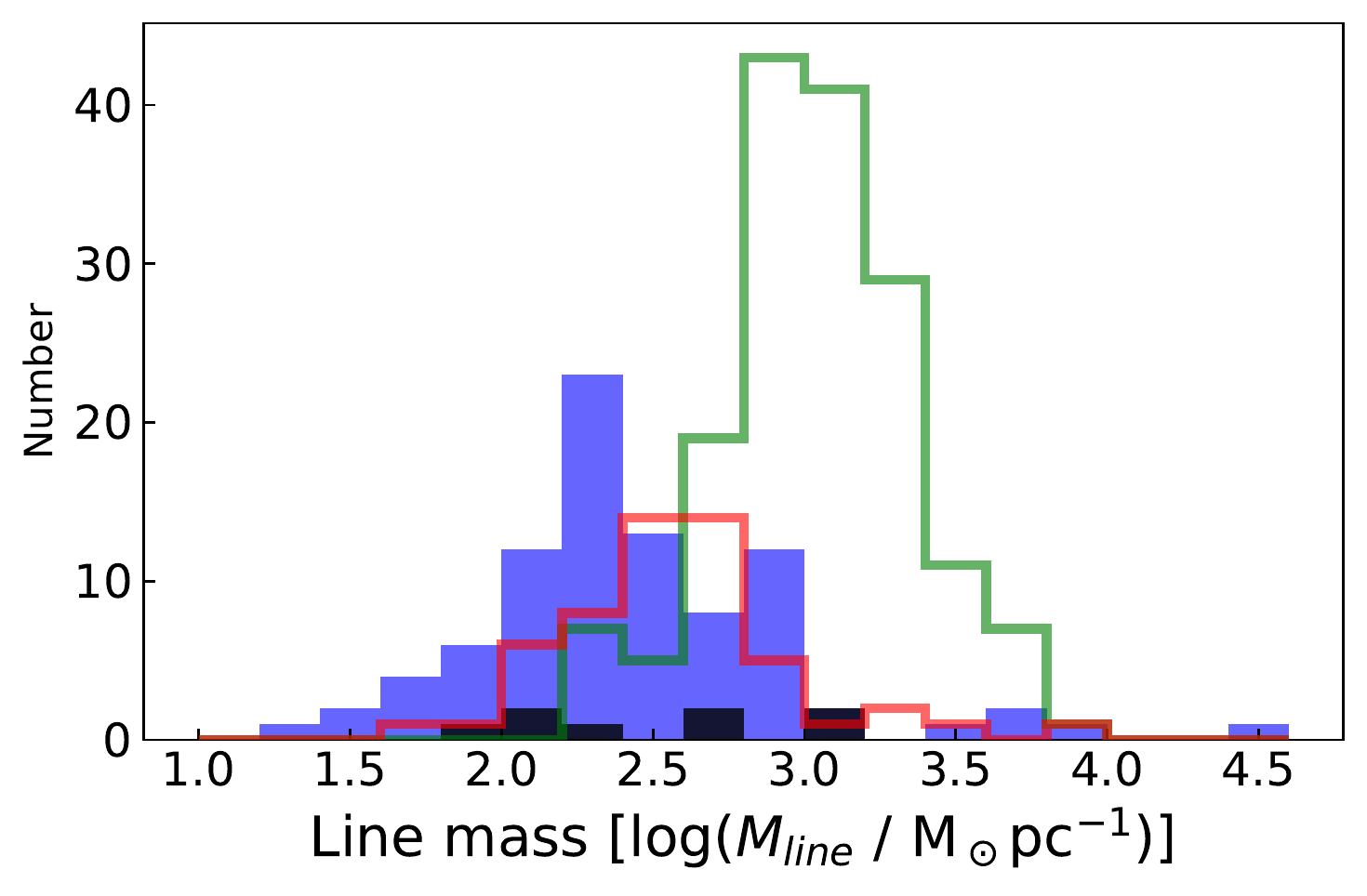}}
    \subfloat[]{
    \includegraphics[width=.33\textwidth]{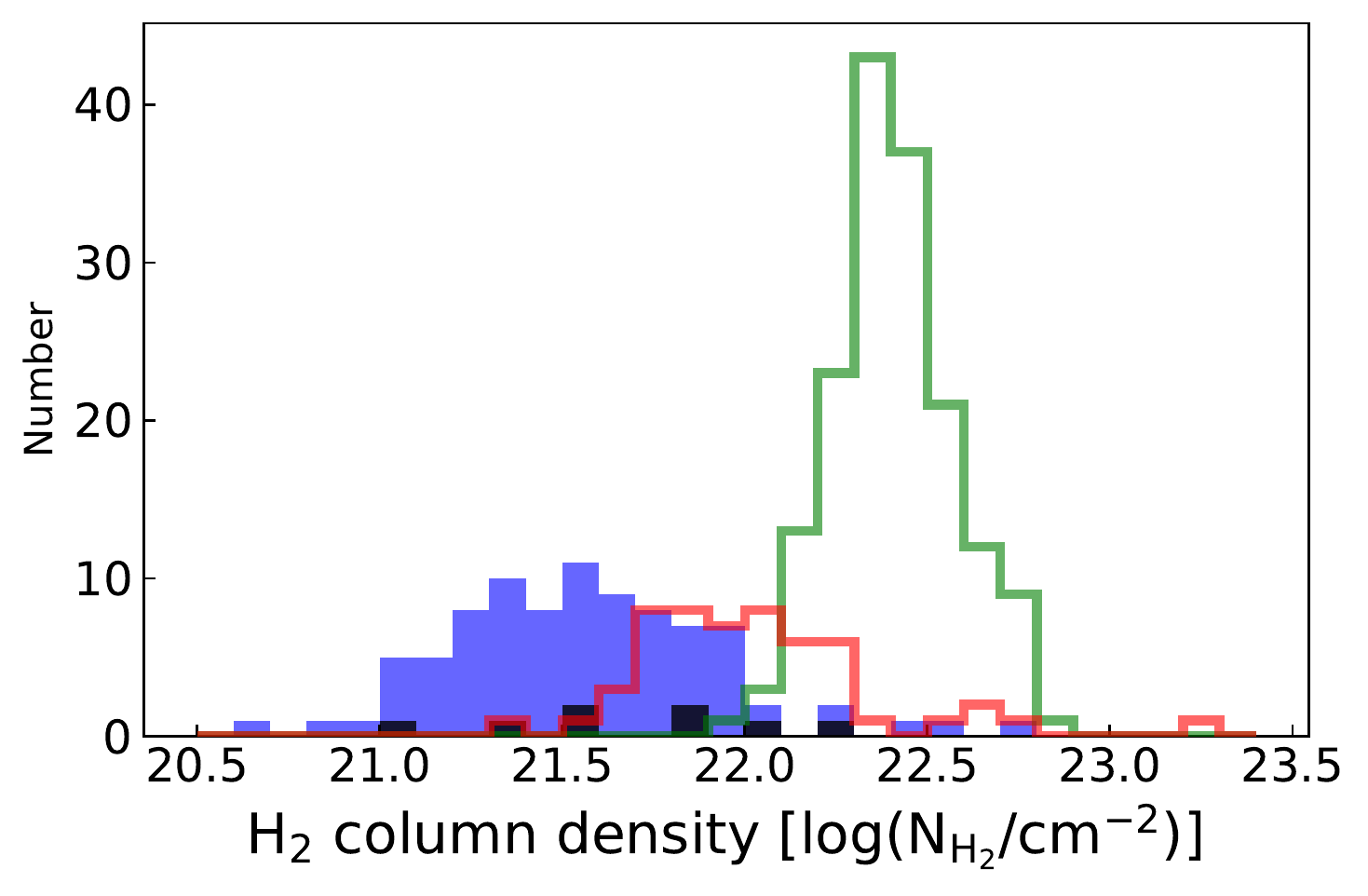}}\\
    \subfloat[]{
    \includegraphics[width=.33\textwidth]{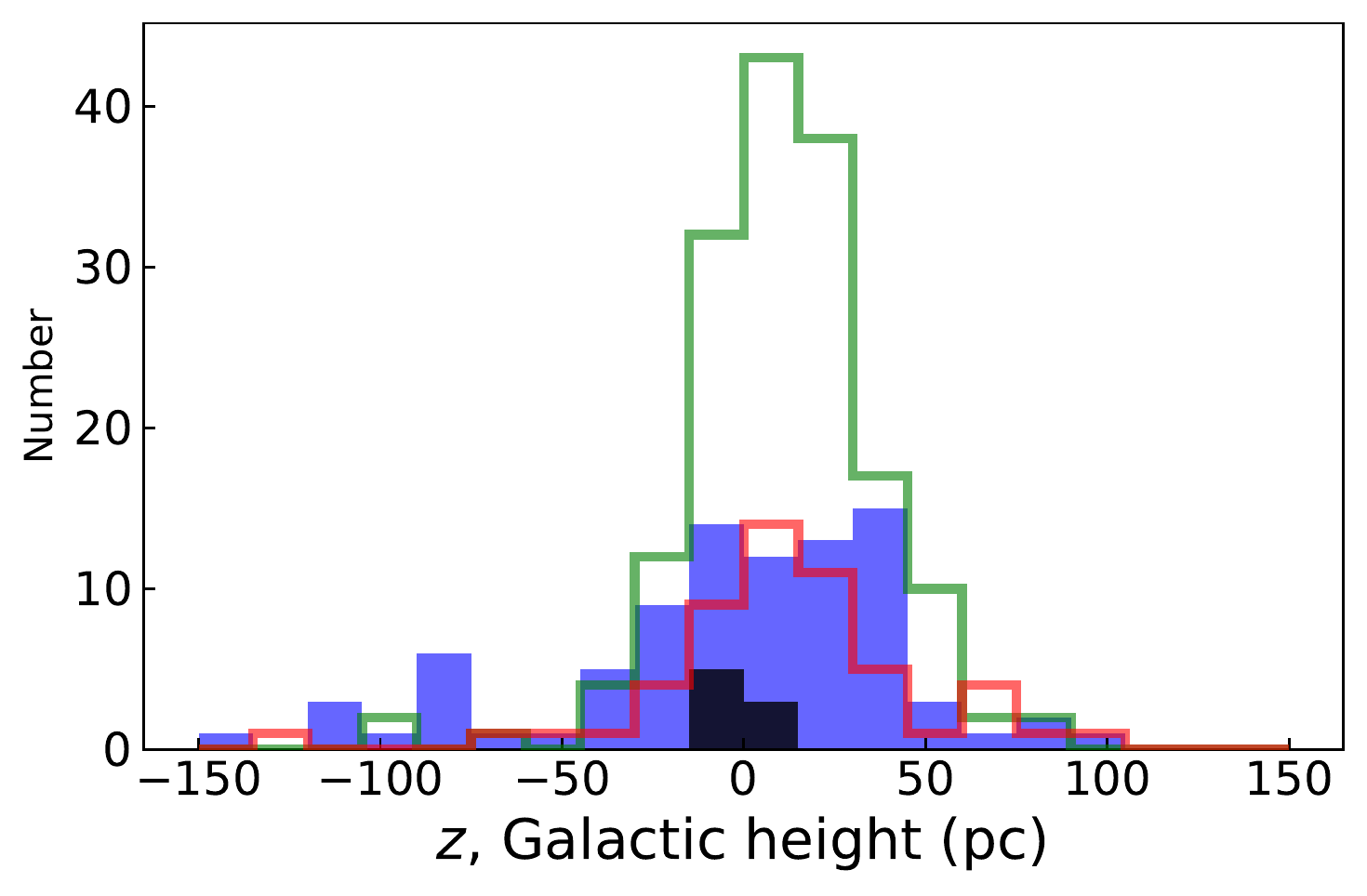}}
    \subfloat[]{
    \includegraphics[width=.33\textwidth]{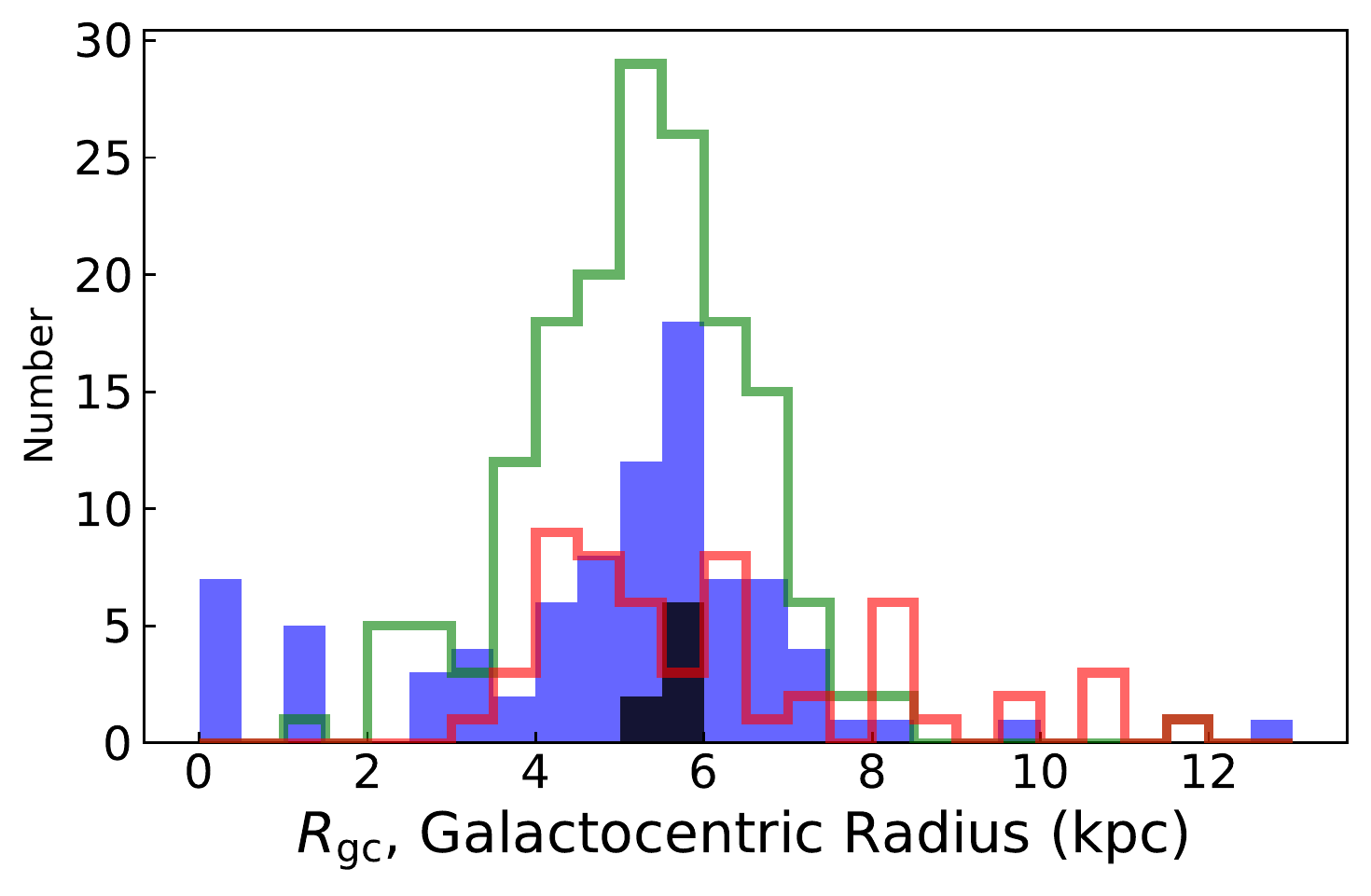}}
    \subfloat[]{
    \includegraphics[width=.33\textwidth]{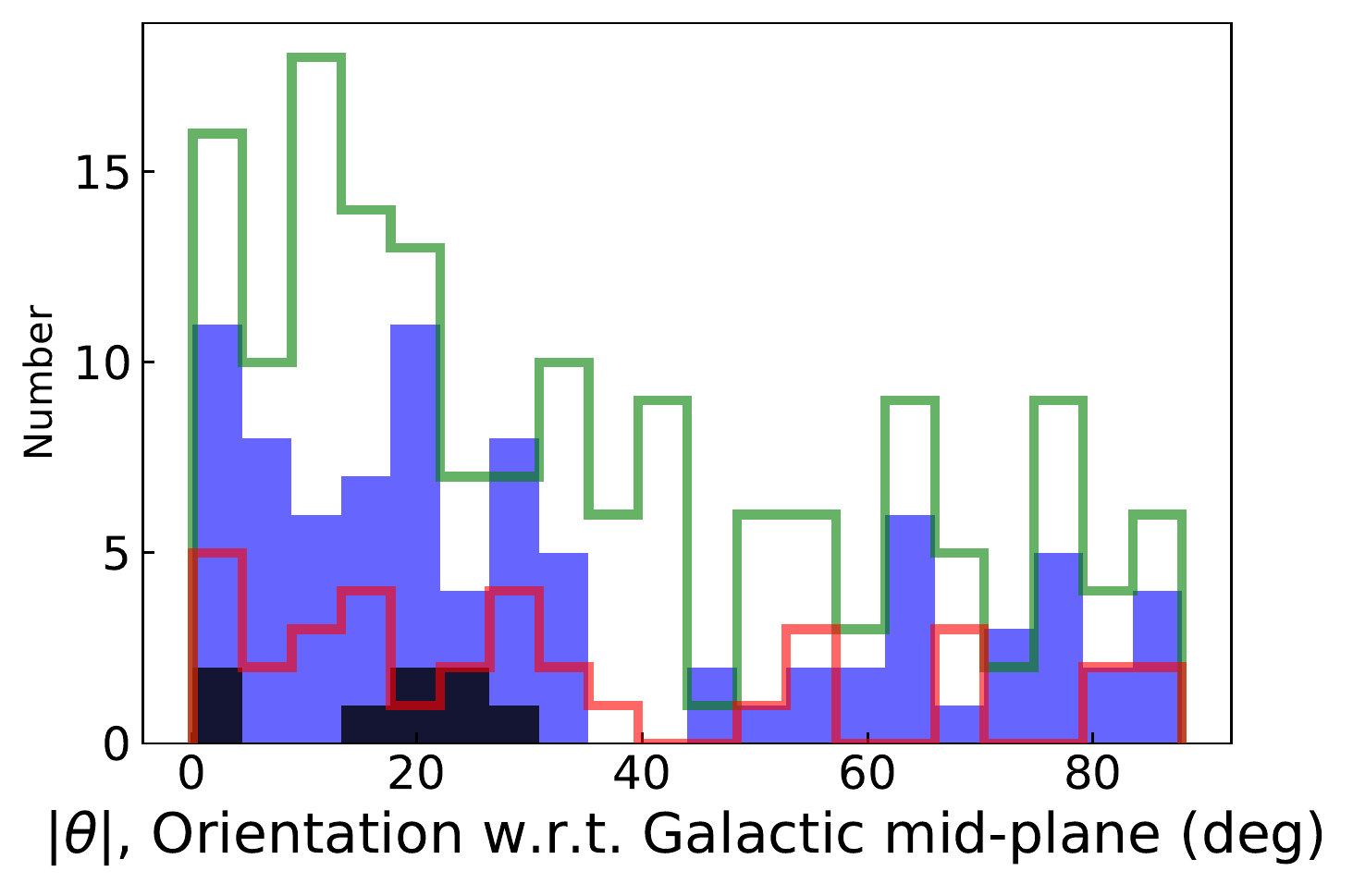}}
\caption{Histograms showing physical properties of the 88 SEDIGISM filaments (blue, this work), as compared to the 54 BGPS filaments \citep[red,][]{Wang2016} and the 163 ATLASGAL filaments \citep[green,][]{Ge2022}. Panels (a)-(k) show distribution in Galactic longitude and latitude, LSR velocity, distance, total mass, length (sum of edges), line mass, column density, Galactic height (distance from the Galactic mid-plane), Galactocentric radius and orientation angle, respectively. ``Bones'' discussed in Sect. \ref{sec:MW} are plotted in black.
}
\label{sta}
\end{figure*}

\noindent
  Histograms of the Galactic location and physical properties of filaments are shown in Fig. \ref{sta}. Panels (a)-(k) show their Galactic longitudes and latitudes, LSR velocities, distances, total mass, lengths (sum of edges), line mass, column density, Galactic height, Galactocentric radius and orientation angles in the projected sky, respectively. For comparison, the properties of filaments from two other categories of MST filaments are also plotted, where red histograms show MST filaments identified through BGPS sources \citep[hereafter ``BGPS filaments'', ][]{Wang2016} and green histograms denote MST filaments identified through ATLASGAL clumps \citep[hereafter ``ATLASGAL filaments'', ][]{Ge2022}.\\
  
  For the Galactic longitude distribution shown in Fig. \ref{sta} (a), there is a peak around $l=335^\circ$. That is because the distribution of Galactic longitude for leaves (SEDIGISM clouds containing only one dendrogram leaf, which we use to identify filaments) also peaks around that longitude, and filaments tend to be found in the region where leaves are crowded. Meanwhile, ATLASGAL filaments also peak around this Galactic longitude. The peak is found in \citet{Mattern2018} as well. Their sample is selected from ATLASGAL filaments \citep{Li2016} identified through DisPerSE with a median length of about 10 pc. They attribute the decrease in the number of filaments towards the Galactic center to the difficulty in identifying filaments in confused structures, and the decrease in the number towards the outer Galaxy to the lack of dense molecular clouds. In Fig. \ref{sta} (b), we can see that the distribution of Galactic latitudes for filaments in this work is less spread (with a standard deviation of $0.26^\circ$, while this value for BGPS filaments and ATLASGAL filaments are $0.38^\circ$ and $0.36^\circ$, respectively) as a result of the relatively narrow Galactic latitude range for SEDIGISM data. No filaments are found in this work in the velocity range between 50 km s$^{-1}$ and 130 km s$^{-1}$ (see Fig. \ref{sta} (c)). Five filaments have a velocity larger than 130 km s$^{-1}$, and they are all close to the Galactic center with Galactic longitudes $0^\circ <l<10^\circ$. This distribution is also consistent with the velocity of SEDIGISM clouds (see Fig. 1 in \citealt{DuarteCabral2021}). As seen in Fig. \ref{sta} (d), unlike BGPS and ATLASGAL where the majority of filaments are relatively nearby (<5 kpc), filaments in this work have distances spread over a wide range. This is due to the fact that the majority of BGPS and ATLASGAL clumps are nearby limited by resolution, while two thirds of the SEDIGISM leaves have distances larger than 5 kpc because their sizes are larger and thus could be resolved at larger distance.\\
  
  As shown in Fig. \ref{sta} (e), the total mass of filaments in this work ranges from $10^3-10^6M_\odot$. The length (sum of edges) distribution of the filaments in this work is shown in Fig. \ref{sta} (f) and has a longer tail than the other two samples. There are 17\% of the filaments in this work that have a length larger than 100 pc, while those for BGPS and ATLASGAL filaments are 2\% and 4\%, respectively. The average length of filaments in this work (65.5 pc) is also much larger than the other two (35.7 pc for BGPS and 38.7 pc for ATLASGAL). This is due to the fact that SEDIGISM leaves are extracted from $^{13}$CO emission, which traces relatively diffused structures with a median effective radius of 2.0 pc, while BGPS sources and ATLASGAL clumps are more condensed continuum sources both with mean/median effective radius of about 0.7 pc. For the same reason, the line mass and column density of the filaments in this work is smaller than the other two as shown in Fig. \ref{sta} (g) and (h). \citet{Zucker2018} also find that one class of large-scale filaments called ``GMFs'' (giant molecular filaments), with boundaries derived from $^{13}$CO, have lower column densities than other classes of filaments.\\
  
  Regarding the Galactic height shown in Fig. \ref{sta} (i), there are more filaments with positive z than negative (skews towards the negative values with $S=-0.8$). That does not mean that large-scale filaments prefer to be located above the Galactic mid-plane. The reason is that the Galactic plane survey is centered at $b=0^\circ$ while the true Galactic mid-plane has a negative Galactic latitude due to our Sun being located above the plane \citep{Ragan2014}. The Galactocentric radius shown in Fig. \ref{sta} (j) is more centrally peaked compared to a normal distribution (with Kurtosis, $K=1.9$). There are also a number of filaments near the Galactic center, where no BGPS or ATLASGAL filaments exist (limited by the coverage of the relevant PPV catalogs).\\
  
  In Fig. \ref{sta} (k), we can see that the majority of filaments are either parallel ($|\theta|<30^\circ$, parallel end) or perpendicular ($|\theta|>60^\circ$, vertical end) with respect to the Galactic mid-plane. This preference does not occur in BGPS (orientations for BGPS filaments are recalculated with PCA) or ATLASGAL filaments. Most of the observed large-scale filaments tend to be more parallel to the Galactic mid-plane than expected randomly \citep{Zucker2018}. This is also the case for our SEDIGISM filaments as $\sim 60\%$ of them distributed in the parallel end. Using a sample of large-scale filaments extracted from a simulation of a Milky Way-like galaxy, \citet{Zucker2019} find that without feedback in the simulations, it is unsurprising that every filament in the sample forms aligned with the gravitational mid-plane of the simulation. When SN feedback exists in the simulation, the shells developed from SN remnant collide to form large filamentary clouds \citep{Joung2006}. Interaction between large-scale filaments and SN bubbles has also been observed \citep{Li2013,Li2022,Chen2023}. So a portion of our vertical filaments might be as a result of SN feedback. A study on HI filament \citep{Soler2022} find that there is a significant number of vertical HI filaments and SN bubbles beyond the Solar circle in the region of the SEDIGISM survey. But we note that the orientation of the filamentary structures traced by HI is not, in general, inherited by the structures traced by CO \citep{Soler2021}.

\section{Discussion}\label{discussion}
\subsection{Large-scale filaments in the Milky Way}\label{sec:MW}

\begin{figure*}[!t]
\centering
\includegraphics[width=.85\textwidth]{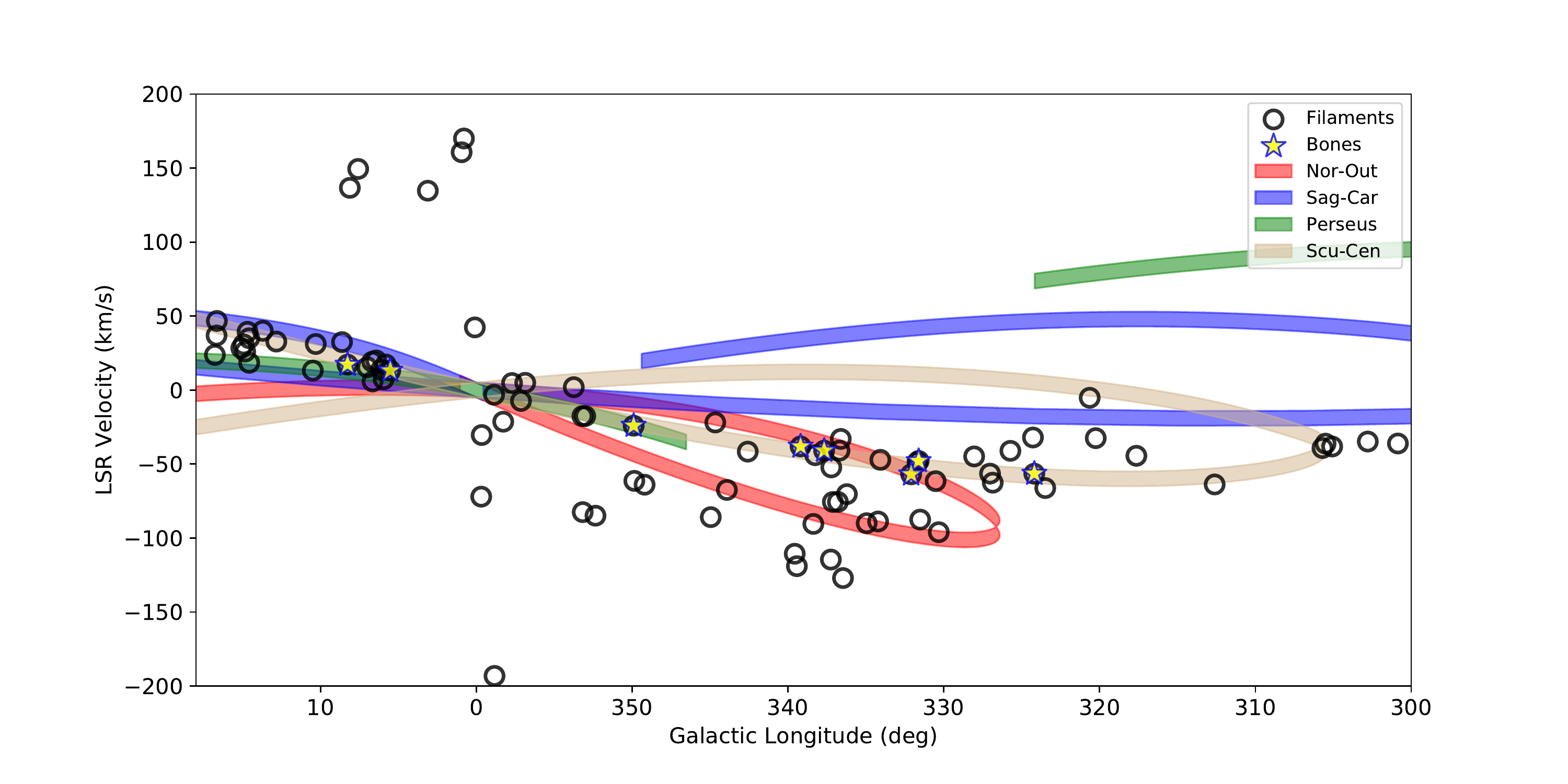}\\
\includegraphics[width=.5\textwidth]{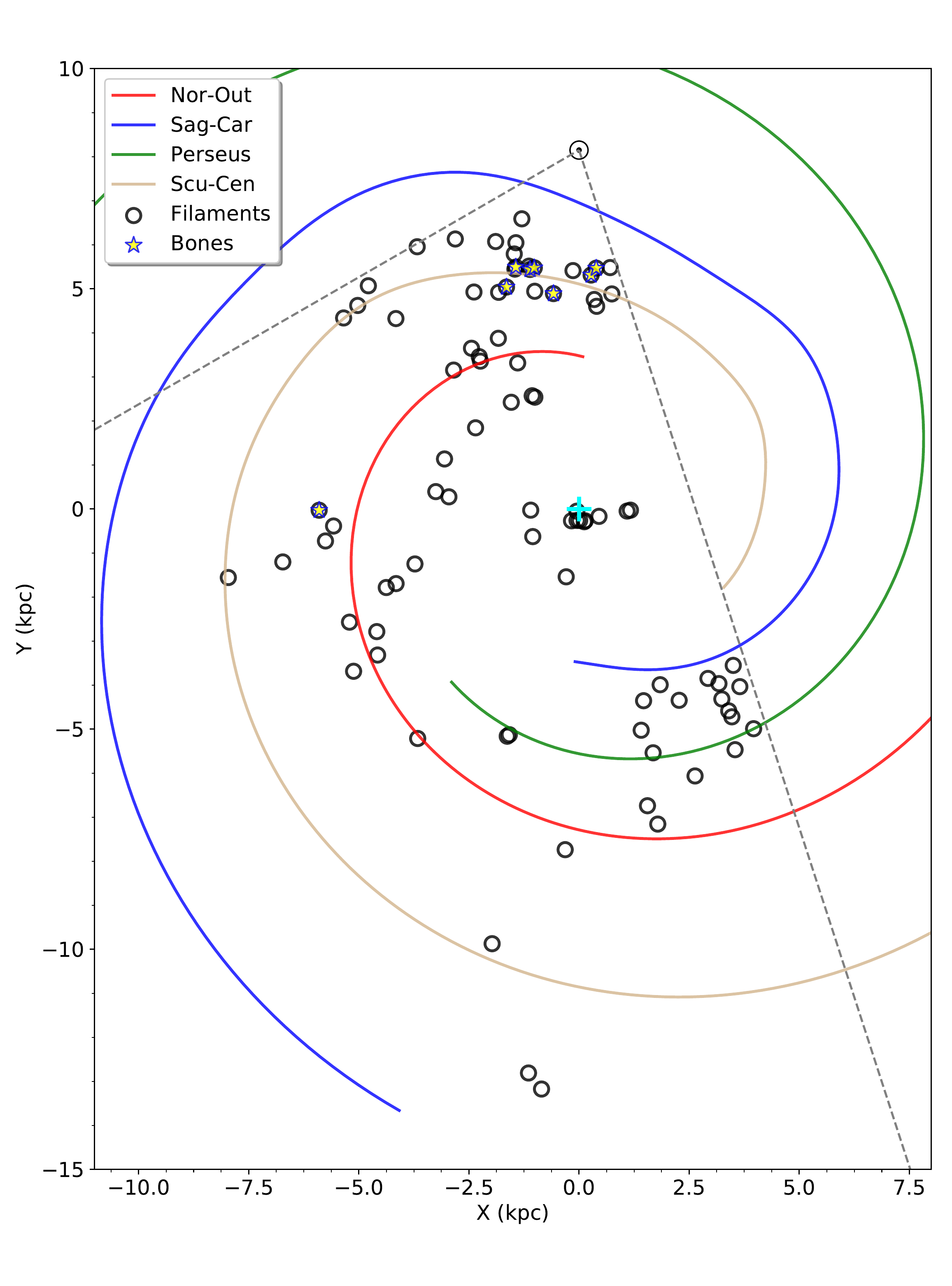}
\caption{Top panel is longitude–velocity view of filaments and spiral arms. Belts with various colors are arms in \citet{TC93} model, and widths are 10 km s$^{-1}$. Red, brown, blue, and green represent the Norma-Outer, Scutum-Centaurus, Sagittarius-Carina, and Perseus arms, respectively. Black circles denote our filaments and stars are ``bones''. The bottom panel shows a face-on view of filaments and spiral arms. The cyan plus symbol represents the Galactic center, and grey dashed lines show the data range. The solar symbol $\odot$ is plotted at (0, 8.15) kpc \citep{Reid2019}.}
\label{arm}
\end{figure*}
\subsubsection{``Bones'' of the Milky Way?}\label{sec:bone}
\noindent
  Some large-scale filaments have been described as ``bones'' of the Milky Way since \citet{Goodman2014} reported the first case. \citet{Zucker2015} established a set of criteria for large-scale filaments to be considered as bones, which was later refined by \citet{Wang2016} and \citet{Ge2022} defining three major aspects that can be adopted to judge whether a filament is a bone. They are: (1) Parallel to the Galactic plane, to within $30^\circ$; (2) Lie very close to the Galactic mid-plane, $|z|\leq 20 $pc; (3) The flux-weighted LSR velocity is within $\pm$ 5 km s$^{-1}$ from the centers of spiral arms. To investigate the third criteria, we employ the spiral arm model from \citet{TC93} as updated by \citet{Cordes2004}. We do not choose another widely used spiral arm model such as that of \citet{Reid2019} because most high-mass star-forming regions used to fit the model are in the first and second Milky Way quadrants, while the majority of our filaments reside in the fourth quadrant where their spiral arm model is determined by extrapolation. The \citet{Reid2019} spiral arm model and the more recent spiral arm model by \citet{Hou2021} are used to test the influence in our results of the chosen model described in Appendix \ref{A}.\\
  
  To investigate the association between large-scale filaments and spiral arms, we examine whether our filaments are close to spiral arms in the position-velocity (PV) space. The x and y positions of \citet{TC93} arms shown in the bottom panel of Fig. \ref{arm} are converted into $l$ and $v$ using rotation curve from \citet{Reid2019}. Solar position and velocity are also from \citet{Reid2019}, which are 8.15 kpc and 236 km s$^{-1}$. The longitude-velocity ($lv$) plot of filaments and spiral arms is shown in the top panel of Fig. \ref{arm}. Spiral arm loci (belts) with various colors represent arms in \citet{TC93} model, and widths are 10 km s$^{-1}$. Black circles denote our filaments. If the center of a circle falls within one of the belts, it means the corresponding filament is associated with this arm in PV space. We find 44 (out of a total of 88) of the filaments are arm filaments, and their PV-associated arms are listed in Col. (20) of Table \ref{t1}. Most arm filaments (25/44) are associated with the Scutum-Centaurus arm. This preference in association is also found by \citet{Abreu2016}, \citet{Zucker2018} and \citet{Mattern2018}. We can not eliminate the possibility that it is just a sensitivity effect because mass of filaments is correlated with distance (the Spearman correlation coefficient is 0.38 with a p-value of 0.0002). It could also be that this is the arm that appears to us mostly in the plane of the sky, which enhances the chance of detecting these large filaments if they are aligned with the arms. Looking through the tangent then these filaments would look a lot less ``stretched''. Five arm filaments are associated with Perseus arm, nine with Norma-Outer, and five with Sagittarius-Carina. Our data do not support a grand design spiral pattern, but neither are they opposed to this pattern. We just simply show this optimistic presumption of a spiral structure.\\
  
  For the filaments associated with spiral arms in the PV space, we consider the remaining two criteria (roughly parallel and close to Galactic mid-plane) and find that eight of them satisfy both. Therefore, eight of our filaments are bones according to the three criteria, and they are listed in Col. (20) of Table \ref{t1}. Their physical properties are shown as black bars in Fig. \ref{sta}, which do not stand out compared to other filaments. Our eight bones are smaller and less massive than those identified in \citet{Abreu2016}. Compared to bones from \citet{Ge2022}, our bones are less massive.
  \\
\subsubsection{Association with spurs}
\begin{figure*}[!t]
\centering
\includegraphics[width=.5\textwidth]{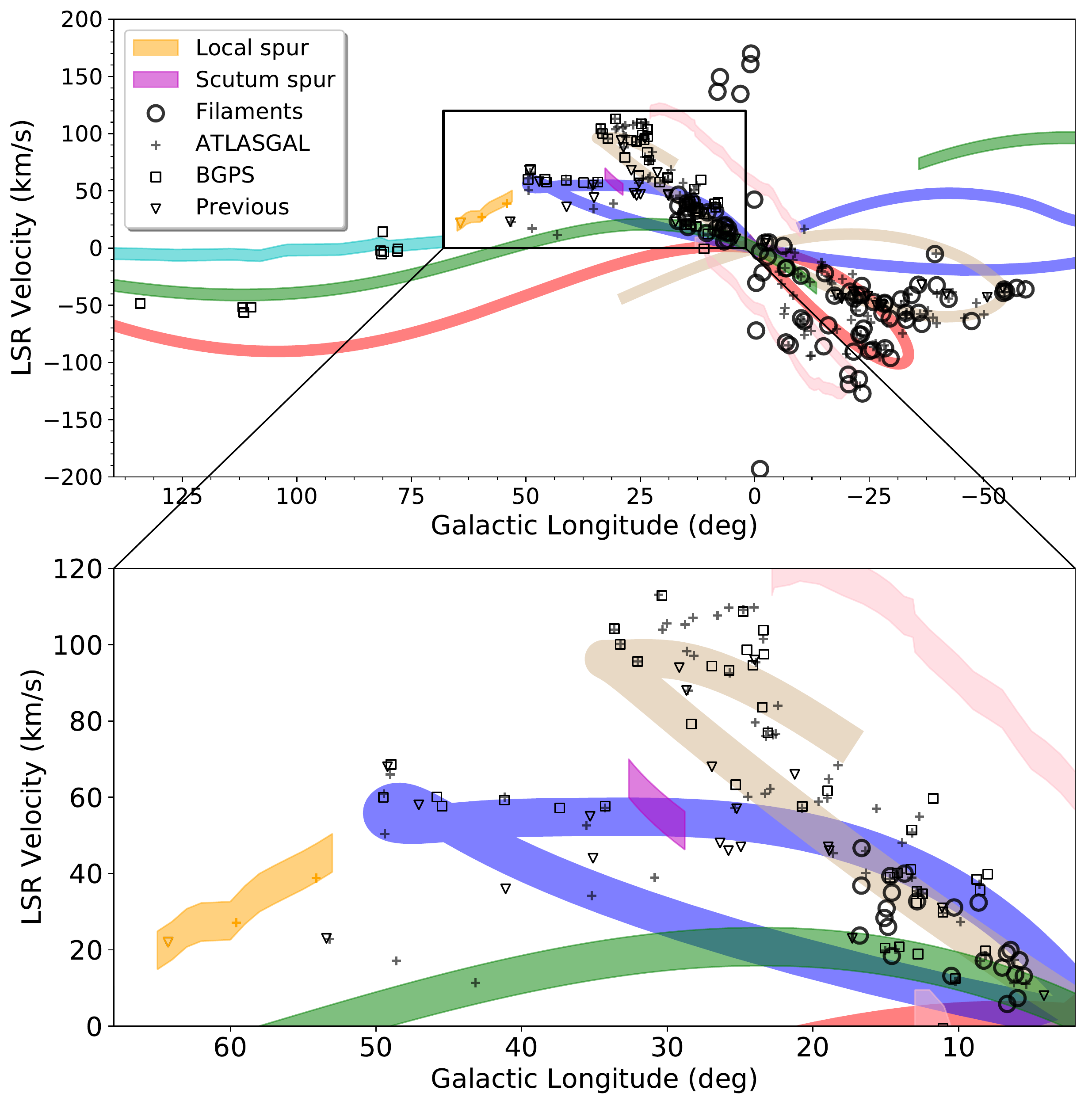}\\
\includegraphics[width=.8\textwidth]{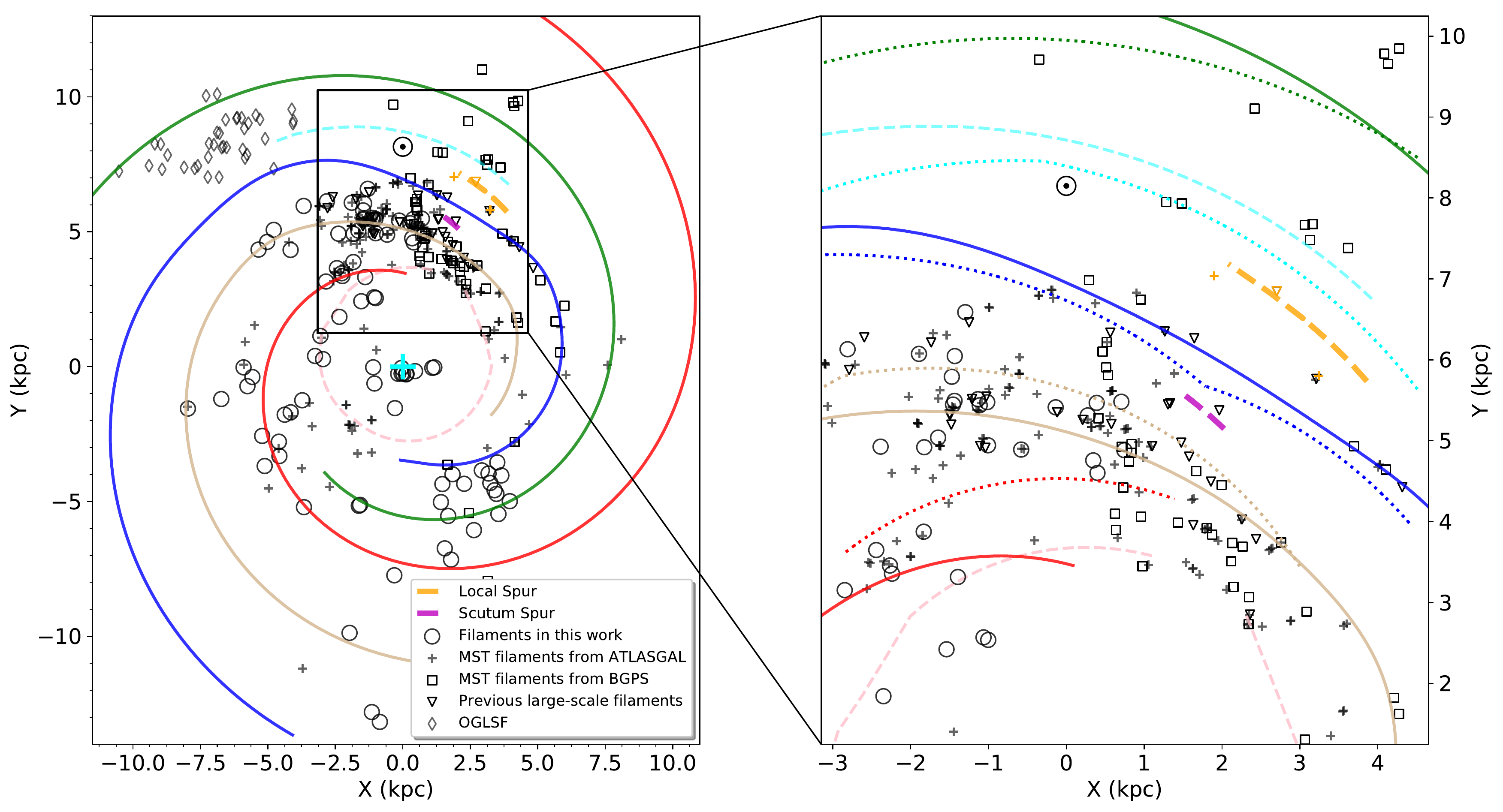}
\caption{Top panel is a longitude–velocity view of filaments and spiral arms as well as spurs. Belts with various colors are arms in \citet{TC93} model or spurs from \citet{Reid2019} with widths of 10 km s$^{-1}$. Red, brown, blue, and green represent the Norma-Outer, Scutum-Centaurus, Sagittarius-Carina, and Perseus arms, respectively. The orange belt marks the Local Spur and the purple belt is the Scutum spur. Black circles denote our SEDIGISM filaments. Black plus symbols, squares, triangles, and diamonds represent ATLASGAL filaments \citep{Ge2022}, BGPS filaments \citep{Wang2016}, previous known filaments \citep{Zucker2018}, and outer Galaxy large-scale filaments \citep[OGLSF,][]{Colombo2021}, respectively. Orange marks denote filaments associated with the Local Spur. We zoom in the region around the two spurs. Bottom panel is a face-on view of filaments and spiral arms as well as spurs. The cyan and pink dashed curve represent Local arm and 3 kpc arm from \citet{Reid2019}. Since the zoom-in region is in the solar neighborhood, as a reference, we plot arms from \citet{Hou2021} model as dotted curves. Other symbols follow the convention of the top panel.}
\label{spurs}
\end{figure*}
\noindent
   Besides the main spiral arms, some substructures named `spurs' (or `bridges' or `branches' or `armlets') are also observed in our Milky Way \citep[e.g.][]{Sofue1976,Rickard1979,Xu2016}. They are much shorter than spiral arms but hugely longer than large-scale filaments. These substructures also occur in simulations and their formation is different for the different spiral arm models, including gravitational instabilities \citep[e.g.][]{Dobbs2006}, magneto-Jeans instabilities \citep[e.g.][]{Kim2006}, wiggle instabilities \citep[e.g.][]{Wada2004,Mandowara2022}, effect of correlated SN feedbacks \citep[e.g.][]{Kim2020}. Spurs could possibly act as a diagnostic tool regarding the origin of spiral arm structure, with different arm generation mechanisms (such as tidal interactions) producing comparatively strong spurs \citep{Pettitt2020}. Study on the association between spurs and large-scale filaments is helpful for our understanding of large-scale filaments and spiral arms. \\
   
   We check whether large-scale filaments are associated with two known spurs. One is a spur identified by \citet{Xu2016} bridging the Local arm to the Sagittarius arm. It is shown as an orange belt labeled ``Local Spur'' in Fig. \ref{spurs}. The other is a spur between the Sagittarius and Scutum arms mentioned by \citet{Reid2019}. It is shown as a purple belt labeled ``Scutum spur'' in Fig. \ref{spurs}. As seen in the upper panel of Fig. \ref{spurs}, there are two ATLASGAL filaments and one previous large-scale filament lying within the Local Spur in the PV space. The two ATLASGAL filaments are shown as orange plus symbols and they are F69 and F70 in \citet{Ge2022}. The one previously identified large-scale filament is shown as an orange triangle and it is G64 from \citet{Wang2015}. As shown in the lower panel of Fig. \ref{spurs}, the three filaments are also located close to the Local Spur in the face-on view of the Galaxy. In fact, the three filaments lie very close to the Galactic mid-plane and satisfy the criteria for bones. So we suggest that these three filaments are potentially bones of the Local Spur. Based on the Local Spur CO survey, \citet{Kohno2022} find three large-scale filaments in the Sh 2-86 high-mass star-forming region. The F70 is part of their Filament C. They suggest that these filaments are formed by galactic-scale dynamics like spiral shocks or shear motions. As for Scutum Spur, no filament is associated with it.
   
\subsection{Dense Gas Mass Fraction}

\begin{figure*}[!htp]
  \centering
  \subfloat[]{
    \label{sfig:DGMFa}
    \includegraphics[width=.45\textwidth]{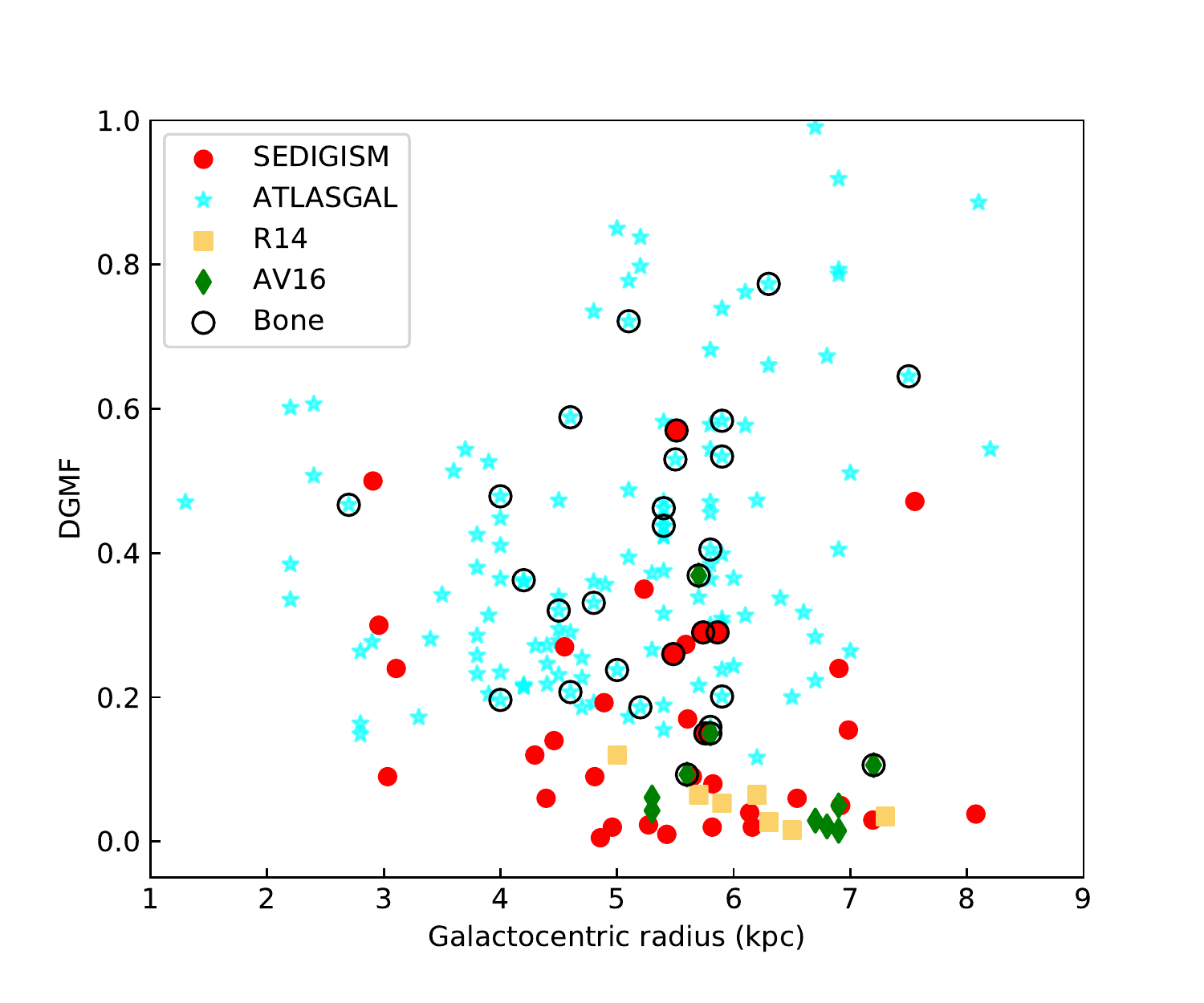}}
  \subfloat[]{
    \label{sfig:DGMFb} 
    \includegraphics[width=.45\textwidth]{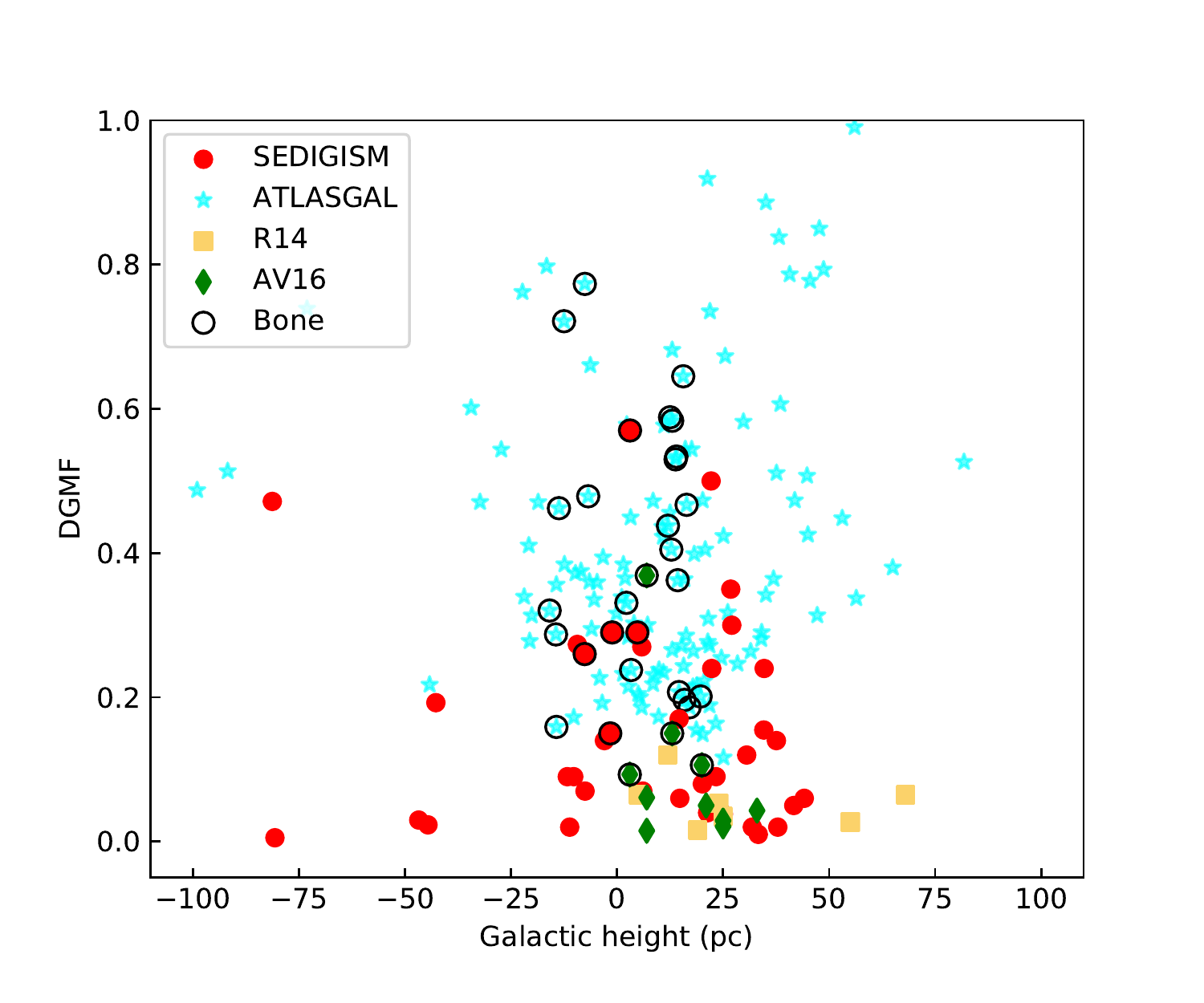}}\\
  \subfloat[]{
    \label{sfig:DGMFc} 
    \includegraphics[width=.45\textwidth]{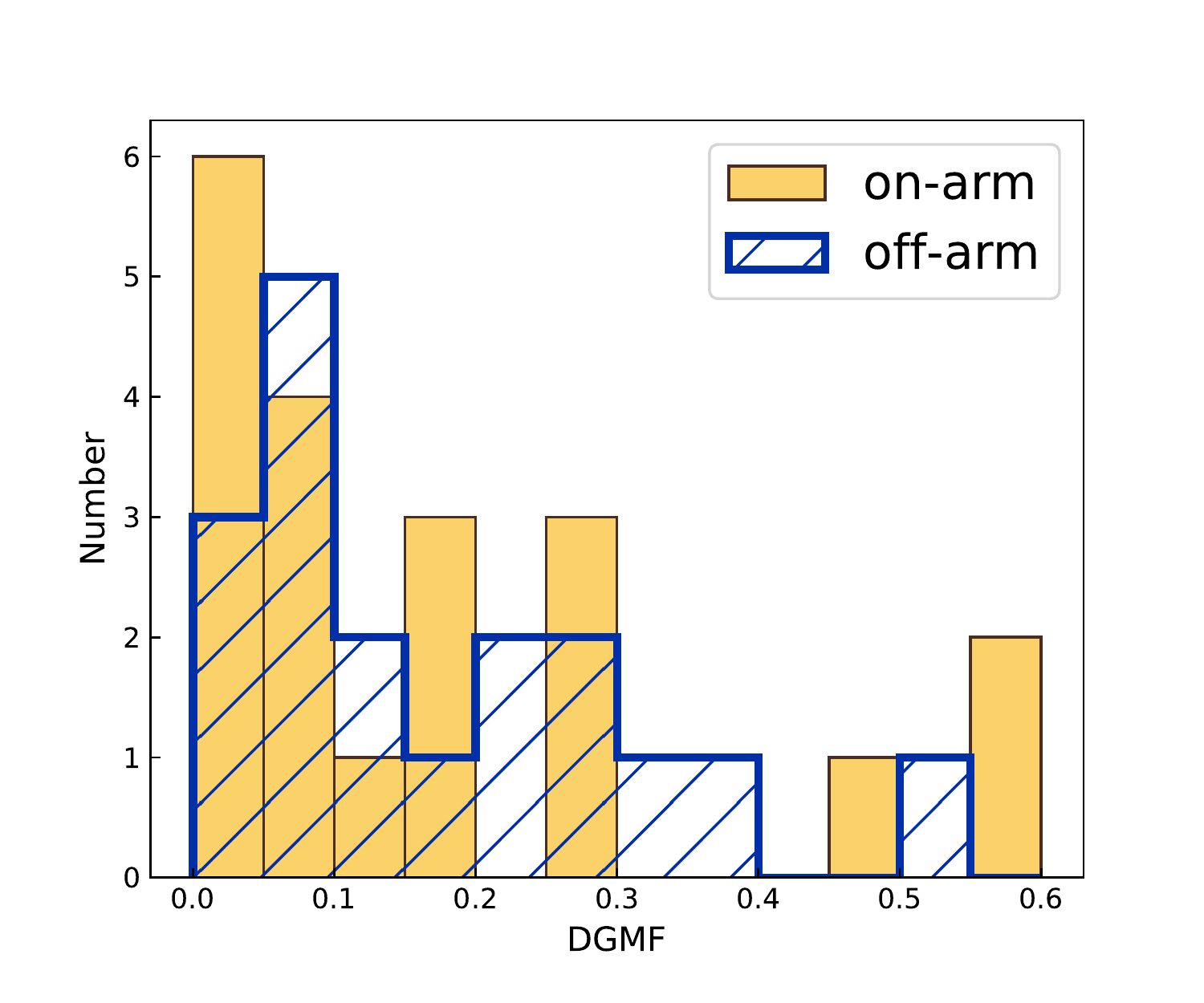}}
  \subfloat[]{
    \label{sfig:DGMFd}
    \includegraphics[width=.45\textwidth]{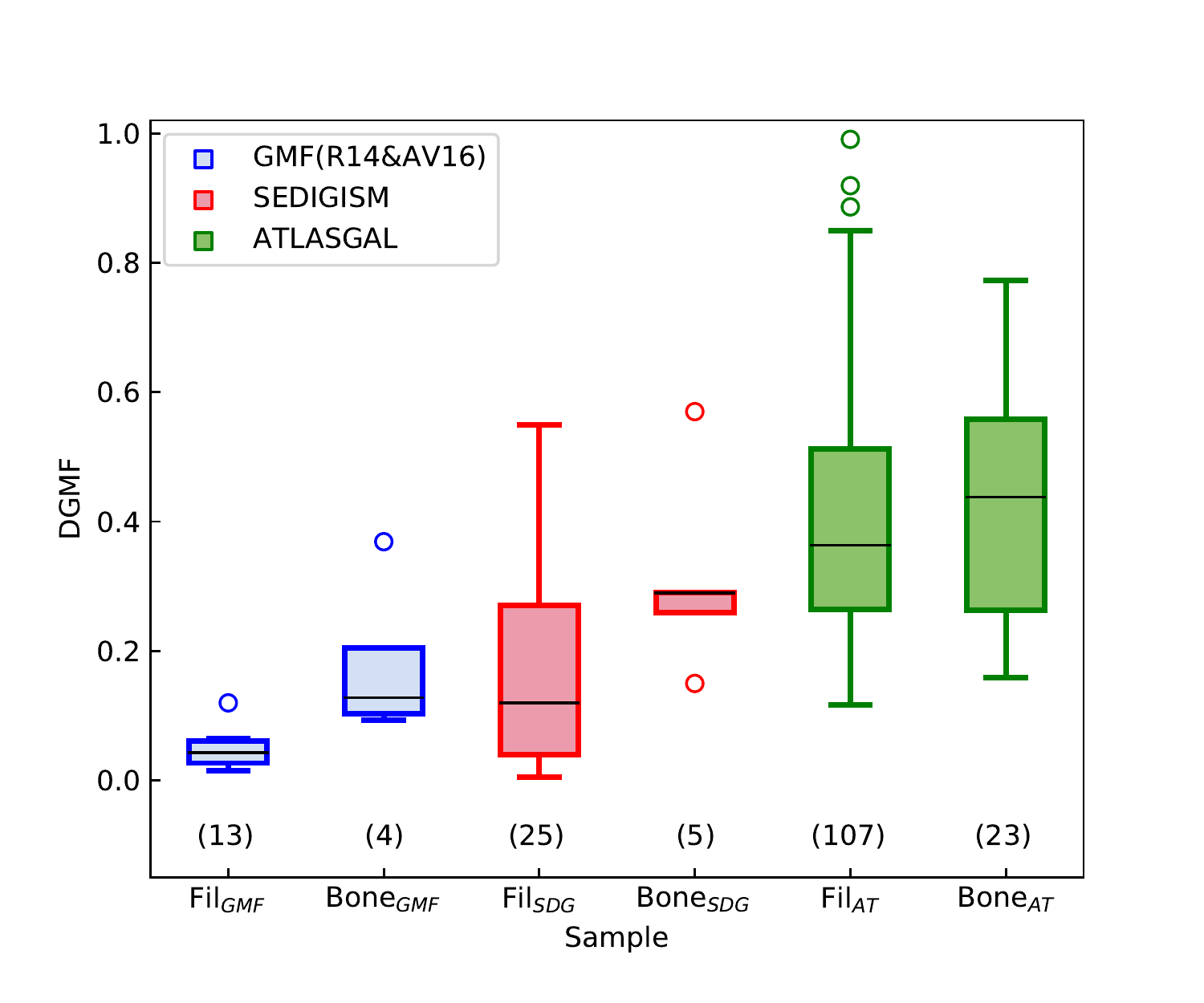}}
\caption{(a) DGMFs of large-scale filaments as a function of Galactocentric radius. Red-filled circles show DGMFs for filaments in this work. Cyan stars represent those from ATLASGAL filaments \citep{Ge2022}. Yellow squares denote filaments from \citet{Ragan2014} and green diamonds are from \citet{Abreu2016}. If a filament also satisfies the bone criteria, a black circle is added to the polygon; (b) DGMF versus Galactic height. Symbols follow the convention of the panel (a); (c) the distribution of DGMFs for on-arm or off-arm filaments. Yellow bars denote filaments in spiral arms and blue hatch marks inter-arm filaments; (d) differences in DGMF between bones and other filaments for three samples. Red boxes are SEDIGISM filaments from this work. Blue boxes represent GMFs from \citet{Ragan2014,Abreu2016}. Green boxes mark ATLASGAL filaments. The right boxes for each color are bones and the left ones are other filaments. A box extends from the first quartile to the third quartile of the data, with a line at the median. The whiskers extending from the boxes enclose all of the data except outliers, which are denoted as dots. Numbers above the bottom border are the numbers of bones or other filaments.}
\label{DGMF}
\end{figure*}

\noindent
  The DGMF (or CFE) is the dense-clump analogue (or precursor) of the SFE and is tightly associated with star formation \citep[e.g.][]{Lada2012,Eden2013,Battisti2014}. There are two key points in the calculation of DGMFs. One is to define dense gas properly, and the other is to unify the dense gas mass and the total mass with the same formulation and parameters. The dense gas in this work is determined as the total mass of ATLASGAL clumps matching a given filament in position and velocity \citep{Urquhart2021}. The DGMF is then the ratio of the dense gas mass to the filament mass as listed in Table 1. Since the masses of ATLASGAL clump and SEDIGISM clouds are calculated with different dust opacity values, \citet{Urquhart2021} unify the them to the same opacity. The ATLASGAL uses 1.85~cm$^2$g$^{-1}$ at 350 GHz as opacity \citep{Schuller2009} while the SEDIGISM employ CO X factor to converse CO to H$_2$. The X factor is derived from an opacity of 0.1~cm$^2$g$^{-1}$ at 1200 GHz \citep{Elia2013}. This opacity will be $8.5\times10^{-3}$~cm$^2$g$^{-1}$ at the ATLASGAL frequency. The result of the opacity difference is that the ATLASGAL clump masses should be multiplied by a factor of 1.61 to match the masses of SEDIGISM clouds \citep{Urquhart2021}. The relating $5\sigma$ column density threshold for the ATLASGAL survey should also change from $7.5\times10^{21}{\rm cm}^{-2}$ to ~$1.5\times10^{22}{\rm cm}^{-2}$. For comparison, the column density sensitivity of SEDIGISM is ~$0.95\times10^{21}{\rm cm}^{-2}$. The advantage of this calculation is that it has the well-defined dense gas mass and the parameters have been calculated following a similar approach. We note that we do not derive or provide a DGMF value for those filaments that are not associated with ATLASGAL clumps.\\ 
  
  Comparing our DGMF calculation to that of other works, the calculation of dense gas mass and total mass of filaments in \citet{Ge2022} are both from Herschel column density maps through the PPMAP approach \citep{Marsh2017}. They take circles guided by clumps in a given filament as the boundary of the dense gas, and a combination of circles and rectangles guided by the filament as the boundary for the total mass calculation. The advantage of this calculation is that unification for data from different observations is not needed, and each pixel in a clump has a temperature measurement rather than a single temperature for the clump. However, the mass of unrelated structures that overlap with filaments in the projected sky may be incorrectly included in the mass calculation because the column density map is 2D. \citet{Ragan2014} and \citet{Abreu2016} restrict their measurement in a mask created by $^{13}$CO emission for both the dense gas mass and the total mass. They use the dust emission from the ATLASGAL 870 \textmu m survey \citep{Schuller2009} to calculate dense gas mass and $^{13}$CO emission to calculate the total mass of filaments. But limited by data available at that time, their temperature for the dense gas calculation had to be assumed, and the excitation temperature of $^{13}$CO was assumed to be the same as that of $^{12}$CO for the total mass calculation. Their dense gas mass and total mass are also not unified to the same dust opacity values. \\
  
  To investigate whether the DGMF of filaments is related to their Galactic location, we examine this quantity at different Galactocentric radii and Galactic heights in Fig. \ref{DGMF}. As we can see, DGMFs in this work (red dots in the top two panels of Fig. \ref{DGMF}) are neither correlated with Galactocentric radius nor Galactic height. An analysis of 163 ATLASGAL filaments (cyan stars in the top two panels of Fig. \ref{DGMF}) provides similar results \citep{Ge2022}. Here we recalculate their DGMFs by changing the dense gas mass to be the mass of ATLASGAL clumps located within the filaments because the other three works we mention use this definition for the dense gas mass. \citet{Abreu2016} also find that the DGMF is not correlated with Galactic height for their ten filaments in the fourth Galactic quadrant (green diamonds in  Fig. \ref{DGMF} (b)). \citet{Ragan2014} suggest that the DGMF decreases with increasing Galactocentric radius for their seven large-scale filaments shown as yellow squares in Fig. \ref{DGMF} (a) and they find that GMFs located closer to the Galactic mid-plane tend to have higher DGMF values than those significantly out of the plane (yellow squares in Fig. \ref{DGMF} (b)). However, their filaments are selected as the densest and longest ones and they also indicate that the trends are weak and not statistically robust due to the small sample size. We note that comparing the absolute values of the DGMF across different works makes little sense unless they are unified carefully.\\

  We also investigate how the DGMF may change between inter-arm filaments and filaments located within spiral arms. As shown in Fig. \ref{DGMF} (c), DGMFs in spiral arms or inter-arm regions have no clear differences (a KS-test gives a p-value of 0.5). The mean values of the DGMF for on-arm and off-arm filaments are 17.3\% and 17.1\%, respectively. Considering the tight relation between DGMF and star formation \citep[e.g.][]{Lada2012}, we suggest that the star-forming potential in large-scale filaments has no distinct association with respect to their location with spiral arms. This result is consistent with \citet{Ge2022} but in contrast with \citet{Ragan2014} and \citet{Abreu2016}. By analysing 8 \citep{Ragan2014} and 16 \citep{Abreu2016} filaments, they found that mean DGMF of spiral arm filaments is higher than that of the inter-arm filaments. Studies of giant molecular clouds also find that the DGMFs of molecular clouds exhibit no differences between the inter-arm and spiral-arm regions \citep{Eden2013}. \citet{Urquhart2021} find that there is no enhancement of dense gas in clouds with respect to their proximity to spiral arms. Other physical properties of molecular clouds such as effective radius, molecular gas mass, molecular gas mass surface density, and virial parameter have no significant differences between spiral arms and inter-arm regions if the distance bias is avoided \citep{Colombo2022}. As for different arms, the mean DGMFs are $23.0\%\pm5.6\%$, $9.4\%\pm4.1\%$ and $7.6\%\pm4.8 \%$ for filaments in the Scutum-Centaurus, Norma-Outer, and Sagittarius-Carina arms, respectively, where the uncertainties are the standard errors. \citet{Abreu2016} also find that the mean DGMF of filaments in the Scutum-Centaurus arm is higher than those in the Sagittarius arm. But they lack a statistically significant sample. Of their nine filaments relating to spiral arms, only two are located within the Sagittarius arm and the other seven are within the Scutum-Centaurus arm. From the study of 1619 molecular clouds, \citet{Eden2021} find that DGMFs are similar for the Scutum–Centaurus, Sagittarius, and Perseus spiral arms. From power-spectrum analysis, they also suggest that the largest variations in DGMF occur from cloud to cloud rather than on larger scales. That study is based upon CHIMPS molecular clouds from \citet{Rigby2016} extracted from $^{13}$CO(3-2), at $27^{\prime \prime}$ resolution, and so closely related to the tracer ($^{13}$CO(2-1)) used and resolution ($28^{\prime \prime}$) in SEDIGISM. In addition, \citet{Rigby2019} find that most properties for their sample are indistinguishable between the Sagittarius and Scutum-Centaurus arms.\\
  
  Spiral arms are important in gathering gas molecular \citep{Wang2020b}. However, the lack of any trends in DGMFs as a function of Galactic location in the disk suggests that the conditions for star formation, which is directly related to DGMFs, is mostly unaffected by Galactocentric radius, Galactic height, and association with spiral arms. Instead, the conditions for star formation is determined by factors at much smaller scales. This reinforces the conclusion by \citet{Wang2016} and is consistent with the result of other studies that the subsequent star formation processes after the cloud formation and HI to H$_2$ conversion, depend more on local environment \cite[e.g.][]{Urquhart2021,Colombo2022}.\\

  Do DGMFs stand out in bones? We investigate this issue by comparing DGMFs for bones and other filaments. As shown in Fig. \ref{DGMF} (d), for SEDIGISM filaments in this work (red boxes), DGMFs of bones are significantly higher than for other filaments. We also classify GMFs \citep{Ragan2014,Abreu2016} into bones and other filaments according to criteria in Sect.\ref{sec:bone}. GMFs from \citet{Ragan2014} are all classified as other non-bone filaments due to their departure from spiral arms or verticality to the Galactic plane. Four GMFs from \citet{Abreu2016} pass the bone criteria and are classified as bones. DGMFs of GMFs for bones and other filaments are plotted as blue boxes in Fig. \ref{DGMF} (d). For bones, DGMFs are also obviously larger than other filaments, which is the same as the result for filaments in this work. For ATLASGAL filaments (green boxes), this difference is not so evident. Only the median DGMF is a bit higher for bones than for other filaments. To summarize, we collect DGMFs of 4 bones and 13 other filaments from \citet{Ragan2014} and \citet{Abreu2016}, 23 bones and 107 other filaments from \citet{Ge2022}, as well as 5 bones and 25 other filaments from SEDIGISM filaments in this work. If we exclude redundant filaments that are identified several times by different works, the total sample size for DGMFs is 168. It seems that bones, filaments located at the very center of spiral arms, potentially stand out in DGMFs when compared to other filaments, and thus promoting star formation activity. However, on one hand, the number of known bones is still small. On the other hand, the dust continuum is possibly not always a good tracer of dense gas due to optical depth effects. Therefore, in the future we will collect more bones to investigate this issue. In addition, we will employ an optically thin dense gas tracer (such as N$_2$H$^+$) to recalculate the dense gas content. 
  
\section{Summary}\label{sec:summary}
\noindent
  We have built a catalog of 88 large-scale $^{13}$CO filaments in the inner Galactic plane ($-60^\circ <l<18^\circ$, $|b|<0^\circ.5$). The catalog is composed of 55 large-scale filaments identified through SEDIGISM leaves using MST and 33 large elongated filaments selected from the SEDIGISM cloud catalog.
  Our main results are summarized as follows.
  
   \begin{enumerate}
      \item Of the 88 filaments, 66 are not included in the former large-scale filament catalogs. The $^{13}$CO emission is able to trace relatively diffuse structures with lower density than continuum surveys like BGPS and ATLASGAL.
      \item We compare the physical properties of our filaments to MST filaments identified through BGPS and ATLASGAL sources. The $^{13}$CO filaments are on average longer than the other two (based on the dust continuum). The column density and line mass of our filaments are lower than the other two.
      \item We investigate the association of our filaments with the Galactic spiral arms to determine whether they satisfy the criteria to be considered as Galactic bones. We find half of the filaments are associated with spiral arms in PV space. We find eight bone candidates of the Milky Way.
      \item From the study of Galactic spur and large-scale filaments, we find that three filaments satisfy the criteria for bones associated with the Local Spur in PPV space. So we suggest that they may be bone candidates of the Local Spur.
      \item By compiling 168 large-scale filaments with unified DGMFs across the Galaxy, an order of magnitude more than previously investigated, we find that DGMFs of large-scale filaments do not correlate with Galactocentric radius or Galactic height. DGMFs of filaments in spiral arms or inter-arm regions have no distinct differences. But a further comparison between bones and other filaments indicates that bones have higher DGMFs than other filaments. 
      
   \end{enumerate}

\begin{acknowledgements}
      We thank Ligang Hou, Siju Zhang, Nannan Yue, Fengwei Xu, Chao Wang, Wenyu Jiao, and Juan Diego Soler for the helpful discussion. We thank an anonymous referee for constructive comments, which has significantly improved the manuscript. We acknowledge support from the National Science Foundation of China (12033005, 11973013), the China Manned Space Project (CMS-CSST-2021-A09), the National Key Research and Development Program of China (2022YFA1603102, 2019YFA0405100), and the High-Performance Computing Platform of Peking University. DC acknowledges support by the German \emph{Deut\-sche For\-schungs\-ge\-mein\-schaft, DFG\/} project number SFB956A. ADC acknowledges the support by the Royal Society University Research Fellowship (URF/R1/191609). HB acknowledges support from the Deutsche Forschungsgemeinschaft in the Collaborative Research Center (SFB 881) “The Milky Way System” (subproject B1). LB gratefully acknowledges support by the ANID BASAL projects ACE210002 and FB210003. CLD acknowledges funding from the European Research Council for the Horizon 2020 ERC consolidator grant project ICYBOB, grant number 818940. 
      This publication is based on data acquired with the Atacama Pathfinder Experiment (APEX) under programmes 092.F-9315 and 193.C-0584. APEX is a collaboration among the Max-Planck-Institut fur Radioastronomie, the European Southern Observatory, and the Onsala Space Observatory. The processed data products are available from the SEDIGISM survey database located at \url{https://sedigism.mpifr-bonn.mpg.de/index.html}, which was constructed by James Urquhart and hosted by the Max Planck Institute for Radio Astronomy.
\end{acknowledgements}

\bibliographystyle{aa.bst}
\bibliography{ref_SED.bib}

\appendix

\renewcommand\thefigure{\Alph{section}\arabic{figure}} 
\section{Influence of Different Spiral Arm Models}\label{A}

\begin{figure*}
\centering
\includegraphics[width=.8\textwidth]{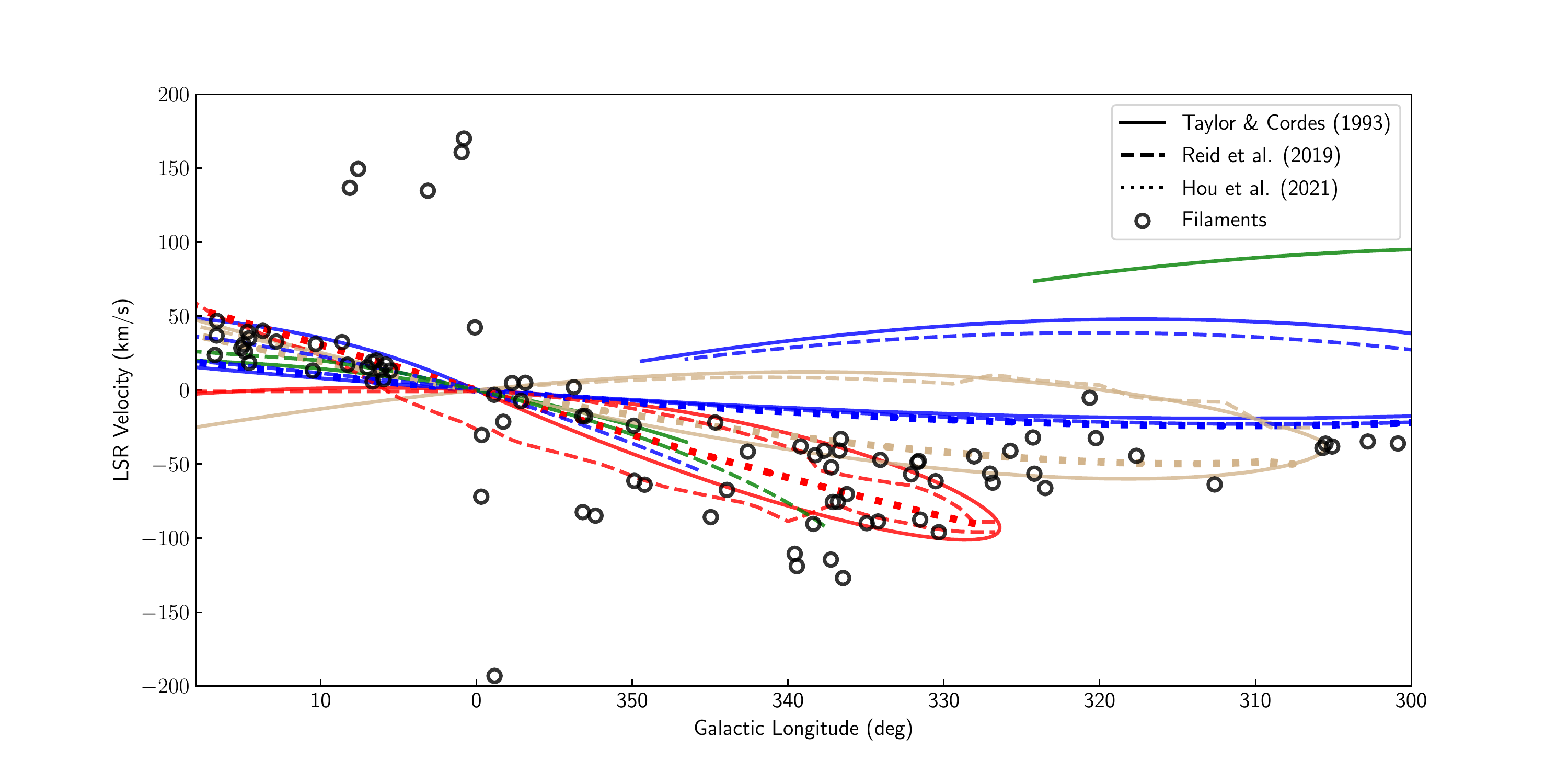}\\
\includegraphics[width=.5\textwidth]{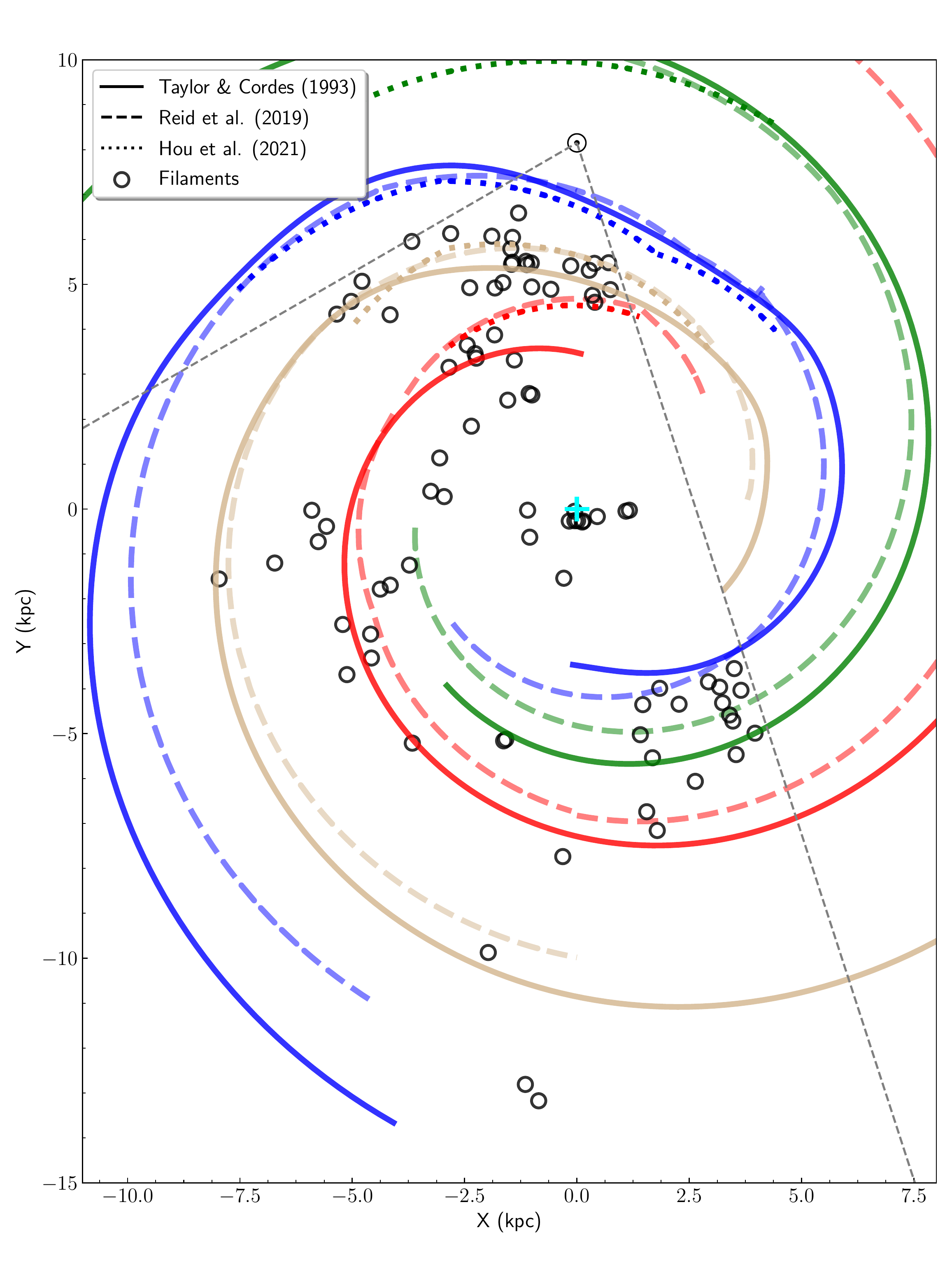}
\caption{Top panel is longitude–velocity view of filaments and spiral arms. Spiral arms from \citet{TC93}, \citet{Reid2019}, and \citet{Hou2021} are plotted as solid lines, dashed lines, and dotted lines, respectively. Red, brown, blue, and green represent the Norma-Outer, Scutum-Centaurus, Sagittarius-Carina, and Perseus arm, respectively. Black circles denote our filaments. Bottom panel is face-on view of filaments and spiral arms. The cyan plus symbol represents the Galactic center, and the grey dashed lines show the data range.}
\label{3arm}
\end{figure*}

\noindent
  To investigate to what degree different spiral arm models will affect our results, we test two other spiral arm models. One is the model in \citet{Reid2019}, R19 model, hereafter. They fit log-periodic spirals to the locations of masers and extend to the fourth Galactic quadrant with arm tangencies. Another recent spiral arm model is constructed by \citet{Hou2021}, H21 model, hereafter. They summarize good spiral tracers such as giant molecular clouds, high-mass star-formation region masers, HII regions, O-type stars and young open clusters to depict the nearby spiral arms. These two spiral arm models are overlaid on the \citet{TC93} model (TC93 model, hereafter) in Fig. \ref{3arm}. On the whole, the three spiral arm models are consistent with each other. But there still exist some differences. In the PV space, the three models have the largest difference for the Norma-Outer arm. On the face-on view, nearby arms in R19 model and H21 model show consistency. At the beginning of the Norma-Outer arm, TC93 model has large separation with R19 model and H21 model. In the fourth Galactic quadrant, the Norma-Outer, Scutum-Centaurus, and Sagittarius-Carina arm have differences for R19 model and TC93 model. The number of filaments associated with each arm in PV space differs with respect to different models, but the number of bones is six for R19 model, which is similar with TC93 model (eight bones). Difference of mean DGMFs for on-arm filaments and off-arm filaments is a factor of 1.1 if we use H21 model, and 1.4 for R19 model. Differences of mean DGMFs for different arms are less than 10\% if we use R19 model. For H21 model, mean DGMF in Scutum-Centaurus Arm is about twice that of Norma Arm. The statistics on different spiral models are summarized in Table\ref{t3}.
\onecolumn
\begin{table}[htb]
\centering
\begin{tabular}{cccccc}
\toprule
Model &                                                                     On-arm filaments & Off-arm filaments & Number of bones &  DGMFs for on-arm & DGMFs for off-arm \\
\midrule
TC93 &                                                              
    44  & 44    & 8     & 0.173  & 0.171 \\
R19 &                                                              
    25  & 63    & 6     & 0.200  & 0.146 \\
H21 &                                                              
    14  & 74    & 3     & 0.155  & 0.178 \\
\bottomrule
\end{tabular}
\caption{Summary of basic statistics based on different spiral models. The first column lists spiral models used. TC93 for \citet{TC93}, R19 for \citet{Reid2019}, and H21 for \citet{Hou2021}. The second column lists the number of filaments associated with spiral arms for each spiral model and the third column lists the number of filaments that are not associated with any arms. The fourth column lists the number of filaments satisfying the criteria for bones. The fifth column lists mean DGMF for filaments associated with spiral arms, while the sixth column lists those off arms.}
\label{t3}
\end{table}
\clearpage
\onecolumn
\section{Original SEDIGISM Cloud IDs of Leaves in MST Filaments }\label{sec:t2}
\begin{table}[!b]
\centering

\begin{tabular}{ll}
\toprule
Filament ID &                                                                     SEDIGISM cloud id \\
\midrule
         M1 &                                                              [40, 44, 47, 48, 54, 55] \\
         M2 &                                                        [101, 110, 115, 155, 156, 158] \\
         M3 &                                                   [248, 254, 255, 257, 259, 264, 333] \\
         M4 &                                                        [272, 317, 348, 349, 350, 351] \\
         M5 &                                                  [1697, 1699, 1701, 1702, 1791, 1798] \\
         M6 &                          [1984, 1986, 1987, 1988, 2098, 2099, 2100, 2101, 2102, 2103] \\
         M7 &                                            [2014, 2017, 2019, 2020, 2125, 2126, 2127] \\
         M8 &                                                  [2199, 2222, 2363, 2367, 2368, 2372] \\
         M9 &                                [2235, 2239, 2393, 2397, 2400, 2401, 2403, 2615, 2637] \\
        M10 &                          [3524, 3542, 3543, 3546, 3547, 3804, 3806, 3808, 3809, 3824] \\
        M11 &                                            [3691, 3693, 3699, 3702, 3708, 4006, 4009] \\
        M12 &                                            [3907, 3908, 3912, 3913, 3914, 4186, 4198] \\
        M13 &                          [4224, 4226, 4230, 4239, 4241, 4543, 4544, 4551, 4560, 4562] \\
        M14 &                                                  [4346, 4348, 4623, 4628, 4629, 4632] \\
        M15 &                                                  [4420, 4422, 4423, 4424, 4673, 4684] \\
        M16 &                                                  [4160, 4161, 4162, 4466, 4467, 4469] \\
        M17 &                                                  [4714, 4904, 4913, 4915, 4918, 4924] \\
        M18 &                                            [4597, 4598, 4600, 4842, 4850, 4852, 4853] \\
        M19 &                                                  [4063, 4067, 4366, 4372, 4374, 4376] \\
        M20 &                                                  [4793, 4803, 4805, 4815, 5070, 5072] \\
        M21 &                                            [5523, 5524, 5531, 5533, 5679, 5680, 5684] \\
        M22 &                                                  [5618, 5623, 5743, 5750, 5826, 5841] \\
        M23 &                                                  [2122, 2123, 2276, 2281, 2282, 2285] \\
        M24 &                                [6215, 6220, 6224, 6322, 6324, 6328, 6337, 6339, 6346] \\
        M25 &                                      [6546, 6547, 6554, 6697, 6698, 6699, 6700, 6704] \\
        M26 &              [6614, 6615, 6618, 6621, 6762, 6767, 6768, 6769, 6770, 6771, 6774, 6777] \\
        M27 &                                [6761, 6766, 6948, 6951, 6952, 6953, 6962, 6970, 6975] \\
        M28 &                                                  [6868, 6873, 6897, 6939, 6942, 6961] \\
        M29 &                                      [7205, 7211, 7215, 7216, 7310, 7315, 7317, 7320] \\
        M30 &                                                  [7399, 7538, 7541, 7542, 7543, 7544] \\
        M31 &                                            [7434, 7435, 7441, 7564, 7567, 7578, 7586] \\
        M32 &                                                  [7369, 7484, 7488, 7490, 7491, 7492] \\
        M33 &                                            [7896, 7897, 7898, 7899, 7901, 8033, 8036] \\
        M34 &                                                  [7891, 8022, 8024, 8026, 8027, 8028] \\
        M35 &                                                  [8252, 8254, 8256, 8257, 8258, 8260] \\
        M36 &  [8484, 8493, 8528, 8595, 8606, 8612, 8616, 8617, 8681, 8682, 8735, 8742, 8760, 8765] \\
        M37 &                    [8413, 8414, 8416, 8430, 8432, 8533, 8558, 8559, 8563, 8579, 8583] \\
        M38 &                                            [8514, 8538, 8564, 8580, 8633, 8767, 8768] \\
        M39 &                                            [8499, 8523, 8544, 8652, 8659, 8667, 8669] \\
        M40 &                                                  [8626, 8707, 8723, 8724, 8876, 8879] \\
        M41 &                                                  [8816, 8817, 8818, 8971, 8972, 8973] \\
        M42 &                                            [8963, 8964, 8965, 9023, 9036, 9077, 9081] \\
        M43 &                                            [8918, 9067, 9068, 9071, 9073, 9074, 9082] \\
        M44 &                                                  [9190, 9198, 9203, 9206, 9214, 9362] \\
        M45 &                                            [9250, 9253, 9266, 9269, 9270, 9272, 9423] \\
        M46 &     [9735, 9746, 9972, 9974, 9981, 9983, 9992, 9993, 9995, 9998, 10018, 10020, 10024] \\
        M47 &                                          [9913, 9917, 9921, 9923, 9928, 10132, 10152] \\
        M48 &                                       [9930, 9931, 10140, 10154, 10161, 10163, 10164] \\
        M49 &             [9875, 9882, 9887, 9888, 10102, 10105, 10107, 10108, 10115, 10116, 10118] \\
        M50 &                               [9940, 9941, 9947, 9950, 9951, 9952, 9953, 9957, 10172] \\
        M51 &                                [9978, 9979, 10180, 10185, 10191, 10196, 10207, 10208] \\
        M52 &                                     [10131, 10139, 10143, 10147, 10150, 10388, 10392] \\
        M53 &                              [10299, 10312, 10336, 10338, 10556, 10568, 10579, 10580] \\
        M54 &                                            [10403, 10404, 10531, 10540, 10550, 10553] \\
        M55 &                              [10472, 10482, 10484, 10485, 10487, 10491, 10497, 10503] \\
 
\bottomrule
\end{tabular}
\caption{The first column lists IDs of our filaments identified by means of the MST approach. The second column shows the origin SEDIGISM cloud ids \citep{DuarteCabral2021} of the leaves that the filament includes.}
\label{t2}
\end{table}
\clearpage
\twocolumn
\section{Comparison of the results from the MST and the SCIMES}\label{sec:ap_twomethod}
\noindent
   As we combine the filaments identified by the MST algorithm and filaments selected from clouds identified by the SCIMES algorithm to obtain a filament sample, the differences between the two methods should be examined. We split the 33 SCIMES filaments in our catalogue (S56-S88) into leaves and apply the MST to the leaves. We find that 11/33 can be recovered by the MST without any changes on parameters. For example, S75 is shown in the top panel of Fig.\ref{fig:ap_twomethod}. We can see, the MST successfully connects the leaves of S75. If we change the parameters for the MST not too much, another 17 SCIMES filaments can be re-connected. An example is S56 shown in the middle panel of Fig.\ref{fig:ap_twomethod}. To re-connect the leaves, the matching velocity is changed from 2 km s$^{-1}$ to 3.4 km s$^{-1}$. For the rest 5 SCIMES filaments, parameters should change much to re-connect them. For example, S59 in the bottom panel of Fig.\ref{fig:ap_twomethod}. One leaf (the bottom pink circle) is too far from the others (the top 4 pink ones) to be connected. As a result, the number of leaves can not reach 5 and thus eliminated by the MST algorithm. So the two methods have similarities as most of filaments identified by the SCIMES can be reconnect by the MST. But they also focus differently and complementary so as to give a more complete sample.

\renewcommand\thefigure{\Alph{section}\arabic{figure}} 
\setcounter{figure}{0} 
\begin{center}
\setlength{\tabcolsep}{1.2mm}{
{
\doublerulesep=5pt
\begin{figure*}[!htp]
  \centering
  \subfloat[]{
    \label{sfig:tma}
    \includegraphics[width=1\textwidth]{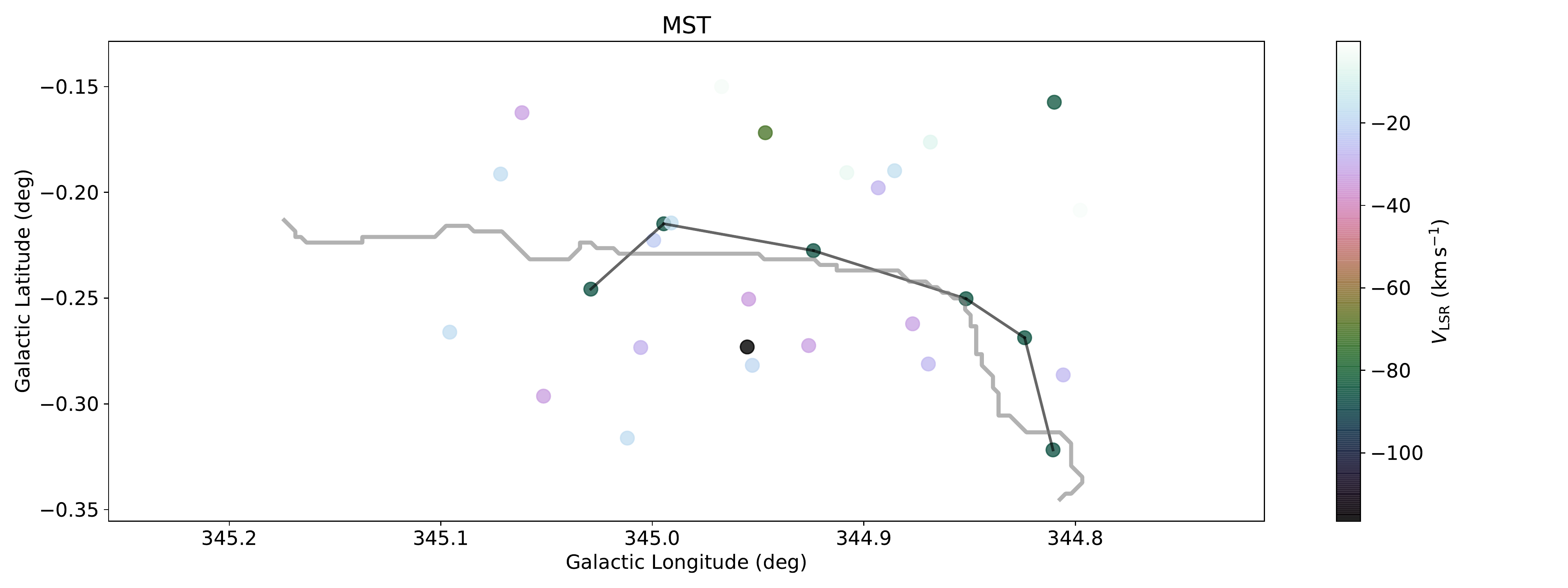}}\\
  \subfloat[]{
    \label{sfig:tmb} 
    \includegraphics[width=1\textwidth]{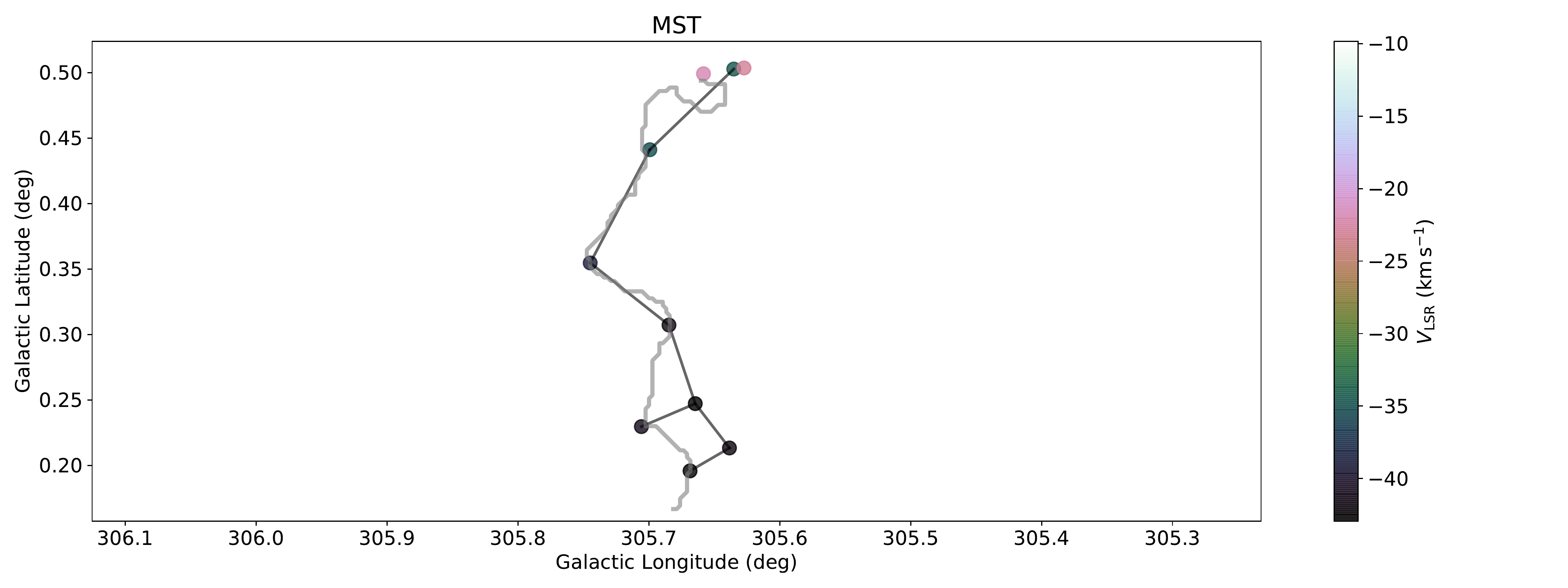}}\\
  \subfloat[]{
    \label{sfig:tmc} 
    \includegraphics[width=1\textwidth]{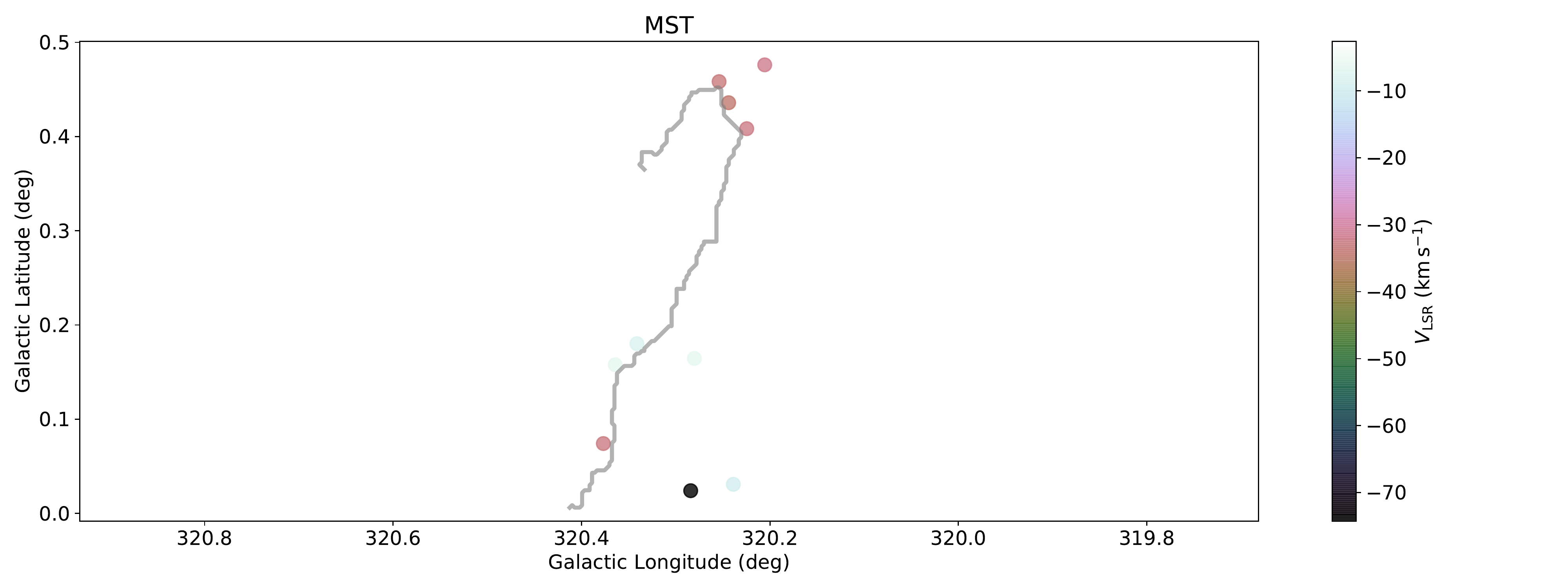}}
\caption{Re-connecting SCIMES filaments with MST. Circles are SEDIGISM leaves color coded with velocity. The color bars show their radial velocities. Black line segments are MST edges and gray curves are SCIMES filament spines.}
\label{fig:ap_twomethod}
\end{figure*}
}}
\end{center}

\section{Two Color Views of Filaments}\label{sec:ap_twocolor}
\noindent
   We have displayed several filaments in Fig. \ref{twocolor}. Two color composite images of all the filaments are shown in Fig. \ref{fig:ap_twocolor}. The color-coded circles denote leaves in filaments with different velocities. For M1-M55, white line segments are edges. For S56-S88, white curves are their medial axis. For backgrounds, cyan represents intermediate infrared 24 $\mu m$ emission on logarithmic scale from MIPSGAL \citep{Carey2009} and red shows integrated $^{13}$CO(2-1) emission in linear scale. 
\renewcommand\thefigure{\Alph{section}\arabic{figure}} 
\setcounter{figure}{0} 
\begin{center}
\setlength{\tabcolsep}{1.2mm}{
{
\doublerulesep=5pt
\begin{figure*}
\includegraphics[width=1\linewidth]{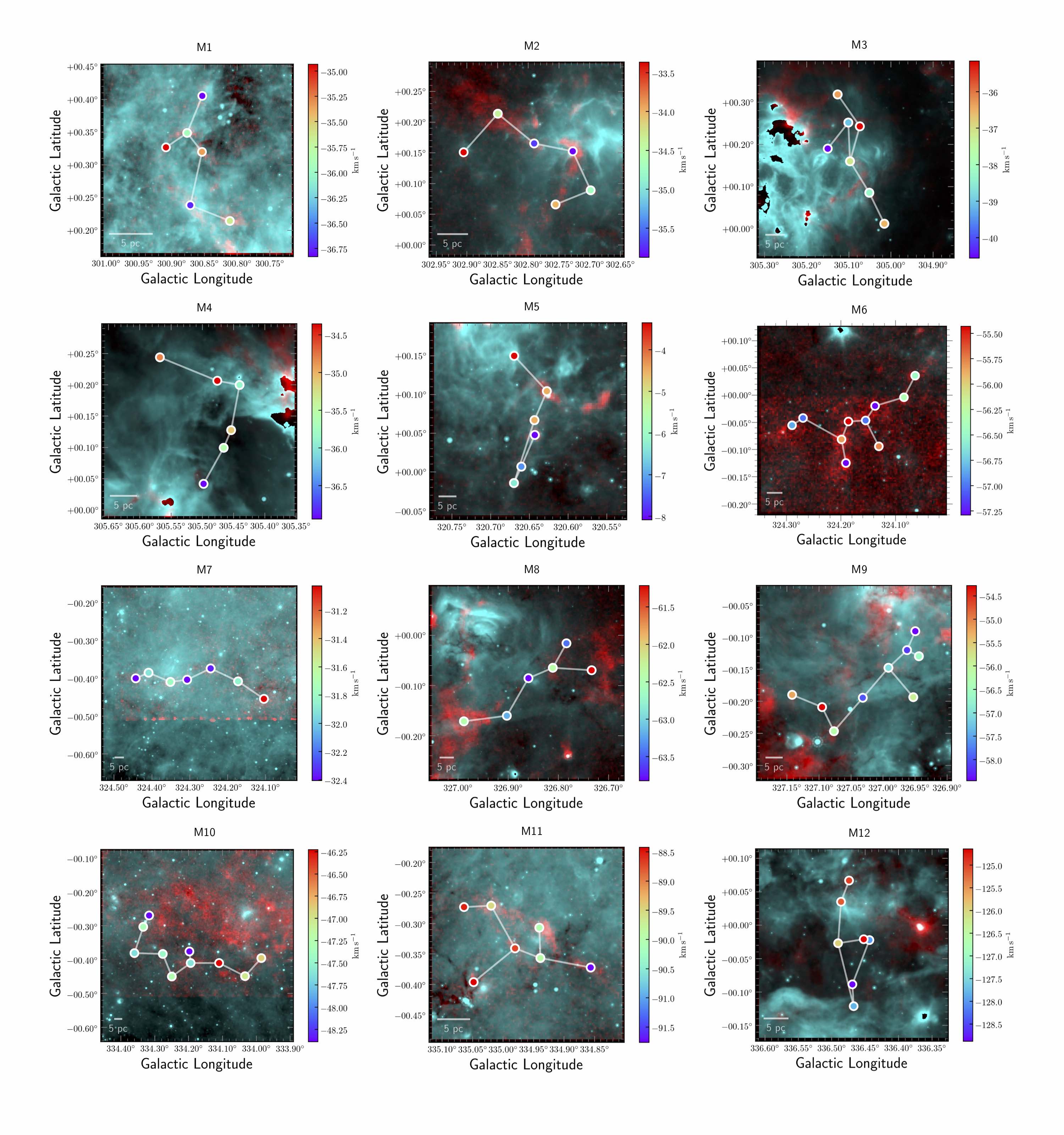}
\caption{Two-color view of filaments. The color-coded circles denote leaves in filaments with different velocities. For M1-M55, white line segments are edges. For S56-S88, white curves are their medial axis. For backgrounds, cyan represents intermediate infrared 24 $\mu$m emission on logarithmic scale from MIPSGAL \citep{Carey2009} and red shows s integrated $^{13}$CO(2-1) emission.}
\label{fig:ap_twocolor}
\end{figure*}
}}
\end{center}

\setcounter{figure}{0} 
\begin{center}
\setlength{\tabcolsep}{1.2mm}{
{
\doublerulesep=5pt
\begin{figure*}
\includegraphics[width=1\linewidth]{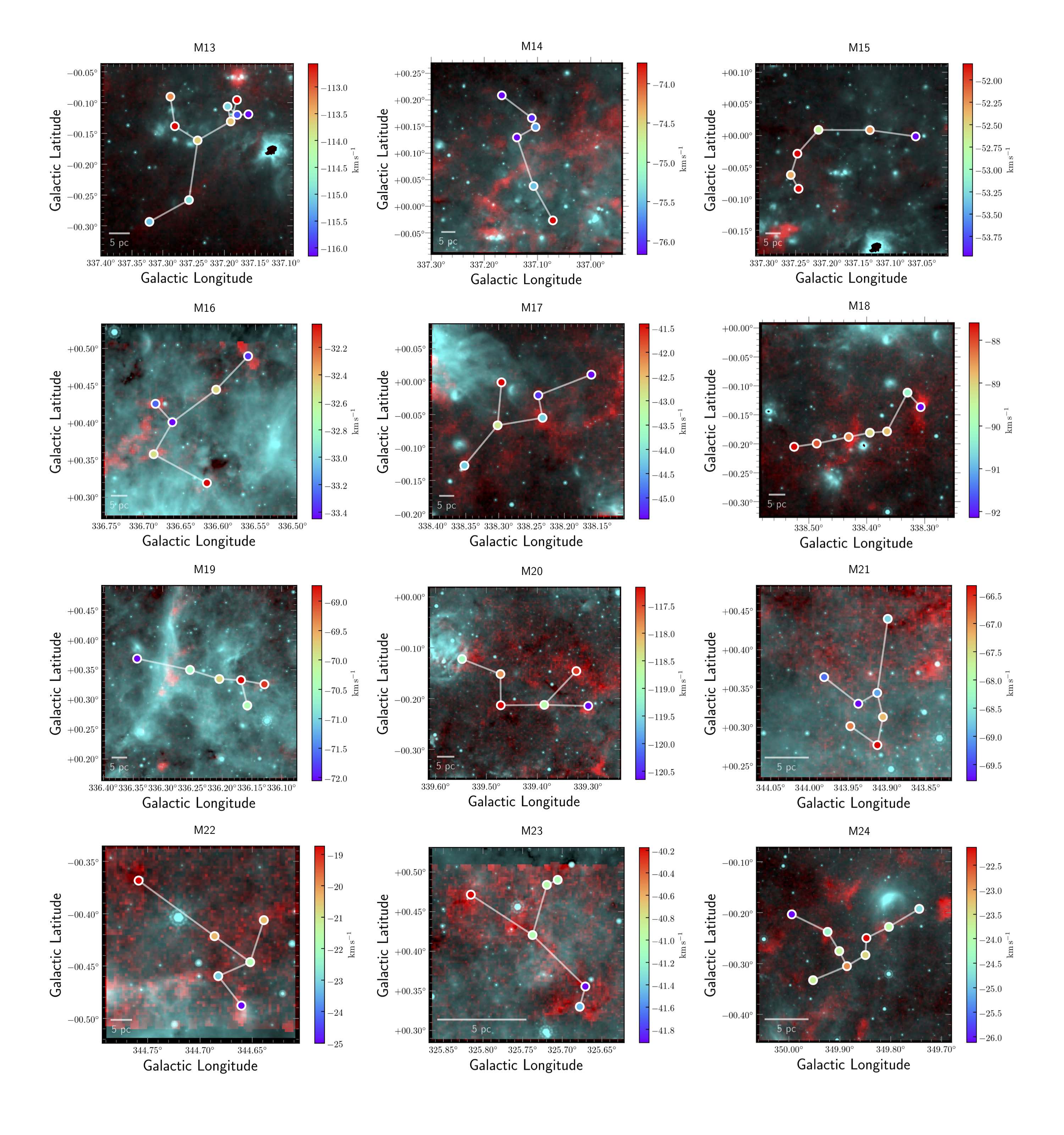}
\caption{Continued: M13-M24}
\end{figure*}
}}
\end{center}

\setcounter{figure}{0} 
\begin{center}
\setlength{\tabcolsep}{1.2mm}{
{
\doublerulesep=5pt
\begin{figure*}
\includegraphics[width=1\linewidth]{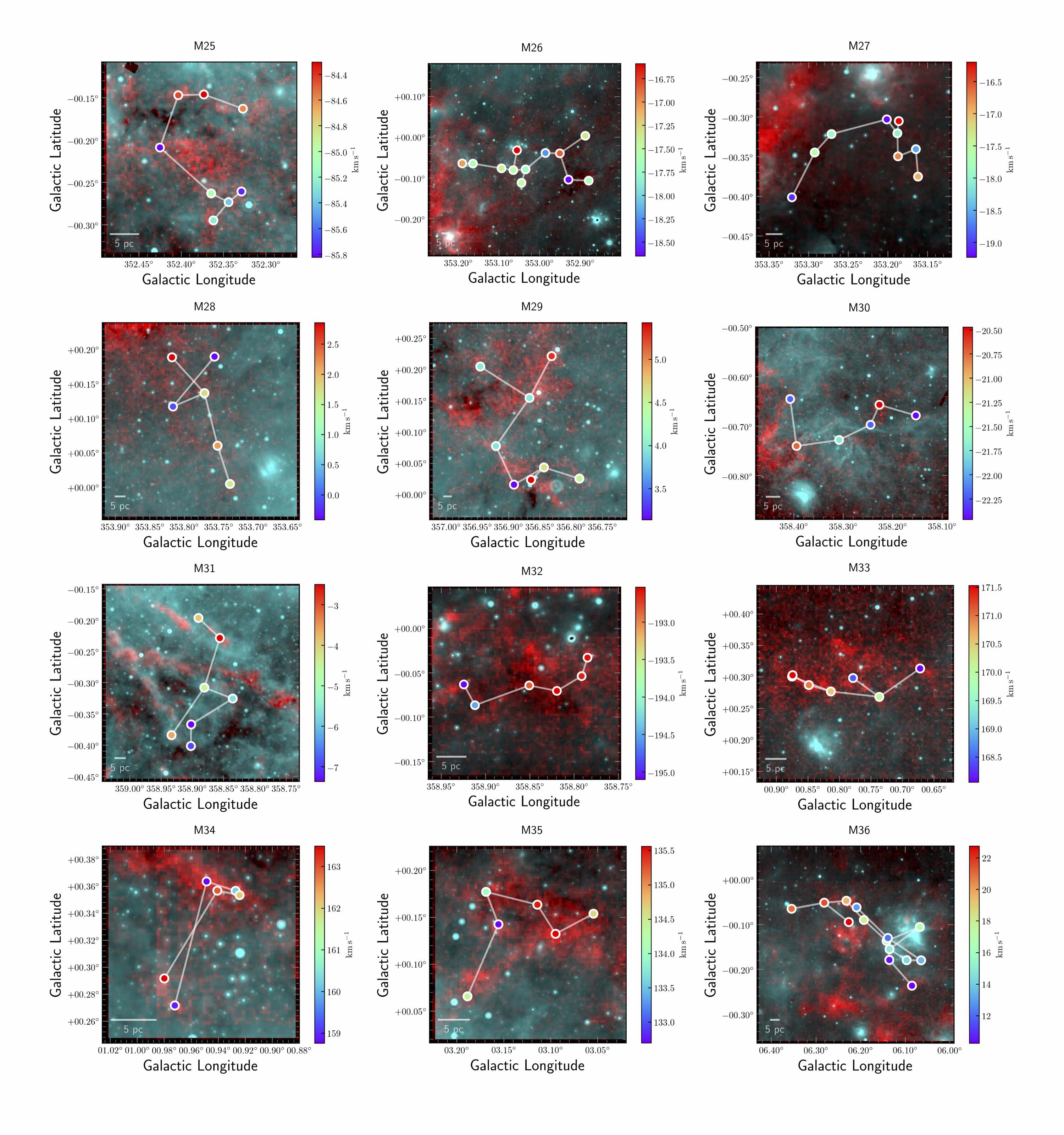}
\caption{Continued: M25-M36}
\end{figure*}
}}
\end{center}

\setcounter{figure}{0} 
\begin{center}
\setlength{\tabcolsep}{1.2mm}{
{
\doublerulesep=5pt
\begin{figure*}
\includegraphics[width=1\linewidth]{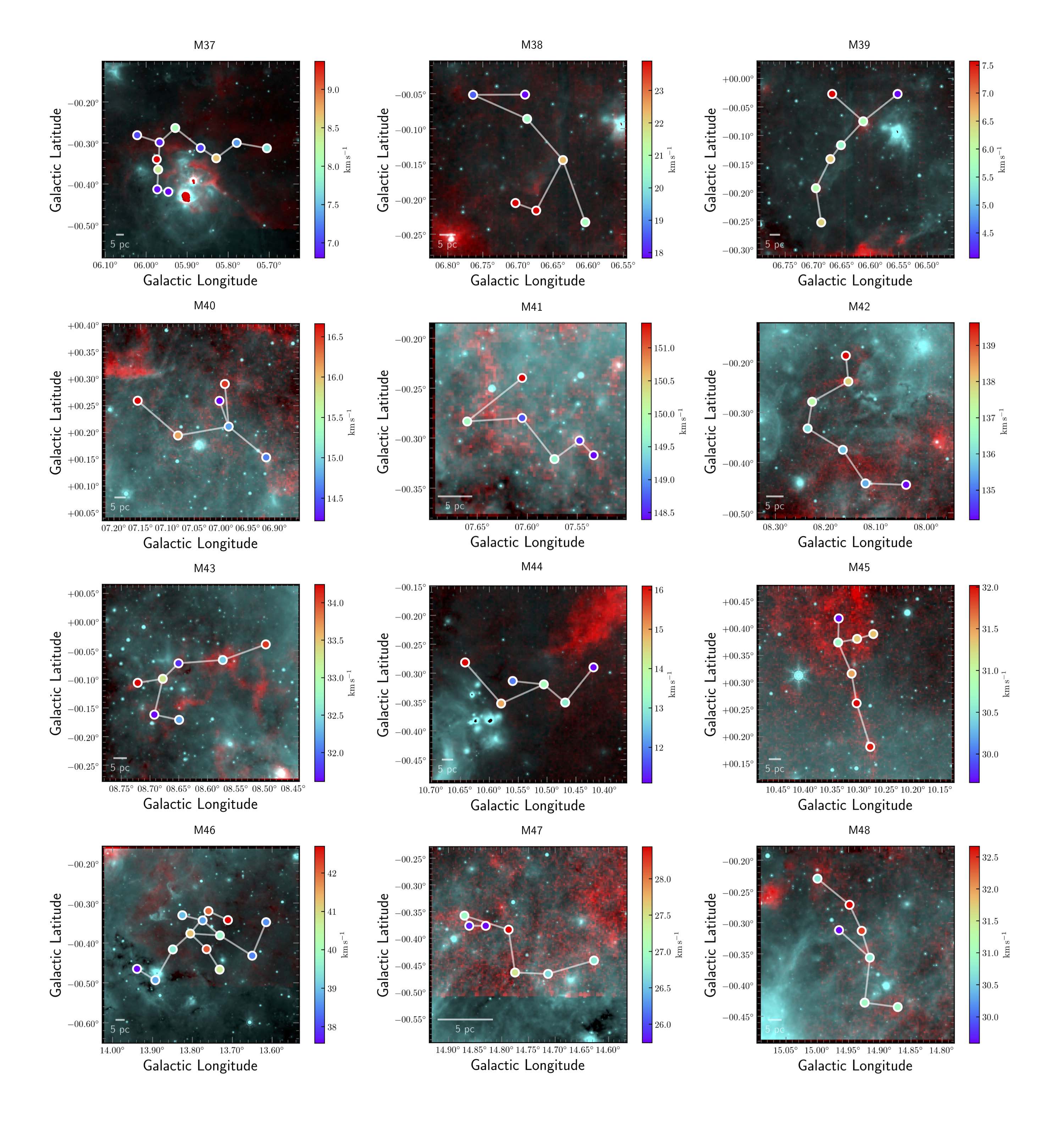}
\caption{Continued: M37-M48}
\end{figure*}
}}
\end{center}

\setcounter{figure}{0} 
\begin{center}
\setlength{\tabcolsep}{1.2mm}{
{
\doublerulesep=5pt
\begin{figure*}
\includegraphics[width=1\linewidth]{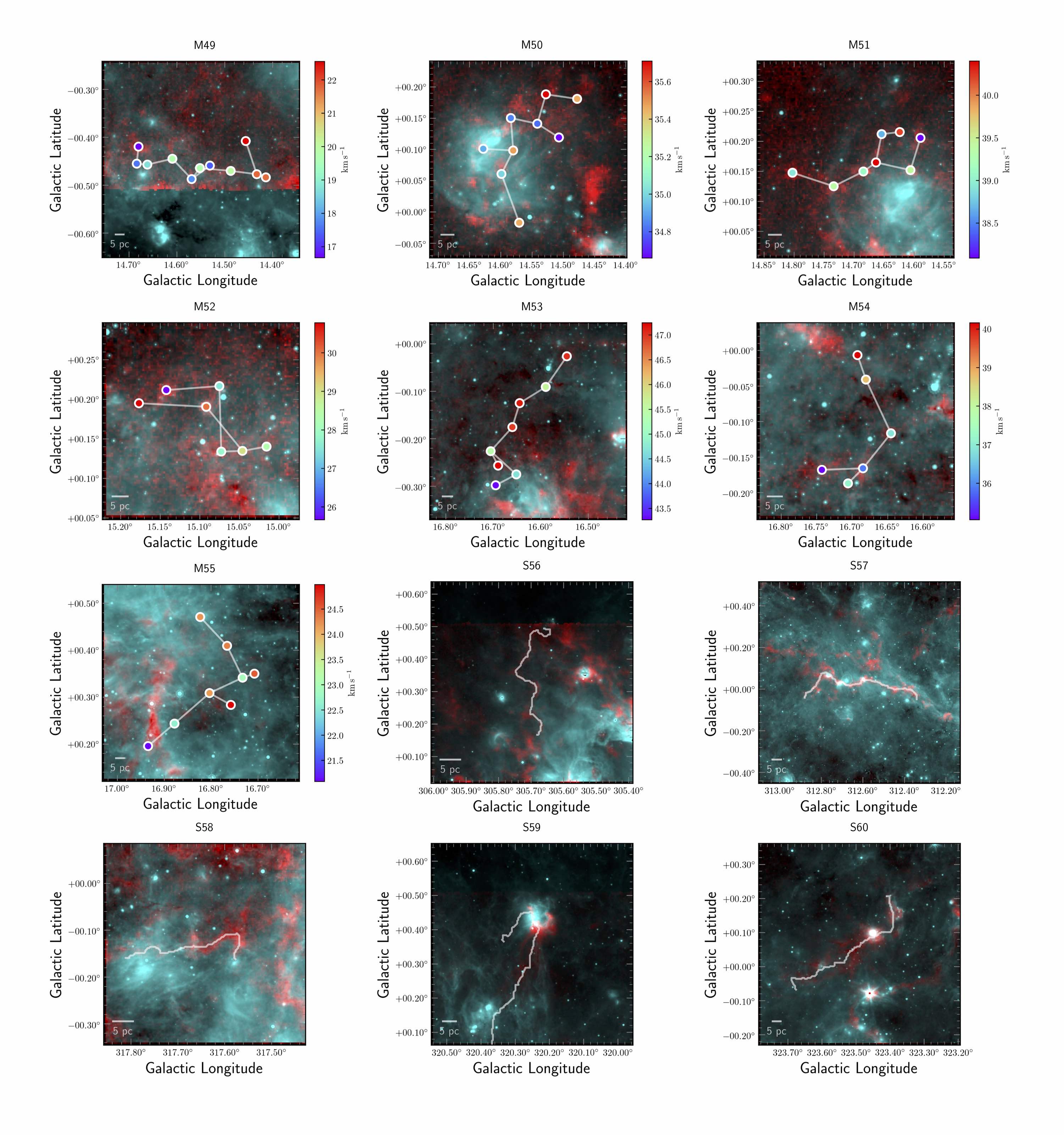}
\caption{Continued: M49-M55 \& S56-S60}
\end{figure*}
}}
\end{center}

\setcounter{figure}{0} 
\begin{center}
\setlength{\tabcolsep}{1.2mm}{
{
\doublerulesep=5pt
\begin{figure*}
\includegraphics[width=1\linewidth]{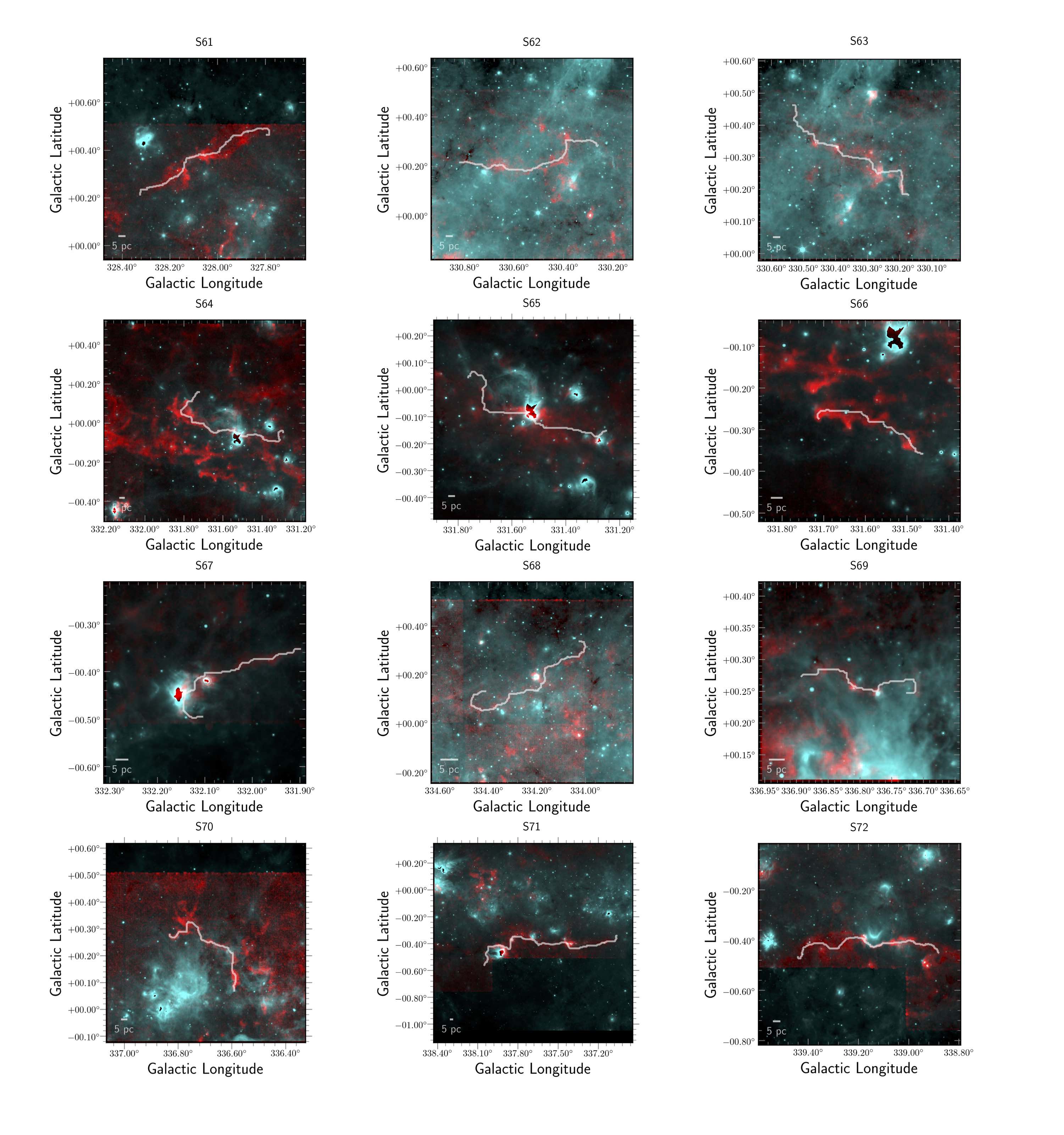}
\caption{Continued: S61-S72}
\end{figure*}
}}
\end{center}

\setcounter{figure}{0} 
\begin{center}
\setlength{\tabcolsep}{1.2mm}{
{
\doublerulesep=5pt
\begin{figure*}
\includegraphics[width=1\linewidth]{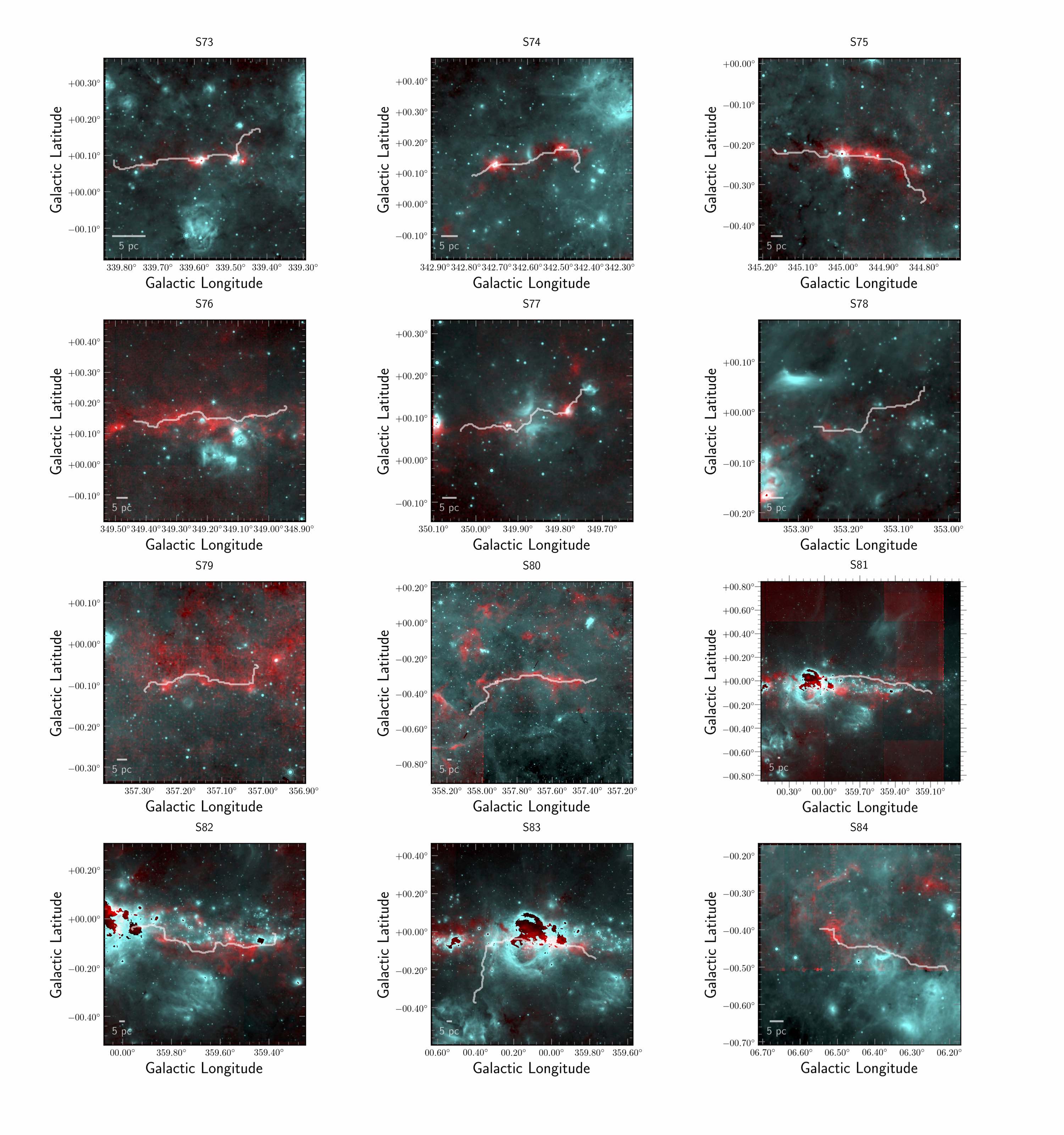}
\caption{Continued: S73-S84}
\end{figure*}
}}
\end{center}

\setcounter{figure}{0} 
\begin{center}
\setlength{\tabcolsep}{1.2mm}{
{
\doublerulesep=5pt
\begin{figure*}
\includegraphics[width=1\linewidth]{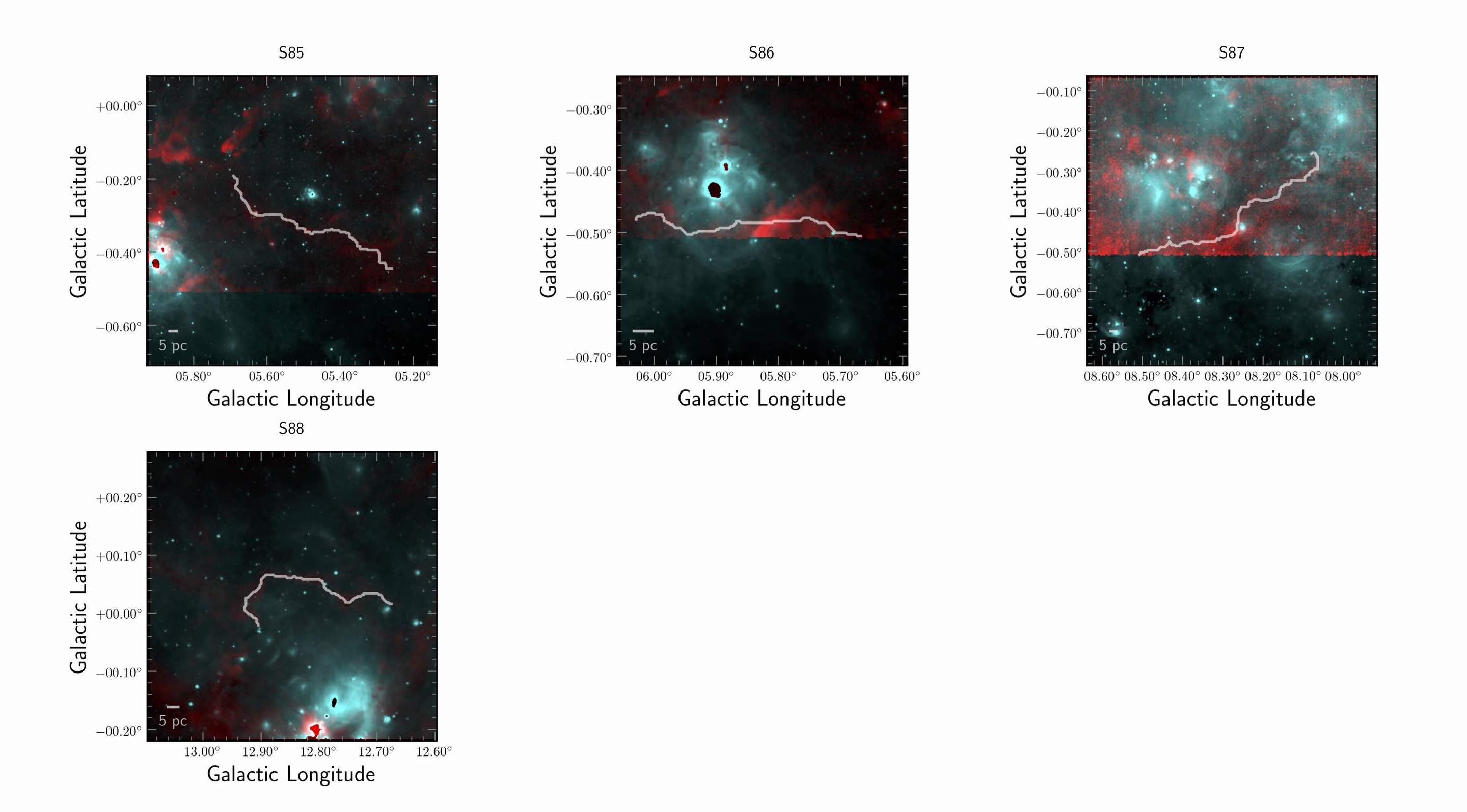}
\caption{Continued: S85-S88}
\end{figure*}
}}
\end{center}

\end{document}